\documentclass{amsart} 
\usepackage{graphicx}
\usepackage{amssymb, amsmath}
\usepackage{amsfonts}
\usepackage{amssymb}
\usepackage{sidecap}

\usepackage{float}

\newtheorem{theorem}{Theorem}

\newtheorem{definition}[theorem]{Definition}

\newcommand{\C}{\mathbb C}

\newcommand{\Z}{\mathbb Z}

\newcommand{\R}{\mathbb R}

\def\T{\mathbb T}
\def\la{\label}
\def\bt{\begin{thm}}
\def\et{\end{thm}}

\def\bl{\begin{lem}}
\def\el{\end{lem}}

\def\bd{\begin{defi}}
\def\ed{\end{defi}}

\def\bc{\begin{cor}}
\def\ec{\end{cor}}

\def\bp{\begin{proof}}
\def\ep{\end{proof}}

\def\br{\begin{rem}}
\def\er{\end{rem}}

\newtheorem{thm}{Theorem}[section]
\newtheorem{lem}{Lemma}[section]
\newtheorem{defi}{Definition}[section]

\newtheorem{rem}{Remark}[section]
\newtheorem{cor}{Corollary}[section]

\numberwithin{equation}{section}
\numberwithin{theorem}{section}
\numberwithin{example}{section}

\numberwithin{figure}{section}

\begin{document}

\title{Cahn-Hilliard Equations  and Phase Transition Dynamics  for Binary Systems}

\author[Ma]{Tian Ma}
\address[TM]{Department of Mathematics, Sichuan University,
Chengdu, P. R. China}

\author[Wang]{Shouhong Wang}
\address[SW]{Department of Mathematics,
Indiana University, Bloomington, IN 47405}
\email{showang@indiana.edu}

\thanks{The work was supported in part by the
Office of Naval Research and by the National Science Foundation.}


\begin{abstract} The process of phase separation of binary systems is described by the Cahn-Hilliard equation.  The main objective of this article is to give a classification on the dynamic phase transitions for binary systems using either the classical Cahn-Hilliard equation or the Cahn-Hilliard equation coupled with entropy, leading to some interesting physical predictions. 
The analysis is based on dynamic transition theory for nonlinear systems and new classification scheme for dynamic transitions, developed recently by the authors. 
\end{abstract}
\maketitle
\tableofcontents

\section{Introduction}
\label{sc1}
Cahn-Hilliard equation  describes the process of phase separation, by which the two components of a binary fluid spontaneously separate and form domains pure in each component. The main objective of this article is to provide a theoretical approach to dynamic phase transitions for binary systems.

Classically, phase transitions are classified by  the Ehrenfest classification scheme,  based on  the lowest derivative of the free energy that is discontinuous at the transition. 
In general, it is a difficult task to classify phase transitions of higher order, which appears in many equilibrium phase transition systems, such as the PVT system, the ferromagnetic system, superfluids as well as the binary systems studied in this article. 

For this purpose, a new  dynamic transition theory is developed recently by the authors. This new theory provides an efficient tool to analyze phase transitions of higher order. With this theory in our disposal, a new  dynamic  classification scheme is obtained, and classifies  phase transitions into three categories: Type-I, Type-II and Type-III, corresponding mathematically  to continuous,  jump, and  mixed transitions, respectively; see the Appendix as well as two recent books by the authors \cite{b-book, chinese-book} for details. 

There have been extensive studies in the past on the dynamics of 
the Cahn-Hilliard equations. However, very little is known about the higher order transitions encountered for this problem, and this article gives a complete classification of the dynamics transitions for binary systems. The results obtained lead in particular to various physical predictions. First,  the order of  phase transitions is precisely determined by the sign of  a nondimensional 
parameter $K$ such that if $K>0$, the transition is first-order with latent heat and if $K <0$, the transition is second-order. Second, a theoretical transition diagram is derived, leading in particular  to  a prediction that there is only second-order transition for molar fraction near $1/2$. This is different from the prediction made by the classical transition diagram. Third, a critical length scale  is derived such that no phase separation occurs at any temperature if the length of the container is smaller than the critical length scale. These physical predictions will be addressed in another article.

This article is organized as follows. In Section 2, both the classical Cahn-Hilliard equation and the Cahn-Hilliard equation coupled with entropy are introduced in a unified fashion using a general principle for equilibrium phase transitions outlined in Appendix B. Sections 3-6 analyze dynamic transitions for 
the Cahn-Hilliard equation in general domain, rectangular domain, with periodic boundary conditions, and for the Cahn-Hilliard equation coupled with entropy. 
 Physical conclusions are given in Section 7, and the dynamic transition theory is recalled in  Appendix A.

\section{Dynamic Phase Transition  Models for Binary Systems}
Materials compounded by two components $A$ and $B$, such as  binary
alloys, binary solutions and polymers, are called binary
systems. Sufficient cooling of a binary system may lead to phase
separations, i.e., at the critical temperature, the concentrations
of both components $A$ and $B$ with homogeneous distribution undergo  changes,
leading to heterogeneous distributions in space. Phase separation
of binary systems observed will be in one of two main ways. The first is by nucleation in which sufficiently large nuclei of the
second phase appear randomly and grow, and this corresponds to Type-II phase transitions. The second is  by spinodal
decomposition in which the systems appear to nuclear at once, and
periodic or semi-periodic structure is seen, and this corresponds to Type-I phase transitions. 

Since binary systems are conserved, the equations describing the
Helmholtz process and the Gibbs process are the same. Hence,
without distinction we use the term "free energy" to discuss this
problem. 

Let $u_A$   and $u_B$ be the concentrations of components $A$  and  $B$ respectively,  then $u_B=1-u_A$.  In a homogeneous state, $u_B=\bar{u}_B$ is a constant, and the
entropy density $S_0=\bar{S}_0$ is also a constant. We take $u, S$
the concentration and entropy density deviations:
$$u=u_B-\bar{u}_B,\ \ \ \ S=S_0-\bar{S}_0$$
By (\ref{7.28}) and (\ref{7.29}), the free energy is given
by
\begin{equation}
F(u,S)=F_0+\int_{\Omega} \Big[\frac{\mu_1}{2}|\nabla u|^2+\frac{\mu_2}{2}|\nabla
S|^2+\frac{\beta_1}{2}|S|^2+\beta_0Su\label{8.49}
 +\beta_2Su^2+\sum^{2p}_{k=1}\alpha_ku^k\Big]dx.
\end{equation}
Since  entropy is increasing as $u\rightarrow 0$, and by
$$
\frac{\delta}{\delta S}F(u,S)=-\mu_2\Delta
S+\beta_1S+\beta_0u+\beta_2u^2=0,
$$ 
we have 
\begin{equation}
\beta_0=0,\ \ \ \ \beta_1>0,\ \ \ \ \beta_2>0,\label{8.50}
\end{equation}
which implies that $S$ is a decreasing function of $|u|$.

According (\ref{7.38}) and (\ref{7.32}),  we derive from (\ref{8.49}) and
(\ref{8.50})  the  following equations governing a binary system:
\begin{equation}
\left.
\begin{aligned} &\frac{\partial S}{\partial t}=k_1\Delta
S-a_1S-a_2u^2,\\
&\frac{\partial u}{\partial t}=-k_2\Delta^2u+b_0\Delta (Su)+\Delta
f(u),\\
&\int_{\Omega}u(x,t)dx=0,
\end{aligned}
\right.\label{8.51}
\end{equation}
where $k_1,k_2,b_0,a_1,a_2$ are positive constants, and
\begin{equation}
f(u)=\sum^{2p-1}_{k=1}b_ku^k,\ \ \ \ b_{2p-1}>0 \ \ \ \ (p\geq
2).\label{8.52}
\end{equation}

Physically sound boundary condition for (\ref{8.51}) is either the Neumann boundary condition:
\begin{equation}
\frac{\partial u}{\partial n}=\frac{\partial\Delta u}{\partial
n}=0,\ \ \ \ \frac{\partial S}{\partial n}=0\ \ \ \ \text{on}\
\partial\Omega ,\label{8.53}
\end{equation}
with  $\Omega\subset \R^n$  $(1\leq n\leq 3)$ being  a bounded domain, or 
the periodic boundary condition:
\begin{equation}
u(x+KL)=u(x),\ \ \ \ S(x+KL)=S(x)\label{8.54}
\end{equation}
with  $\Omega =[0,L]^n,K=(k_1,\cdots,k_n),1\leq n\leq 3.$

\medskip

For simplicity, in this section we always assume that $p=2$. Thus,
function (\ref{8.52}) is rewritten as
\begin{equation}
f(u)=b_1u+b_2u^2+b_3u^3,\ \ \ \ b_3>0.\label{8.55}
\end{equation}
Based on Theorem \ref{t5.1}, we have to assume that there exists a
temperature $T_1>0$ such that $b_1=b_1(T)$ satisfies
\begin{equation}
b_1(T)\left\{
\begin{aligned}
&  >0&&  \quad \text{  if } T>T_1,\\
& =0&&  \quad \text{ if } T=T_1,\\
& <0&&  \quad \text{ if } T<T_1.
\end{aligned}
\right.\label{8.56}
\end{equation}


If we ignore the coupled action of entropy in (\ref{8.49}), then
the free energy $F$ is in the following form
$$F(u)=F_0+\int_{\Omega}\left[\frac{\mu_1}{2}|\nabla
u|^2+\frac{\alpha_1}{2}u^2+\frac{\alpha_2}{3}u^3+\frac{\alpha_3}{4}u^4\right]dx,$$
and the equations (\ref{8.51}) are the following classical  Cahn-Hilliard
equation:
\begin{equation}
\left.
\begin{aligned} 
&\frac{\partial u}{\partial
t}=-k\Delta^2u+\Delta f(u),\\
&\int_{\Omega}u(x,t)dx=0,
\end{aligned}
\right.\label{8.57}
\end{equation}
where $f(u)$ is as in (\ref{8.55}).

\section{Phase Transition in General Domains}
\la{s8.2.3}
In this section, we shall discuss 
the Cahn-Hilliard equation from the mathematical point of view. We start with  the nondimensional form of equation. Let
\begin{align*}
&x=lx^{\prime}, &&  t=\frac{l^4}{k}t^{\prime}, && u=u_0u^{\prime},\\
&\lambda =-\frac{l^2b_1}{k},&& \gamma_2=\frac{l^2b_2u_0}{k},
&& \gamma_3=\frac{l^2b_3u^2_0}{k},
\end{align*}
where $l$ is a given length,  $u_0=\bar{u}_B$ is the constant
concentration of $B$, and $\gamma_3>0$. Then the equation (\ref{8.57}) can be
rewritten as follows (omitting the primes)
\begin{equation}
\left.
\begin{aligned} 
&\frac{\partial u}{\partial
t}=-\Delta^2u-\lambda\Delta u+\Delta (\gamma_2u^2+\gamma_3u^3),\\
&\int_{\Omega}u(x,t)dx=0,\\
&u(x,0)=\varphi .
\end{aligned}
\right.\label{8.58}
\end{equation}

Let
$$
 H=\left\{u\in L^2(\Omega )\ |\ \int_{\Omega}udx=0\right\}.$$
For the Neumann boundary condition (\ref{8.53}) we define
$$
H_1=\left\{u\in H^4(\Omega )\cap H\ \Big|\ \frac{\partial u}{\partial
n}=\frac{\partial\Delta u}{\partial
n}=0\ \text{ on } \partial \Omega \right\},$$ 
and for the periodic boundary
condition (\ref{8.54}) we define
$$H_1=\left\{u\in H^4(\Omega )\cap H\ \Big|\ u(x+KL)=u(x) \ \ \forall
K\in \Z^n\right\}.$$

Then we define the operators $L_{\lambda}=-A+B_{\lambda}$ and
$G:H_1\rightarrow H$ by
\begin{equation}
\left.
\begin{aligned} 
& Au=\Delta^2u,\\
& B_{\lambda}u=-\lambda\Delta u,\\
& G(u)=\gamma_2\Delta u^2+\gamma_3\Delta u^3.
\end{aligned}
\right.\label{8.59}
\end{equation}

Thus, the Cahn-Hilliard equation (\ref{8.58}) is equivalent to the
following operator equation
\begin{equation}
\left.
\begin{aligned} &\frac{du}{dt}=L_{\lambda}u+G(u), \\
&u(0)=\varphi
\end{aligned}
\right.\label{8.60}
\end{equation}
It is known that the operators defined by (\ref{8.59}) satisfy the
conditions (\ref{5.2}) and (\ref{5.3}).

We first consider the case where $\Omega\subset \R^n(1\leq n\leq
3)$ is a general bounded and smooth domain. Let $\rho_k$ and $e_k$ be the
eigenvalues and eigenfunctions of the following eigenvalue problem:
\begin{equation}
\left.
\begin{aligned} 
 &-\Delta e_k=\rho_ke_k,\\
&\frac{\partial e_k}{\partial n}|_{\partial\Omega}=0,\\
&\int_{\Omega}e_kdx=0.
\end{aligned}
\right.\label{8.61}
\end{equation}
The eigenvalues of (\ref{8.61}) satisfy
$ 0<\rho_1\leq\rho_2\leq\cdots\leq\rho_k\leq\cdots,$  and $ \lim_{k \to \infty} \rho_k= \infty$. The eigenfunctions $\{e_k\}$ of (\ref{8.61})
constitute an orthonormal  basis of $H$. Furthermore,  the
eigenfunctions of (\ref{8.61}) satisfy
$$\frac{\partial\Delta e_k}{\partial n}|_{\partial\Omega}=0,\ \ \
\ k=1,2,\cdots .$$ 
Hence, $\{e_k\}$ is also an orthogonal basis of
$H_1$ under the following equivalent norm
$$\|u\|_1=\left[\int_{\Omega}|\Delta^2u|^2dx\right]^{{1}/{2}}.$$

We are now in position to give a phase transition theorem 
for the problem (\ref{8.58}) with the following
Neumann boundary condition:
\begin{equation}
\frac{\partial u}{\partial n} =0,\ \ \ \
\frac{\partial\Delta u}{\partial
n}=0 \qquad \text{ on } \partial \Omega.\label{8.62}
\end{equation}

\bt\la{t8.3}
Assume that $\gamma_2=0$ and $\gamma_3>0$ in
(\ref{8.58}), then the following assertions hold true:

\begin{enumerate}

\item[(1)] If the first eigenvalue $\rho_1$ of (\ref{8.61}) has
multiplicity $m\geq 1$, then the problem (\ref{8.58}) with
(\ref{8.62}) bifurcates from $(u,\lambda )=(0,\rho_1)$ on $\lambda
>\rho_1$ to an attractor $\Sigma_{\lambda}$, homeomorphic to an
$(m-1)$-dimensional sphere $S^{m-1}$, and $\Sigma_{\lambda}$
attracts $H \setminus \Gamma$,  where $\Gamma$ is the stable manifold of $u=0$
with codimension $m$. 

\item[(2)] $\Sigma_{\lambda}$ contains at
least $2m$ singular points. If $m=1, \Sigma_{\lambda}$ has exactly
two steady states $\pm u_{\lambda}$, and if $m=2,
\Sigma_{\lambda}=S^1$ has at most eight singular points.

\item[(3)]  Each singular point $u_{\lambda}$ in
$\Sigma_{\lambda}$ can be expressed as
$$u_{\lambda}=\left(\lambda
-\rho_1\right)^{{1}/{2}} w +o(|\lambda
-\rho_1|^{{1}/{2}}),$$ 
where $w$ is an eigenfunction corresponding to the first eigenvalue of  (\ref{8.61}).
\end{enumerate}
\et

\bp
We proceed in several steps as follows.

\medskip

{\sc Step 1.} It is clear that the eigenfunction $\{e_k\}$ of
(\ref{8.61}) are also eigenvectors of the linear operator
$L_{\lambda}=-A+B_{\lambda}$ defined by (\ref{8.59}) and the
eigenvalues of $L_{\lambda}$ are given by
\begin{equation}
\beta_k(\lambda )=\rho_k(\lambda -\rho_k),\ \ \ \ k=1,2,\cdots
.\label{8.63}
\end{equation}
It is easy to verify the conditions (\ref{5.4}) and (\ref{5.5})
in our case  at $\lambda_0=\rho_1$. We shall prove this theorem
using the attractor bifurcation theory introduced in \cite{b-book}.

We need to verify that $u=0$ is a global asymptotically stable
singular point of (\ref{8.60}) at $\lambda =\rho_1$. By
$\gamma_2=0$, from the energy integration of (\ref{8.58}) we can
obtain
\begin{equation}
\frac{1}{2}\frac{d}{dt}\int_{\Omega}u^2dx=\int_{\Omega}\left[-|\Delta
u|^2+\rho_1|\nabla u|^2-3\gamma_3u^2|\nabla
u|^2\right]dx\label{8.64}
\end{equation}
$$\leq -C\int_{\Omega}|\Delta
v|^2dx-3\gamma_3\int_{\Omega}u^2|\nabla u|^2dx,$$ where $C>0$ is a
constant, $u=v+ w $, and $\int_{\Omega}v wdx=0$, and $w$ is a first engenfunction. It follows from
(\ref{8.64}) that $u=0$ is global asymptotically stable. 
Hence, 
for Assertion (1), we only have to prove that  
$\Sigma_{\lambda}$ is homeomorphic to $S^{m-1}$, as the rest of  this assertion follows directly from the attractor bifurcation theory introduced in \cite{b-book}.

\bigskip

{\sc Step 2.} Now we prove that the bifurcated attractor $\Sigma_{\lambda}$
from $(0,\rho_1)$ contains at least $2m$ singular points.

Let $g(u)=-\Delta u-\lambda u+\gamma_3u^3$. Then the stationary
equation of (\ref{8.58})  is given by 
\begin{align*}
&\Delta g(u)=0,\\
&\int_{\Omega}udx=0,
\end{align*}
which is equivalent, by the maximum principle,  to
\begin{equation}
\left.
\begin{aligned} 
&-\Delta u-\lambda
u+\gamma_3u^3=\text{constant},\\
&\int_{\Omega}udx=0,\\
&\frac{\partial u}{\partial n}|_{\partial\Omega}=0.
\end{aligned}
\right.\label{8.65}
\end{equation}
By the Lagrange multiplier theorem, (\ref{8.65}) is the Euler
equation of the following functional with zero average constraint:
\begin{align}
& F(u)=\int_{\Omega}\left[\frac{1}{2}|\nabla
u|^2-\frac{\lambda}{2}u^2+\frac{\gamma_3}{4}u^4\right]dx,\label{8.66}
\\
& u\in \left\{u\in H^1(\Omega )\cap H\ \Big|\ \
\frac{\partial u}{\partial n}|_{\partial\Omega}=0\quad \int_{\Omega}udx=0\right\}. \nonumber 
\end{align}
 Since $F$ is an even functional, 
 by the classical Krasnoselskii bifurcation theorem for even functionals, 
 (\ref{8.66}) bifurcates from
$\lambda >\rho_1$ at least to $2m$ mini-maximum points, i.e., 
equation (\ref{8.65}) has at least $2m$ bifurcated solutions on
$\lambda >\rho_1$. Hence, the attractor $\Sigma_{\lambda}$
contains at least $2m$ singular points.

\bigskip

{\sc Step 3.} 
To complete the proof, we reduce the equation (\ref{8.60}) to the
center manifold near $\lambda =\rho_1$. 
By the approximation formula given in \cite{b-book},  
the reduced equation of (\ref{8.60}) is given by:
\begin{equation}
\frac{dx_i}{dt}=\beta_1(\lambda
)x_i-\gamma_3\rho_1\int_{\Omega}v^3e_{1i}dx  + o(|x|^3)\qquad \text{ for }  1\leq i\leq
m,\label{8.67}
\end{equation}
where $\beta_1(\lambda )=\rho_1(\lambda -\rho_1)$, 
$v=\sum^m_{i=1}x_ie_{1i}$, 
and $\{e_{11},\cdots,e_{1m}\}$ are the first eigenfunctions
of (\ref{8.61}). Equations (\ref{8.67}) can be rewritten as
\begin{equation}
\frac{dx_i}{dt}=\beta_1(\lambda
)x_i-\gamma_3\rho_1\int_{\Omega}\left(\sum^m_{j=1}x_je_{1j}\right)^3e_{1i}dx+o(|x|^3).\label{8.68}
\end{equation}
Let
$$ g(x)=\left(\int_{\Omega}v^3e_{11}dx,\cdots,\int_{\Omega}v^3e_{1m}dx\right),
\qquad  v(x)=\sum^m_{j=1}x_je_{1j}.
$$
 Then for any $x\in \R^m$,
\begin{equation}\label{8.69}
<g(x),x>=\sum^m_{i=1}x_i\int_{\Omega}v^3e_{1i}dx
=\int_{\Omega}v^4dx \geq C|x|^4,
\end{equation}
for some constant $C>0$.

Thus by the attractor bifurcation theorem \cite{b-book}, 
it follows from (\ref{8.68}) and (\ref{8.69}) 
 that the attractor $\Sigma_{\lambda}$ is homeomorphic to
$S^{m-1}$. Hence, Assertion (1) is proved. The other conclusions
in Assertions (2) and (3) can be derived from (\ref{8.68}) and
(\ref{8.69}). The proof is complete.
\qed
\ep

Physically, the coefficients $\gamma_2$ and $\gamma_3$ depend on
$u_0=\bar{u}_B$, the temperature $T$, and the pressure $p$:
$$
\gamma_k=\gamma_k(u_0,T,p), \qquad k=2,3.
$$
The set of points satisfying
$\gamma_2(u_0,T,p)=0$ has measure zero in $(u_0,T,p)\in \R^3$.
Hence, it is more interesting to consider the case where
$\gamma_2\neq 0$.

For this purpose, let the multiplicity of the first eigenvalue $\rho_1$ of
(\ref{8.61}) be $m\geq 1$, and $\{e_1,\cdots,e_m\}$ be the first
eigenfunctions. We introduce the following quadratic equations
\begin{equation}
\left.
\begin{aligned} 
& \sum^m_{i,j=1}a^k_{ij}x_ix_j=0\quad &&\text{ for } 1\leq
k\leq m,\\
& a^k_{ij}=\int_{\Omega}e_ie_je_kdx.
\end{aligned}
\right.\label{8.70}
\end{equation}

\bt\la{t8.4}
Let $\gamma_2\neq 0$,  $\gamma_3>0$, and
$x=0$ be an isolated singular point of (\ref{8.70}). Then the
phase transition of (\ref{8.58}) and (\ref{8.62}) is either
Type-II or Type-III. Furthermore, the problem (\ref{8.58}) with
(\ref{8.62}) bifurcates to at least one singular point on each side
of $\lambda =\rho_1$, and has a saddle-node bifurcation on
$\lambda <\rho_1$. In particular,  if $m=1$, then the following
assertions hold true:

\begin{enumerate}

\item[(1)] The phase transition is Type-III, and a neighborhood
$U\subset H$ of $u=0$ can be decomposed into two sectorial regions
$\bar{U}=\bar{D}_1(\pi )+\bar{D}_2(\pi )$ such that the phase
transition in $D_1(\pi )$ is the first order, and in $D_2(\pi )$
is the $n$-th order with $n\geq 3$. 

\item[(2)] The bifurcated
singular point $u_{\lambda}$ on $\lambda >\rho_1$ attracts
$D_2(\pi )$, which can be expressed as
\begin{equation} 
 u_{\lambda}=(\lambda
-\rho_1)e_1/\gamma_2a+o(|\lambda -\rho_1|),\label{8.71}
\end{equation}
where, by assumption,   
$a=\int_{\Omega}e^3_1dx  \neq 0$.

\item[(3)] When $|\gamma_2a|=\varepsilon$ is small, the
assertions in the transition perturbation theorems (Theorems \ref{t6.10}
and \ref{t6.11}) hold true.
\end{enumerate}
\et

\br\la{r8.4}
{\rm
We shall see later that when $\Omega$ is a
rectangular domain,  i.e., 
$$\Omega =\Pi^n_{k=1}(0,L_k)\subset \R^n
\qquad (1\leq n\leq 3),$$ 
then $a=\int_{\Omega} e_1^3 dx =0$. However, for almost all non-rectangular
domains $\Omega$,  the first eigenvalues are
simple and $a\neq 0$. Hence, the Type-III phase transitions for
general domains are generic.\qed
}
\er

\bp[Proof of Theorem \ref{t8.4}]
Assertions (1)-(3) can be directly
proved using Theorems \ref{t5.9}, \ref{t6.10}, and \ref{t6.11}. 
By assumption, $u=0$ is a
second order non-degenerate singular point of (\ref{8.60}) at
$\lambda =\rho_1$, which implies that $u=0$ is not locally
asymptotically stable. Hence, it follows from Theorems \ref{t5.3}  and
the steady state bifurcation theorem for  even-order
nondegenerate singular points \cite{b-book}  that the phase transition of (\ref{8.58}) with (\ref{8.62})
is either Type-II or Type-III, and there is at least one singular
point bifurcated on each side of $\lambda =\rho_1$.

Finally, we shall apply Theorem \ref{t6.4} to prove that there exists a
saddle-node bifurcation on $\lambda <\rho_1$.

It is known that
$$\text{ind}(L_{\lambda}+G,0)
= \left\{
\begin{aligned}
& \text{even}&& \text{ if }  \lambda =\rho_1,\\
& 1&& \text{ if }  \lambda <\rho_1.
\end{aligned}
\right.$$ 
Moreover, since $L_{\lambda}+G$ defined by (\ref{8.59})
is a gradient-type operator,  we can
derive  that
$$\text{ind}(L_{\rho_1}+G,0)\leq 0.$$
Hence there is a bifurcated branch $\Sigma_{\lambda}$ on $\lambda
<\rho_1$ such that
$$
\text{ind}(L_{\lambda}+G,u_{\lambda})=-1
\qquad \forall u_{\lambda}\in\Sigma_{\lambda},\ \ \ \ \lambda <\rho_1.
$$ 
It is
clear that the eigenvalues (\ref{8.63}) of $L_{\lambda}$ satisfy
(\ref{5.4}) and (\ref{5.5}). For any
$\lambda\in \R^1$ (\ref{8.60}) possesses a global attractor.
Therefore, for bounded $\lambda ,b<\lambda\leq\rho_1$, the
bifurcated branch $\Sigma_{\lambda}$ is bounded:
$$\|u_{\lambda}\|_H\leq C\qquad \forall
u_{\lambda}\in\Sigma_{\lambda},\ \ \ \ -\infty
<b<\lambda\leq\rho_1.$$ 
We need to prove that there exists
$\widetilde{\lambda}<\rho_1$ such that for all $\lambda
<\widetilde{\lambda}$ equation (\ref{8.60}) has no nonzero
singular points.

By the energy estimates of (\ref{8.60}),   
for any $\lambda <\widetilde{\lambda}=-{\gamma^2_2}/{2\gamma_3}$  
and  $u\neq 0$ in $ H$, 
\begin{align*}
&\int_{\Omega}\left[|\Delta u|^2-\lambda |\nabla
u|^2+2\gamma_2u|\nabla u|^2+3\gamma_3u^2|\nabla u|^2\right]dx\\
&\geq \int_{\Omega}|\Delta u|^2dx+\int_{\Omega}|\nabla
u|^2(-\lambda -2|\gamma_2u|+3\gamma_3u^2)dx\\
&\geq \int_{\Omega}|\nabla u|^2dx+\int_{\Omega}|\nabla
u|^2(-\lambda
+\gamma_3u^2+2\gamma_3(u-\frac{\gamma_2}{\gamma_3})^2-\frac{\gamma^2_2}{2\gamma_3})dx\\
&\geq \int_{\Omega}|\nabla u|^2dx+\int_{\Omega}|\nabla
u|^2(-\lambda -\frac{\gamma^2_2}{2\gamma_3})dx\\
&> 0.
\end{align*}
Therefore, when $\lambda <\widetilde{\lambda}$,
(\ref{8.60}) has no nontrivial singular points in $H$. Thus we
infer from Theorem \ref{t6.4} that there exists a saddle-node bifurcation
on $\lambda <\rho_1$. This proof is complete. 
\ep

\section{Phase Transition in Rectangular Domains}
\la{s8.2.4}
The dynamical properties of phase separation of a binary system in
a rectangular container is very different from that in a general
container. We see in the previous section that the phase
transitions in general domains are Type-III,  and  we shall show
in the following that the phase transitions in rectangular domains
are either Type-I or Type-II, which are distinguished by a critical
size of the domains.

Let $\Omega =\Pi^n_{k=1}(0,L_k)\subset \R^n$  $ (1\leq n\leq 3)$ be a
rectangular domain. 
We first consider the case where 
\begin{equation}
L=L_1>L_j\qquad  \forall 2\leq j\leq n.\label{8.72}
\end{equation}

\bt\la{t8.5}
Let $\Omega =\Pi^n_{k=1}(0,L_k)$ satisfy
(\ref{8.72}). The following assertions hold true:

\begin{enumerate}

\item[(1)]  If $$\gamma_3<\frac{2L^2}{9\pi^2}\gamma^2_2,$$ then  the
phase transition of (\ref{8.58}) and (\ref{8.62}) at $\lambda
=\lambda_0=\pi^2/L^2$ is Type-II. In particular, the problem
(\ref{8.58}) with (\ref{8.62}) bifurcates from $(u,\lambda
)=(0,\pi^2/L^2)$ on $\lambda <\pi^2/L^2$ to exactly two equilibrium
points which are saddles, and there are two saddle-node
bifurcations on $\lambda < {\pi^2}/{L^2}$ as shown in Figure
\ref{f8.8-1}. 

\item[(2)] If $$\gamma_3>\frac{2L^2}{9\pi^2}\gamma^2_2,$$  
then the transition is Type-I. In particular, 
the problem bifurcates on $\lambda >\pi^2/L^2$ to exactly two
attractors $u^T_1$ and $u^T_2$ which can be expressed as
\begin{equation}
 u^T_{1,2}=\pm\frac{\sqrt{2}(\lambda
-\frac{\pi^2}{L^2})^{{1}/{2}}}{\sigma^{{1}/{2}}}\cos\frac{\pi
x_1}{L}+o(|\lambda -\frac{\pi^2}{L^2}|^{{1}/{2}}),\label{8.73}
\end{equation}
where 
$\sigma =\frac{3\gamma_3}{2}-\frac{L^2\gamma^2_2}{3\pi^2}.
$
\end{enumerate}
\et
\begin{SCfigure}[25][t]
  \centering
  \includegraphics[width=0.5\textwidth]{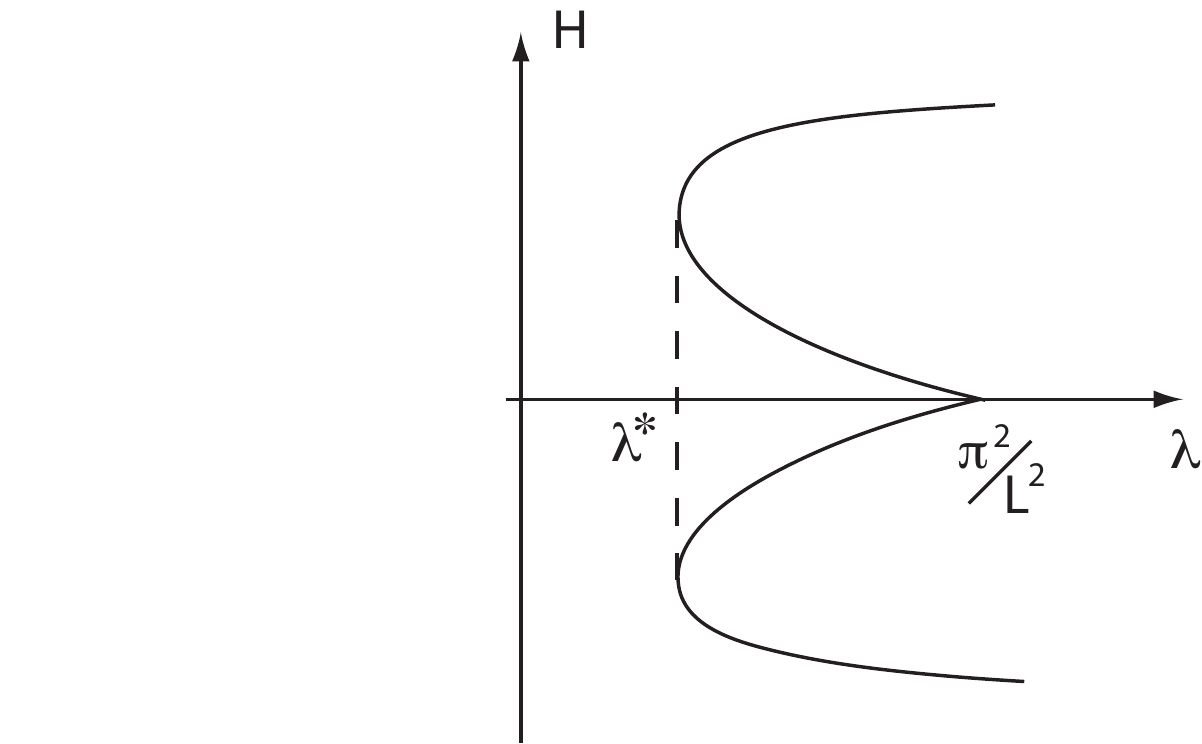}
  \caption{Type-II transition as given by Theorem~\ref{t8.5}.}\la{f8.8-1}
 \end{SCfigure}

\bp
With the spatial domain as given,   the first eigenvalue and eigenfunction of
(\ref{8.61}) are given by
$$\rho_1=\pi^2/L^2,\ \ \ \ e_1=\cos\frac{\pi x_1}{L}.$$
The eigenvalues and eigenfunctions of $L_{\lambda}=-A+B_{\lambda}$
defined by (\ref{8.59}) are as follows:
\begin{align}
& \beta_K=|K|^2(\lambda -|K|^2),\label{8.74}\\
& e_K=\cos\frac{k_1\pi x_1}{L_1}\cdots\cos\frac{k_n\pi x_n}{L_n},\label{8.75}
\end{align}
where
\begin{align*}
&K=\left(\frac{k_1\pi}{L_1},\cdots,\frac{k_n\pi}{L_n}\right)&&  \forall 
  k_i\in \Z,\ \ \ \ 1 \leq i\leq n, \\
&|K|^2=\pi^2\sum^n_{i=1}k^2_i/L^2_i&& |K|^2\neq 0.
\end{align*}

By the approximation of the center manifold obtained in \cite{b-book},  the  reduced equation of
(\ref{8.60}) to the center manifold is given by
\begin{equation}
\frac{dy}{dt}= \beta_1(\lambda )y 
 -\frac{2\pi^2}{(L_1\cdots
L_n)L^2_1}
\int_{\Omega}\left[\gamma_2y^2e^3_1 +\gamma_3y^3 e^4_1
+\gamma_2 (ye_1+\Phi (y))^2e_1\right]dx,
\label{8.76}
\end{equation}
where $y\in \R^1$, $\Phi (y)$ is the center manifold function, and
\begin{equation}
\beta_1(\lambda )=\frac{\pi^2}{L^2}(\lambda -\frac{\pi^2}{L^2}).\label{8.77}
\end{equation}
Direct calculation implies that 
\begin{eqnarray}
&&\int_{\Omega}e^3_1dx=\int^{L_1}_0\cdots\int^{L_n}_0\cos^3\frac{\pi
x_1}{L}dx=0,\label{8.78}\\
&&\int_{\Omega}e^4_1dx=L_2\cdots L_n\int^L_0\cos^4\frac{\pi
x_1}{L}dx_1=\frac{3}{8}L_1\cdots L_n.\label{8.79}
\end{eqnarray}
By (\ref{8.78}) and $\Phi (y)=O(|y|^2)$ we obtain
\begin{equation}
\int_{\Omega}(ye_1+\Phi )^2e_1dx=2y\int_{\Omega}\Phi
(y)e^2_1dx+o(|y|^3)\label{8.80}
\end{equation}
It follows that
\begin{align*}
& 
\Phi (y) = \sum^{\infty}_{|K|^2>\pi^2/L^2}\phi_K(y)e_K+o(|y|^2)\\
& \phi_K(y)= \frac{\gamma_2y^2}{-\beta_K\|e_K\|^2_H}\int_{\Omega}\Delta
e^2_1\cdot e_Kdx = 
\frac{|K|^2\gamma_2y^2}{\beta_K\|e_K\|^2_H}\int_{\Omega}e_Ke^2_1dx.
\end{align*}
Notice that
$$\int_{\Omega}e_Ke^2_1dx=
\left\{
\begin{aligned}
&0&& \forall  K\neq\left(\frac{2\pi}{L_1},0,\cdots,0\right),\\
&L_1\cdots L_n/4&& \forall 
K=\left(\frac{2\pi}{L_1},0,\cdots,0\right).
\end{aligned}
\right.
$$ 
Then we have
$$
\Phi (y)=\frac{\gamma_2y^2}{2(\lambda
-4\pi^2/L^2)}\cos\frac{2\pi x_1}{L}+o(|y|^2).
$$ 
Inserting $\Phi
(y)$ into (\ref{8.80}) we find
\begin{align}
\int_{\Omega}(ye_1+\Phi )^2e_1dx
=&\frac{\gamma_2y^3}{\lambda -\frac{4\pi^2}{L^2}}\int_{\Omega}\cos\frac{2\pi
x_1}{L}e^2_1dx+o(|y|^3)\label{8.81}\\
=&\frac{L_1\cdots L_n}{4}\cdot\frac{\gamma_2}{\lambda
-\frac{4\pi^2}{L^2}}y^3+o(|y|^3).\nonumber
\end{align}
Finally, by (\ref{8.78}), (\ref{8.79}) and (\ref{8.81}),   we derive from (\ref{8.76})
 the following reduced equation of (\ref{8.60}): 
 \begin{equation}
\frac{dy}{dt}=\beta_1(\lambda
)y-\frac{\pi^2}{2L^2}\left(\frac{3\gamma_3}{2}
+\frac{\gamma^2_2}{\lambda -\frac{4\pi^2}{L^2}}\right)y^3+o(|y|^3).\label{8.82}
\end{equation}
Near the critical point $\lambda_0=\pi^2/L^2$, the coefficient
$$\frac{3\gamma_3}{2}+\frac{\gamma^2_2}{\lambda
-\frac{4\pi^2}{L^2}}=\frac{3\gamma_3}{2}-\frac{L^2\gamma^2_2}{3\pi^2}.$$

Thus, by Theorem \ref{t5.2} we derive from (\ref{8.82}) the assertions of
the theorem except the claim for the saddle-node bifurcation in
Assertion (1), which can be proved in the same fashion as used
in Theorem \ref{t8.4}. The proof is complete. 
\ep

We now consider the case where $\Omega =\Pi^n_{k=1}(0,L_k)$
satisfies  that 
\begin{equation}
 L=L_1=\cdots =L_m > L_j  \qquad \text{for } 2\leq m\leq 3, m < j \le n.
 \label{8.83}
\end{equation}

\bt\la{t8.6}
 Let $\Omega =\Pi^n_{k=1}(0,L_k)$ satisfy
(\ref{8.83}). Then the following assertions hold true:

\begin{enumerate}
\item  If 
$$\gamma_3>\frac{26L^2}{27\pi^2}\gamma^2_2,$$ 
then the phase transition of the problem (\ref{8.58}) with (\ref{8.62}) at $\lambda_0=\pi^2/L^2$ is Type-I, satisfying the following properties:

\begin{enumerate}

\item  The problem bifurcates on $\lambda >\pi^2/L^2$ to an attractor $\Sigma_{\lambda}$, containing exactly $3^m-1$ non-degenerate singular points, 
and $\Sigma_{\lambda}$ is homeomorphic to an $(m-1)$-dimensional 
sphere $S^{m-1}$.

\item For $m=2$, the attractor $\Sigma_{\lambda}=S^1$
contains 4 minimal attractors, as shown in Figure \ref{f8.9-1}. 

\item For
$m=3, \Sigma_{\lambda}=S^2$ contains  $8$ minimal attractors as shown in Figure~\ref{f8.10}(a),  if 
$$\gamma_3 < \frac{22 L^2}{9\pi^2} \gamma_2^2,$$
and contains $6$ minimal attractors as shown  in Figure \ref{f8.10}(b)
if 
$$\gamma_3 > \frac{22 L^2}{9\pi^2} \gamma_2^2.$$
\end{enumerate}

\item If 
$$\gamma_3<\frac{26L^2}{27\pi^2}\gamma^2_2,$$
then the transition is Type-II. 
In particular, the problem has
a saddle-node bifurcation on $\lambda <\lambda_0=\pi^2/L^2$, and
bifurcates on both side of $\lambda =\lambda_0$ to exactly $3^m-1$
singular points which are non-degenerate.
\end{enumerate}
\et
\begin{SCfigure}[25][t]
  \centering
  \includegraphics[width=0.4\textwidth]{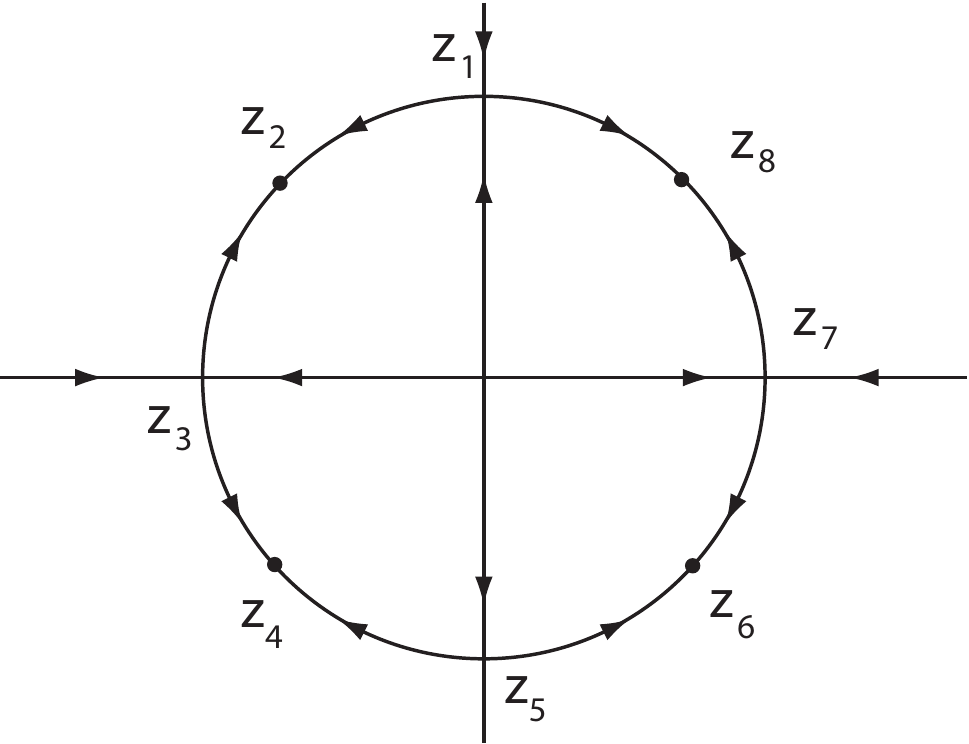}
  \caption{For  $m=2, \Sigma_{\lambda}=S^1$ and
$Z_{2k}$  $(1\leq k\leq 4)$ are attractors.}\la{f8.9-1}
 \end{SCfigure}
\begin{figure}[hbt]
  \centering
  \includegraphics[width=0.38\textwidth]{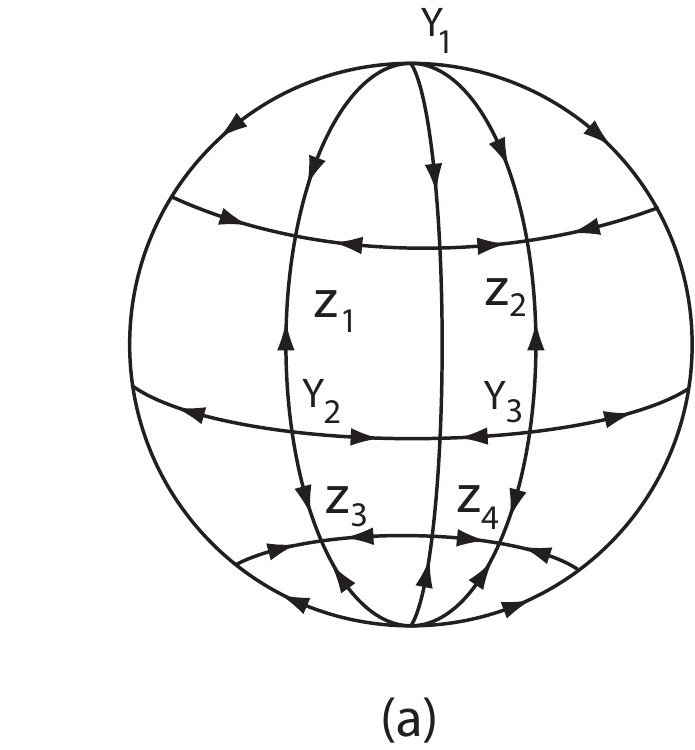}\qquad 
  \includegraphics[width=0.3\textwidth]{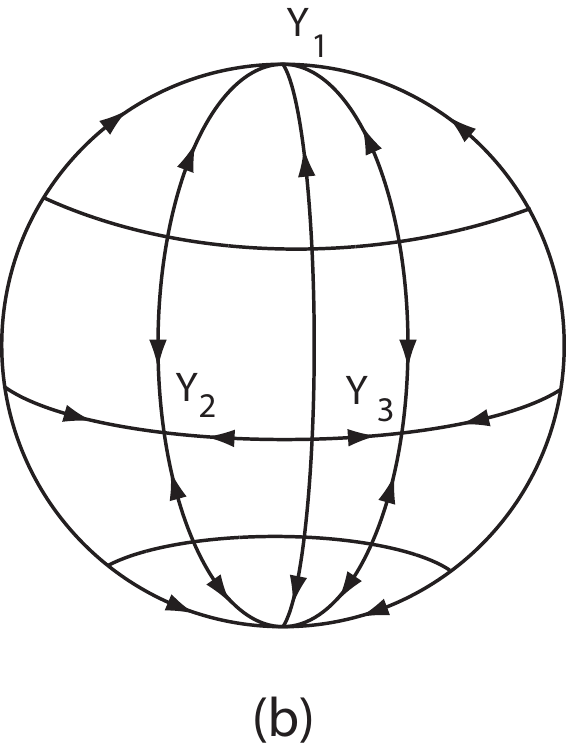}
  \caption{For $m=3$, $\Sigma_{\lambda}=S^2$, (a) $\pm
Z_k$  $(1\leq k\leq 4)$ are attractors, (b) $\pm Y_k$  $(1\leq k\leq 3)$
are attractors.}\la{f8.10}
 \end{figure}

\bp 
 We proceed in several steps as follows.

\medskip

{\sc Step 1.} Consider the center manifold reduction. It is known
that the eigenvalues and eigenfunctions of
$L_{\lambda}=-A+B_{\lambda}$ are given by (\ref{8.74}) and
(\ref{8.75}) with $L_1=\cdots =L_m$.  
As before, the reduced
equations of (\ref{8.60}) are  given by 
\begin{equation}
\frac{dy}{dt}=\beta_1(\lambda)y+g(y)+o(|y|^3),\label{8.84}
\end{equation}
where $y=(y_1,\cdots,y_m)\in \R^m$, $\beta_1(\lambda )$ is as in
(\ref{8.77}), and
\begin{align*}
&g(y)=\frac{2\pi^2}{L_1\cdots L_nL^2_1}(G_2(y)+G_3(y)+G_{23}(y)),\\
& G_2(y)=-\gamma_2\left(\int_{\Omega}v^2e_1dx,\cdots,\int_{\Omega}v^2e_mdx\right),\\
& G_3(y)=-\gamma_3\left(\int_{\Omega}v^3e_1dx,\cdots,\int_{\Omega}v^3e_mdx\right),\\
&G_{23}(y)=-\gamma_2\left(\int_{\Omega}u^2e_1dx,\cdots,\int_{\Omega}u^2e_mdx\right).
\end{align*}
Here 
$ e_i=\cos\pi x_i/L $  for   $ 1\leq i\leq m$,  $L$  is given by  (\ref{8.83}), 
$ v=\sum^m_{i=1}y_ie_i,$   $u=v+\Phi (y)$   and $\Phi$ is the center manifold function.
Direct computation shows that  
\begin{align}
& 
\left.\begin{aligned}
 \int_{\Omega}v^2e_idx
      =&  \int_{\Omega}\left(\sum^m_{j=1}y_j\cos\pi
            x_j/L\right)^2\cos\pi x_i/Ldx=0, \\
 \int_{\Omega}v^3e_idx=& 
     \int_{\Omega}\left(\sum^n_{j=1}y_k\cos\pi x_j/L\right)^3\cos\pi x_i/Ldx\\
 =& \frac{3}{4}L_1\cdots
L_n\left(\frac{1}{2}y^3_i+y_i\sum_{j\neq i}y^2_j\right),
\end{aligned}
\right.\label{8.85} \\
&
\int_{\Omega}u^2e_idx=\int_{\Omega}\left(\sum^m_{j=1}y_je_j+\Phi
(y)\right)^2e_idx \nonumber \\
& \qquad \qquad = 2\sum^m_{j=1}y_j\int_{\Omega}\Phi (y)e_je_idx
+ \int_\Omega \Phi^2(y) e_idx. \label{8.86}
\end{align}

We need to compute the center manifold function $\Phi (y)$. 
As in \cite{b-book}, we have
\begin{equation}
\Phi
(y)=\sum^{\infty}_{|K|>\pi^2/L^2}\phi_K(y)e_K+o(|y|^2)+O(|y|^2|\beta_1|),\label{8.87}
\end{equation}
where
\begin{align*}
\phi_K(y) =&\frac{\gamma_2}{-\beta_K(\lambda)<e_K,e_K>}
\int_{\Omega}\Delta v^2e_Kdx\\
=&\frac{|K|^2\gamma_2}{\beta_K(\lambda)<e_K,e_K>}
\int_{\Omega}v^2e_Kdx\\
=&\frac{|K|^2\gamma_2}{\beta_K<e_K,e_K>}\sum^m_{i,j=1}y_iy_i\int_{\Omega}e_ie_ie_Kdx.
\end{align*}
It is clear that
$$\int_{\Omega}e_ie_je_Kdx=\int_{\Omega}\cos\frac{\pi
x_i}{L}\cos\frac{\pi x_j}{L}e_kdx=0,$$ 
if 
$$ K\neq K_i+K_j, \qquad  K_i=\left(\frac{\pi}{L_1}\delta_{1i},\cdots,\frac{\pi}{L_n}\delta_{ni}\right).
$$
By (\ref{8.74}) and (\ref{8.75}) we have
\begin{align*}
\phi_K(y) =
&
\frac{\gamma_2y_iy_j(2-\delta_{ij})}{(\lambda -|K|^2)<e_K,e_K>}\int_{\Omega}e_ie_je_Kdx\\
=
&\left\{
\begin{aligned} 
&  \frac{\gamma_2y^2_i}{2(\lambda  -\frac{4\pi^2}{L^2})}&& \text{ if }  i=j,\\
& \frac{2\gamma_2y_iy_j}{(\lambda -\frac{2\pi^2}{L^2})}&& \text{ if }   i\neq j,
\end{aligned}
\right.
\end{align*}
for $K=K_i+K_j$. 
Thus, by (\ref{8.87}), we obtain 
$$
\Phi (y)=\sum^m_{i=1}\frac{\gamma_2}{2(\lambda
-\frac{4\pi^2}{L^2})}y^2_i\cos\frac{2\pi x_i}{L}
+\sum^m_{l>r}\frac{2\gamma_2}{(\lambda
-\frac{2\pi^2}{L^2})}y_ly_r\cos\frac{\pi x_l}{L}\cos\frac{\pi x_r}{L}.
$$
Inserting $\Phi (y)$ into (\ref{8.86}) we derive
\begin{align*}
\int_{\Omega}u^2e_idx= &
 \frac{\gamma_2}{\lambda-\frac{4\pi^2}{L^2}} 
\int_{\Omega} \left[y^3_i  \cos^2\frac{\pi
x_i}{L}\cos\frac{2\pi x_i}{L}+ 4y_i\sum_{j\neq
i}y^2_j \cos^2\frac{\pi x_j}{L}\cos^2\frac{\pi
x_i}{L}\right]dx \\
& + o(|y|^3).
\end{align*}
 Direct computation gives that
\begin{equation}
\int_{\Omega}u^2e_idx=\frac{\gamma_2L_1\cdots
L_n}{4}\left[\frac{y^3_i}{\lambda -\frac{4\pi^2}{L^2}}+\frac{4y_i}{\lambda
-\frac{2\pi^2}{L^2}}\sum_{j\neq i}y^2_j\right] + o(|y|^3).\label{8.88}
\end{equation}
Putting (\ref{8.85}) and (\ref{8.88}) in (\ref{8.84}), we get the
reduced equations in the following form
\begin{equation}
\frac{dy_i}{dt}=\beta_1(\lambda
)y_i-\frac{\pi^2}{2L^2}\left[\sigma_1y^3_i+\sigma_2y_i\sum_{j\neq
i}y^2_j\right]+o(|y|^3) \quad \forall 1\leq i\leq m,\label{8.89}
\end{equation}
where
\begin{equation}
\sigma_1=\frac{3\gamma_3}{2}+\frac{\gamma^2_2}{\lambda
-\frac{4\pi^2}{L^2}},\qquad 
\sigma_2=3\gamma_3+\frac{4\gamma^2_2}{\lambda -\frac{2\pi^2}{L^2}}.
\label{8.90}
\end{equation}

{\sc Step 2.}
 It is known that the transition type of (\ref{8.60}) at
the critical point $\lambda_0=\pi^2/L^2$ is completely determined
by (\ref{8.89}), i.e., by the following equations
\begin{equation}
\frac{dy_i}{dt}=-\frac{\pi^2}{2L^2}\left[\sigma^0_1y^3_i+\sigma^0_2y_i\sum_{k\neq
i}y^2_k\right]\qquad \forall  1\leq i\leq m,\label{8.91}
\end{equation}
where
$$\sigma^0_1=\frac{3\gamma_3}{2}-\frac{L^2\gamma^2_2}{3\pi^2},\qquad \sigma^0_2=3\gamma_3-\frac{4L^2\gamma^2_2}{\pi^2}.
$$ 
It is easy to see that
\begin{equation}
\left.
\begin{aligned}
&\sigma^0_1+\sigma^0_2>0\Leftrightarrow\gamma_3>\frac{26L^2}{27\pi^2}\gamma^2_2,\\
&\sigma^0_1+\sigma^0_2<0\Leftrightarrow\gamma_3<\frac{26L^2}{27\pi^2}\gamma^2_2.
\end{aligned}
\right.\label{8.92}
\end{equation}

{\sc Step 3.} We consider the case where $m=2$. Thus, the transition
type of (\ref{8.91}) is equivalent to that of the following equations
\begin{equation}
\left.
\begin{aligned}
&\frac{dy_1}{dt}=-y_1[\sigma^0_1y^2_1+\sigma^0_2y^2_2],\\
&\frac{dy_2}{dt}=-y_2[\sigma^0_1y^2_2+\sigma^0_2y^2_1].
\end{aligned}
\right.\label{8.93}
\end{equation}
We can see that on the straight lines
\begin{equation}
y^2_1=y^2_2, \label{8.94}
\end{equation}
equations (\ref{8.93}) satisfy that
$$\frac{dy_2}{dy_1}=\frac{y_2}{y_1}\qquad \text{ for }\
\sigma^0_1+\sigma^0_2\neq 0,\ \ \ \ (y_1,y_2)\neq 0.$$
Hence the straight lines (\ref{8.94}) are orbits of
(\ref{8.93})   if  $\sigma^0_1+\sigma^0_2\neq 0$. Obviously, the
straight lines
\begin{equation}
y_1=0 \text{  and }  y_2=0 \label{8.95}
\end{equation}
are also orbits of (\ref{8.93}).

There are four straight lines determined by (\ref{8.94}) and
(\ref{8.95}), and each of them contains two orbits. Hence, the
system (\ref{8.93}) has at least eight straight line orbits.  
Hence it is not hard to see that the number of straight line orbits of
(\ref{8.93}), if finite, is eight.

Since (\ref{8.60}) is a gradient-type equation, 
there are no elliptic regions at $y=0$; see \cite{b-book}. Hence, when
$\sigma^0_1+\sigma^0_2>0$ all the straight line orbits  on
(\ref{8.94}) and (\ref{8.95}) tend to $y=0$, as shown in Figure
\ref{f8.11} (a), which implies that the regions are parabolic and
stable, therefore $y=0$ is asymptotically stable for (\ref{8.93}).
Accordingly, by the attractor bifurcation theorem in \cite{b-book}, the transition of (\ref{8.89}) at
$\lambda_0=\pi^2/L^2$ is Type-I.

When $\sigma^0_1+\sigma^0_2<0$ and $\sigma^0_1>0$, namely
$$\frac{2}{9}\frac{L^2\gamma^2_2}{\pi^2}<\gamma_3<\frac{26}{27}\frac{L^2\gamma^2_2}{\pi^2},$$
the four straight line orbits on (\ref{8.94}) are outward from
$y=0$, and other four on (\ref{8.95}) are toward $y=0$, as shown
in Figure \ref{f8.11} (b), which implies that all regions at $y=0$ are
hyperbolic. Hence, by Theorem \ref{t5.3},  the transition of (\ref{8.89})
at $\lambda_0=\pi^2/L^2$ is Type-II.

When $\sigma^0_1\leq 0$, then $\sigma^0_2<0$ too. In this case, no
orbits of (\ref{8.93}) are toward $y=0$, as shown in Figure \ref{f8.11}
(c), which implies by Theorem \ref{t5.3} that the transition is Type-II.

Thus by (\ref{8.92}), for $m=2$ we prove that the transition is
Type-I if $\gamma_3>\frac{26L^2}{27\pi^2}\gamma^2_2$, and Type-II
if  $\gamma_3<\frac{26 L^2}{27\pi^2}\gamma^2_2$.
\begin{figure}[hbt]
  \centering
  \includegraphics[width=0.3\textwidth]{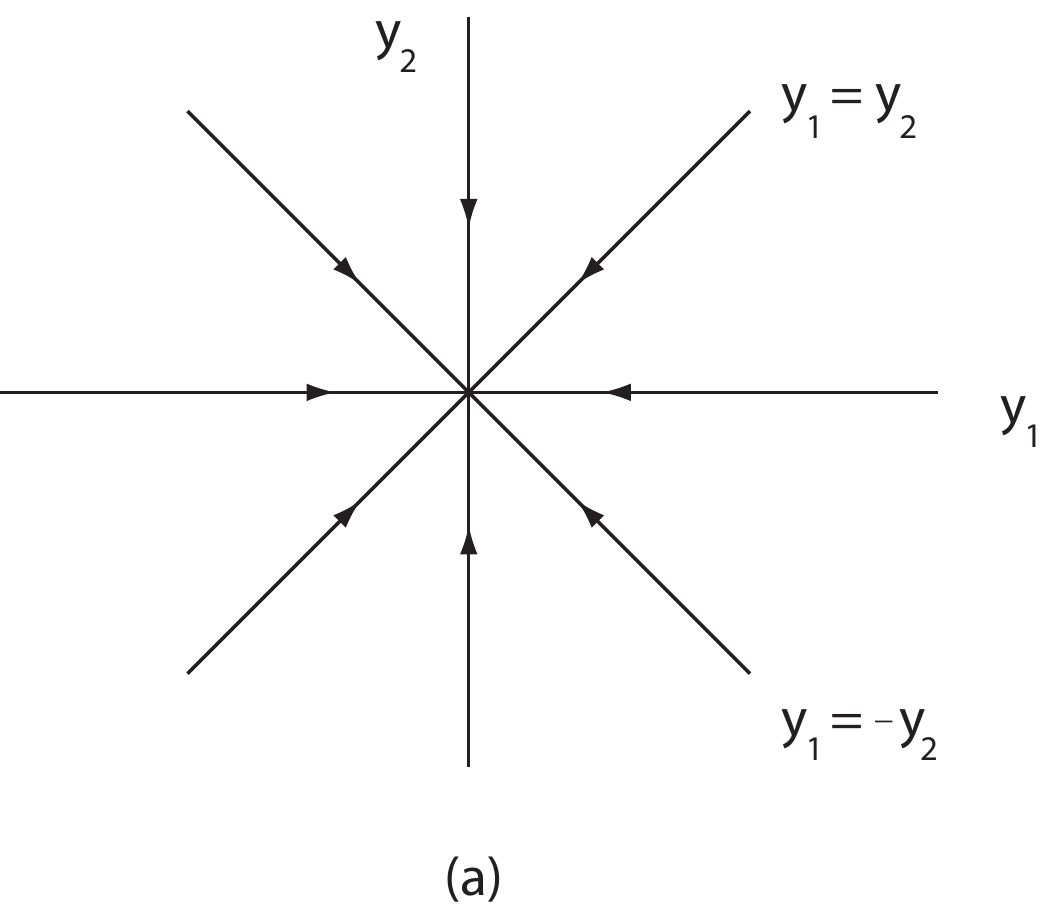} 
  \includegraphics[width=0.3\textwidth]{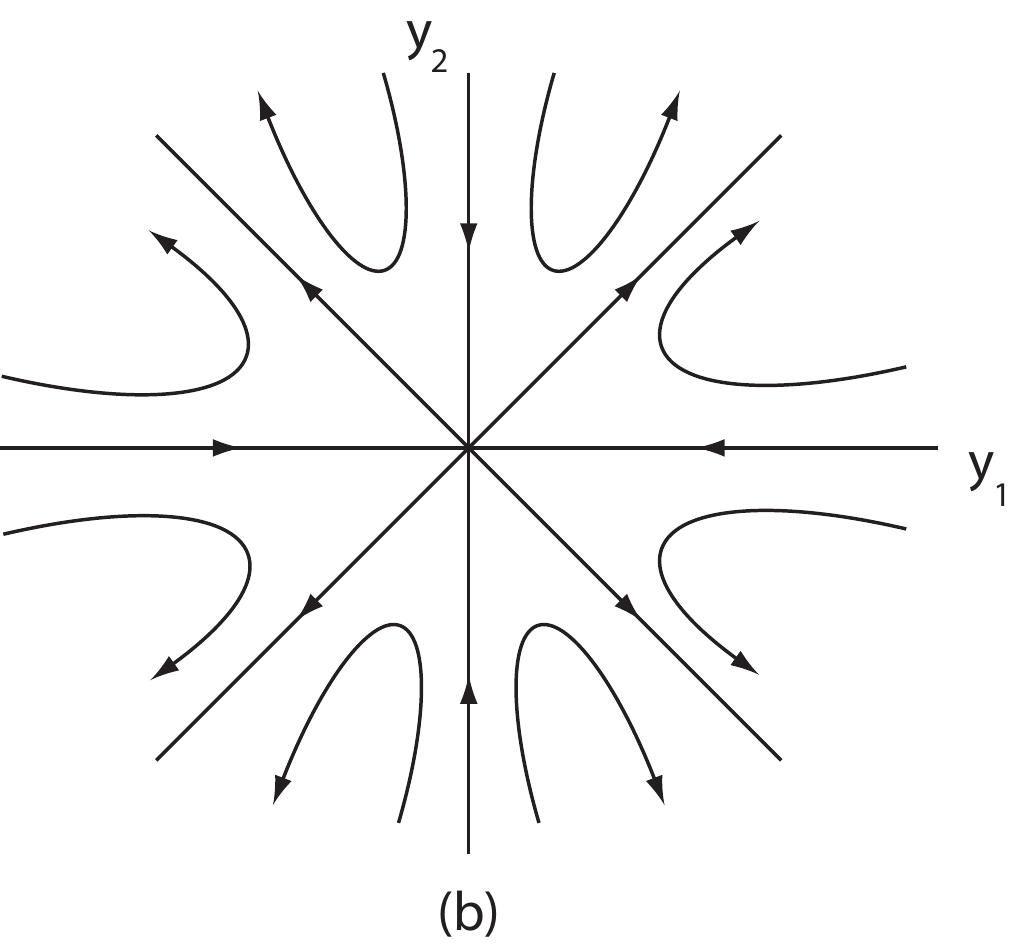} 
  \includegraphics[width=0.28\textwidth]{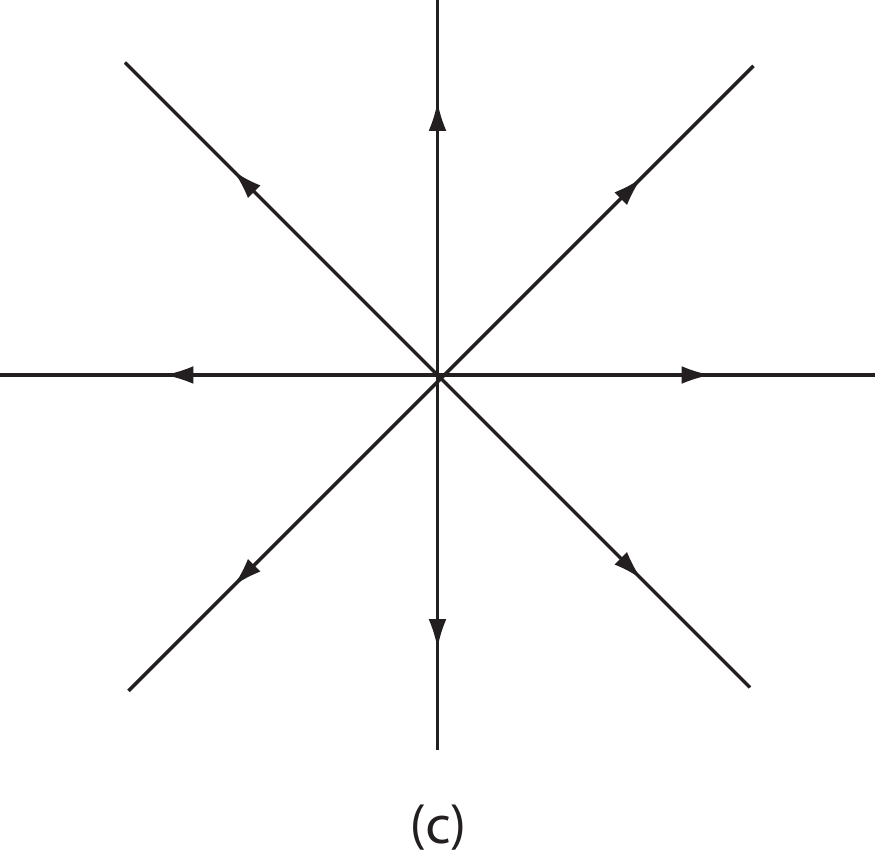}
  \caption{The topological structure of flows of
(\ref{8.89}) at $\lambda_0=\pi^2/L^2$, (a) as
$\gamma_3>\frac{26L^2}{27\pi^2}\gamma^2_2$; (b) as
$\frac{2L^2}{9\pi^2}\gamma^2_2<\gamma_3<\frac{26L^2}{27\pi^2}\gamma^2_2$;
and (c) $\gamma_3\leq\frac{2L^2}{9\pi^2}\gamma^2_2$.}
  \la{f8.11}
 \end{figure}

{\sc Step 4.} Consider the case where $m=3$. Thus, (\ref{8.91}) are
written as
\begin{equation}
\left.
\begin{aligned}
&\frac{dy_1}{dt}=-y_1[\sigma^0_1y^2_1+\sigma^0_2(y^2_2+y^2_3)],\\
&\frac{dy_2}{dt}=-y_2[\sigma^0_1y^2_2+\sigma^0_2(y^2_1+y^2_3)],\\
&\frac{dy_3}{dt}=-y_3[\sigma^0_1y^2_3+\sigma^0_2(y^2_1+y^2_2)].
\end{aligned}
\right.\label{8.96}
\end{equation}
It is clear that the straight lines
\begin{align}
& y_i=0, \quad  y_j=0\qquad  \text{ for } i\neq j,\ 1\leq i,\ j\leq
3,\label{8.97} \\
&
\left\{
\begin{aligned} 
& y^2_i=y^2_j, \quad y_k=0&& \text{ for } i\neq j,i\neq
k,j\neq k,1\leq i,j,k\leq 3,\\
& y^2_1=y^2_2=y^2_3,
\end{aligned}
\right.\label{8.98}
\end{align}
consist of orbits of (\ref{8.96}). There are total 13 straight
lines in (\ref{8.97}) and (\ref{8.98}), each of which consists of
two orbits. Thus, (\ref{8.96}) has at least 26 straight line
orbits. We shall show that (\ref{8.96}) has just the straight
line orbits given by (\ref{8.97}) and (\ref{8.98}). In fact, we
assume that the line
$$y_2=z_1y_1,\ \ \ \ y_3=z_2y_1\ \ \ \ (z_1,z_2\ \text{are\ real\
numbers})$$ is a straight line orbit of (\ref{8.96}). Then
$z_1,z_2$ satisfy
\begin{equation}
\left.
\begin{aligned}
&\frac{dy_2}{dy_1}=z_1=z_1\frac{\sigma^0_1z^2_1+\sigma^0_2(1+z^2_2)}{\sigma^0_1+\sigma^0_2(z^2_1+
z^2_2)},\\
&\frac{dy_3}{dy_1}=z_2=z_2\frac{\sigma^0_1z^2_2+\sigma^0_2(1+z^2_1)}{\sigma^0_1+\sigma^0_2(z^2_1+z^2_2)}.
\end{aligned}
\right.\label{8.99}
\end{equation}
It is easy to see that when $\sigma^0_1\neq\sigma^0_2$ the
solutions $z_1$ and $z_2$ of (\ref{8.99}) take only the values
$$z_1=0,\pm 1;\ \ \ \ z_2=0,\pm 1.$$
In the same fashion, we can prove that the straight line orbits of
(\ref{8.96}) given by
$$y_1=\alpha_1y_3,\ \ \ \ y_2=\alpha_2y_3,\ \ \ \ \text{and}\
y_1=\beta_1y_2,\ \ \ \ y_3=\beta_2y_2$$ have to satisfy that
$$\alpha_i=0,\pm 1\ \ \ \ \text{and}\ \ \ \ \beta_i=0,\pm 1\ \ \ \
(i=1,2).$$ Thus, we prove that when $\sigma^0_1\neq\sigma^0_2$,
the number of straight line orbits of (\ref{8.96}) is exactly 26.

When $\sigma^0_1=\sigma^0_2$, we have that
$\gamma_3=\frac{22}{9}\frac{L^2}{\pi^2}\gamma^2_2$ which implies
that $\sigma^0_1=\sigma^0_2>0$. In this case, it is clear that
$y=0$ is an asymptotically stable singular point of (\ref{8.96}).
Hence, the transition of (\ref{8.89}) at $\lambda_0=\pi^2/L^2$ is
I-type.

When $\sigma^0_1+\sigma^0_2>0$ and $\sigma^0_1\neq\sigma^0_2$, all
straight line orbits of (\ref{8.96}) are toward $y=0$, which
implies  that the regions at $y=0$, are stable, and
$y=0$ is asymptotically stable; see \cite{b-book}. Thereby the transition of
(\ref{8.89}) is Type-I.

When $\sigma^0_1+\sigma^0_2<0$ with $\sigma^0_1>0$, we can see, as
in the case of $m=2$, that the regions at $y=0$ are hyperbolic,
and when $\sigma^0_1+\sigma^0_2<0$ with $\sigma_1\leq 0$ the
regions at $y=0$ are unstable. Hence, the transition is Type-II.

\medskip

{\sc Step 5.} We prove Assertion (1). By Steps 3 and 4, if 
$\gamma_3>\frac{26L^2}{27\pi^2}\gamma^2_2$,  
the reduced equation
(\ref{8.89}) bifurcates on $\lambda >\lambda_0=\pi^2/L^2$ to an
attractor $\Sigma_{\lambda}$. All bifurcated equilibrium points of
(\ref{8.60}) are one to one correspondence  to the bifurcated
singular points of (\ref{8.89}). Therefore,  we only have to consider the stationary equations:
\begin{equation}
\beta_1(\lambda
)y_i-\frac{\pi^2}{2L^2}[\sigma_1y^3_i+\sigma_2y_i\sum_{j\neq
i}y^2_j]+o(|y|^3)=0,\ \ \ \ 1\leq i\leq m,\label{8.100}
\end{equation}
where $\sigma_1$ and $\sigma_2$ are as in (\ref{8.90}).

Consider the following approximative equations of (\ref{8.100})
\begin{equation}
\beta_1(\lambda )y_i-y_i(a_1y^2_i+a_2\sum_{j\neq i}y^2_j)=0\qquad 
\text{ for }  1\leq i\leq m,\label{8.101}
\end{equation}
where $a_1=\pi^2\sigma_1/2L^2, a_2=\pi^2\sigma_2/2L^2$. It is
clear that each regular bifurcated solution of (\ref{8.101})
corresponds to a regular bifurcated solution of (\ref{8.100}).

We first prove that (\ref{8.101}) has $3^m-1$ bifurcated solutions
on $\lambda >\lambda_0$. For each $k$  $(0\leq k\leq m-1)$,
(\ref{8.101}) has $C^k_m\times 2^{m-k}$ solutions as follows: 
\begin{equation}
\left.
\begin{aligned} 
& y_{j_1}=0,\cdots,y_{j_k}=0 &&\text{ for } 1\leq j_l\leq
m,\\
& y^2_{r_1}=\cdots=y^2_{r_{m-k}}=\beta_1(a_1+(m-k-1)a_2)^{-1}
&&\text{ for } r_i\neq  j_l.
\end{aligned}
\right.\label{8.102}
\end{equation}
Hence, the number of all bifurcated solutions of (\ref{8.101}) is
$$\sum^{m-1}_{k=0}C^k_m\times 2^{m-k}=(2+1)^m-1=3^m-1.$$

We need to prove that all bifurcated solutions of (\ref{8.101})
are regular. The Jacobian matrix of (\ref{8.101}) is given by
\begin{equation}
Dv=
\begin{pmatrix}
\beta_1-h_1(y)  &  2a_2y_1y_2      &\cdots   &    2a_2y_1y_m\\
2a_2y_2y_1      &   \beta_1-h_2(y) &\cdots   &    2a_2y_2y_m\\
\vdots                &\vdots                  &             &    \vdots\\
2a_2y_my_1     & 2a_2y_my_2      &\cdots   &  \beta_1-h_m(y)
\end{pmatrix},
\label{8.103}
\end{equation}
where
$$h_i(y)=3a_1y^2_i+a_2\sum_{j\neq i}y^2_j.$$
For the solutions in (\ref{8.102}), without loss of generality, we
take
\begin{align*}
&y_0=(y^0_1,\cdots,y^0_m),\\
&y^0_i=0\qquad   \qquad 1\leq i\leq k, \\
&y^0_{k+1}=\cdots=y^0_m=\beta^{{1}/{2}}_1(a_1+(m-k-1)a_2)^{-{1}/{2}}
\end{align*}
Inserting them into (\ref{8.103}) we find
\begin{equation}
Dv(y_0)=
\left(\begin{matrix} \beta I_k&0\\
0&A_{m-k}
\end{matrix}
\right), \label{8.104}
\end{equation}
where
\begin{align*}
& 
\beta
=\beta_1\left(1-\frac{(m-k)a_2}{a_1+(m-k-1)a_2}\right)=\frac{(\sigma_1-\sigma_2)\beta_1}{\sigma_1+
(m-k-1)\sigma_2}, \\
& A_{m-k} =
\left(\begin{matrix}
\beta_1-(3a_1+(m-k-1)a_2)(y^{0}_{k+1})^2  &  \cdots & 2a_2(y^{0}_{k+1})^2\\
\vdots&&\vdots\\
2a_2(y^{0}_{k+1})^2&\cdots&\beta_1-(3a_1+(m-k-1)a_2)(y^{0}_{k+1})^2
\end{matrix}
\right).
\end{align*}
Direct computation shows that 
$$\text{det}A_{m-k}=\frac{\pi^{2m}\beta^m_1}{(a_1+(m-k-1)a_2)^mL^{2m}}\text{det}
\left(\begin{matrix}
-\sigma_1&\sigma_2&\cdots&\sigma_2\\
\sigma_2&-\sigma_1&\cdots&\sigma_2\\
\vdots&\vdots& & \vdots\\
\sigma_2&\sigma_2&\cdots&-\sigma_1
\end{matrix}
\right),$$ 
where $\sigma_1,\sigma_2$ are given by (\ref{8.90}).

Obviously, there are only finite number of $\lambda >\pi^2/L^2$
satisfying
\begin{align*}
& \beta (\lambda )=\frac{(\sigma_1(\lambda )-\sigma_2(\lambda
))\beta_1(\lambda )}{\sigma_1(\lambda )+(m-k-1)\sigma_2(\lambda
)}=0,\\
& \text{det}A_{m-k}(\lambda )=0.
\end{align*}
Hence, for any $\lambda -\pi^2/L^2>0$ sufficiently small
the Jacobian matrices (\ref{8.103}) at the singular points
(\ref{8.102}) are non-degenerate. Thus, the bifurcated solutions of
(\ref{8.101}) are regular.

Since all bifurcated singular points of (\ref{8.58}) with
(\ref{8.62}) are non-degenerate, and when $\Sigma_{\lambda}$ is
restricted on $x_ix_j$-plane $(1\leq i,j\leq m)$ the singular
points are connected by their stable and unstable manifolds. 
Hence all singular points in $\Sigma_{\lambda}$ are
connected by their stable and unstable manifolds. Therefore,
$\Sigma_{\lambda}$ must be homeomorphic to a sphere $S^{m-1}$.

Assertion (1) is proved.

\medskip

{\sc Step 6. Proof of Assertions (2) and (3).} When $m=2$, by
Step 5, $\Sigma_{\lambda}=S^1$ contains 8 non-degenerate singular
points. By a theorem on minimal attractors in \cite{b-book},  $4$ singular points must be 
attractors and the others are repellors, as shown in Figure \ref{f8.9-1}.

When $m=3$, we take the six singular points
$$\pm Y_1=(\pm\beta_1a^{-1}_1,0,0),\pm
Y_2=(0,\pm\beta_1a^{-1}_1,0),\pm Y_3=(0,0,\pm\beta_1a^{-1}_1)$$
Then the Jacobian matrix (\ref{8.103}) at $Y_i$  $(1\leq i\leq 3)$ is
$$Dv(\pm Y_i)=\left(\begin{array}{lll}
\rho_1&&0\\
&\rho_2\\
0&&\rho_3
\end{array}
\right),
$$ 
where
$\rho_j=\beta_1\left(1-\frac{\sigma_2}{\sigma_1}\right)$ 
as $j\neq i$ and $\rho_i=-2\beta_1$. Obviously, as $\sigma_2<\sigma_1$, 
$0< \rho_j$ $(j\neq i)$ and $\rho_i<0$, in this case, $\pm Y_k$ $(1\leq
k\leq 3)$ are repellors in $\Sigma_{\lambda}=S^2$, which implies
that $\Sigma_{\lambda}$ contains $8$ attractors $\pm z_k$  $(1\leq k\leq
4)$ as shown in Figure \ref{f8.10} (a). As $\sigma_2>\sigma_1,
\rho_j<0$ $(1\leq j\leq 3)$, the six singular point $\pm Y_k$  $(1\leq
k\leq 3)$ are attractors, which implies that $\Sigma_{\lambda}$
contains only six minimal attractors as shown in Figure \ref{f8.10} (b).
Thus Assertion (2) is proved.

The claim for the saddle-node bifurcation in Assertion (3) can be
proved by using the same method as in the proof of  Theorem \ref{t8.4}, and the claim
for the singular point bifurcation can be proved by the same
fashion as used in Step 5.

The proof of this theorem is complete.
\ep

\br\la{r8.5}
{\rm
For the domain $\Omega =[0,L]^m\times D\subset
\R^n(1\leq m<n)$ where $n\geq 2$ is arbitrary and $D\subset
\R^{n-m}$ a bounded open set, Theorems \ref{t8.5} and \ref{t8.6} are also valid
provided $\pi^2/L^2<\lambda_1$, where $\lambda_1$ is the first
eigenvalue of the equation
$$\left.
\begin{aligned}
&-\Delta e=\lambda e,&\ \ \ \ x\in D\subset \R^{n-m},\\
&\frac{\partial e}{\partial n}|_{\partial D}=0,\\
&\int_Dedx=0.
\end{aligned}
\right.
$$ 
}\er

\br\la{r8.6}
{\rm
In Theorem \ref{t8.6}, the minimal
attractors in the bifurcated attractor $\Sigma_{\lambda}$ can be
expressed as
\begin{equation}
u_{\lambda}=(\lambda -\pi^2/L^2)^{{1}/{2}}e+o(|\lambda
-\pi^2/L^2|^{{1}/{2}}),\label{8.105}
\end{equation}
where $e$ is a first eigenfunction of (\ref{8.61}). The expression
(\ref{8.105}) can be derived from the reduced equations
(\ref{8.89}).

We address here that the exponent $\beta ={1}/{2}$ in
(\ref{8.105}), called the critical exponent in physics, is an
important index in the phase transition theory in statistical
physics, which arises only in the Type-I or the continuous phase
transitions. It is interesting to point out that the critical
exponent $\beta =1$ in (\ref{8.71}) is different from these $\beta
={1}/{2}$ appearing in (\ref{8.73}) and (\ref{8.105}). The
first one occurs when  the container $\Omega\subset \R^3$ is a 
non rectangular region, and the second one occurs when $\Omega$ is a
rectangle or a cube. We shall continue to discuss this problem later
from the physical viewpoint. \qed
}\er

\section{Phase Transitions Under Periodic Boundary Conditions}
\la{s8.2.5}

When the sample or container $\Omega$ is a loop, or a torus, or 
bulk in size, then the periodic boundary conditions are necessary.
In this section, we shall discuss the problems in a loop domain
and in the whole space $\Omega =\R^n$.

Let $\Omega =S^1\times (r_1,r_2)\subset \R^2$ be a loop domain,
$0<r_1<r_2$. Then the boundary condition is given by
\begin{equation}
\left.
\begin{aligned} 
&u(\theta +2k\pi ,r)=u(\theta ,r)&& \text{ for } 
0\leq\theta\leq 2\pi ,\  r_1<r<r_2,\   k\in \Z,\\
&\frac{\partial u}{\partial r}=0,\ \ \ \
\frac{\partial^3u}{\partial r^3}=0&& \text{ at } r=r_1,r_2.
\end{aligned}
\right.\label{8.106}
\end{equation}

Assume that the gap $r_2-r_1$ is small in comparison with the mean
radius $r_0=(r_1+r_2)/2$. With proper scaling, we take
the gap $r_2-r_1$ as $r_2-r_1=1$. Then this assumption is
\begin{equation}
r_0=(r_1+r_2)/2\gg 1, \qquad r_2-r_1 =1.\label{8.107}
\end{equation}
With this  condition, the Laplacian operator can be
approximately expressed as
\begin{equation}
\Delta =\frac{\partial^2}{\partial
r^2}+\frac{1}{r^2_0}\frac{\partial^2}{\partial\theta^2}.\label{8.108}
\end{equation}

With the boundary condition (\ref{8.106}) and the operator
(\ref{8.108}), the eigenvalues and eignfunctions of the linear
operator $L_{\lambda}=-A+B_{\lambda}$ defined by (\ref{8.59}) are
given by
\begin{align*}
& 
\beta_K(\lambda )=K^2(\lambda -K^2), \\
& e^1_K=\cos k_1\theta\cos k_2\pi (r-r_1),\\
& e^2_K=\sin k_1\theta\cos k_2\pi (r-r_1), \\
& K^2=\frac{k^2_1}{r^2_0}+k^2_2 &&  k_1,k_2\in \Z.
\end{align*}

By (\ref{8.107}), the first eigenvalue of $L_{\lambda}$
is
$$\beta_1(\lambda )=\frac{1}{r^2_0}(\lambda -\frac{1}{r^2_0}),$$
which has multiplicity 2, and the first eigenfunctions are
$e^1_1=\cos\theta $  and $e^2_1=\sin\theta .$

\bt\la{t8.7}
Let $\Omega =S^1\times (r_1,r_2)$ 
satisfy (\ref{8.107}). Then the following assertions hold true:

\begin{enumerate}

\item 
If 
$$\gamma_3>\frac{2r^2_0}{9}\gamma^2_2, $$
then 
the phase transition of (\ref{8.58}) with (\ref{8.106}) at
$\lambda_0=1/ r^2_0$ is Type-I. Furthermore, the problem
(\ref{8.58}) with (\ref{8.106}) bifurcates on $\lambda
>\lambda_0=r^{-2}_0$ to a cycle attractor $\Sigma_{\lambda}=S^1$
which consists of singular points, as shown in Figure \ref{f8.12}, and
the singular points in $\Sigma_{\lambda}$ can be expressed as
\begin{align*}
& u_{\lambda}=\sigma^{-{1}/{2}}\left(\lambda
-\frac{1}{r^2_0}\right)^{{1}/{2}}\cos (\theta
+\theta_0)+o\left(|\lambda -\frac{1}{r^2_0}|^{{1}/{2}}\right),\\
& \sigma =\frac{3}{4}\gamma_3-\frac{1}{6} r^2_0\gamma^2_2,
\end{align*}
where $\theta_0$ is the angle of $u_{\lambda}$ in
$\Sigma_{\lambda}$. 

\item If
$$\gamma_3<\frac{2r^2_0}{9}\gamma^2_2, $$  
then the transition  is Type-II. Moreover,  the problem bifurcates on
$\lambda <\lambda_0$ to  a cycle invariant set $\Gamma_{\lambda}=S^1$
consisting of singular points, and there is a singularity
separation at $\lambda^*<\lambda_0$,  where $\Gamma_{\lambda}$
and an $S^1$ attractor $\Sigma_{\lambda}=S^1$ are generated such
that the system undergoes a transition at $\lambda =\lambda_0$ from $u=0$ to
$u_{\lambda}\in\Sigma_{\lambda}$, as shown in Figure \ref{f8.13}.
\end{enumerate}
\et

\begin{SCfigure}[25][t]
  \centering
  \includegraphics[width=0.4\textwidth]{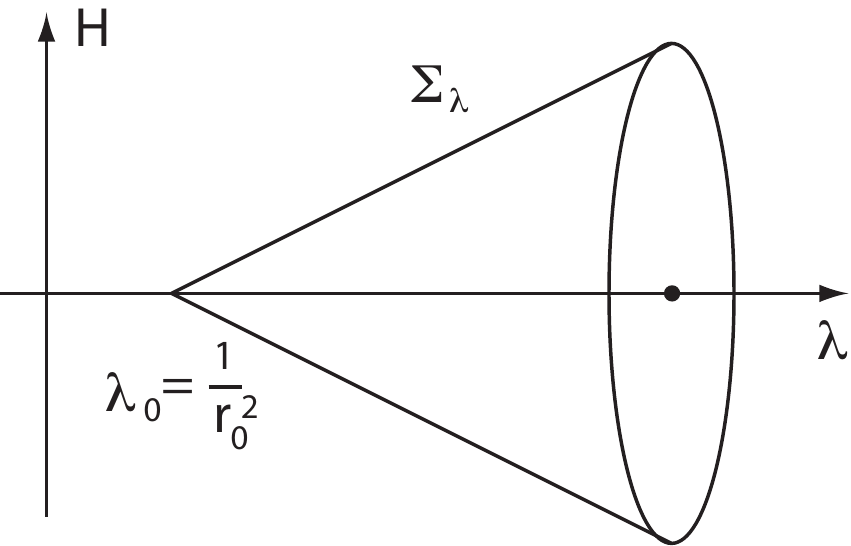}
  \caption{All points in $\Sigma_{\lambda}$ are singular
points.}\la{f8.12}
 \end{SCfigure}
\begin{SCfigure}[25][t]
  \centering
  \includegraphics[width=0.4\textwidth]{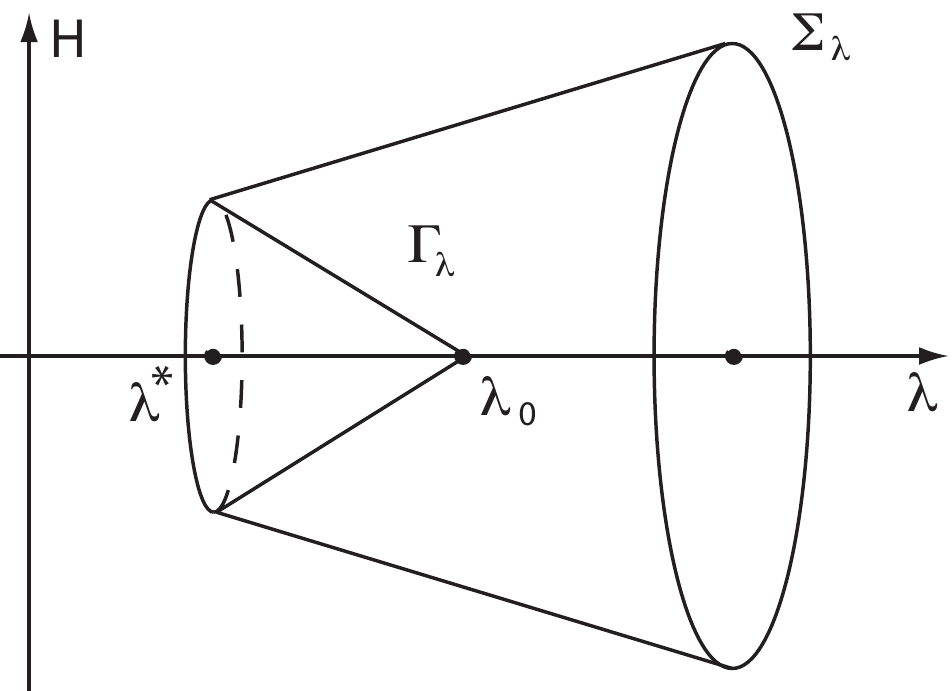}
  \caption{The point $\lambda^*$ is a singularity separation point
from where two invariant sets $\Sigma_{\lambda}$ and
$\Gamma_{\lambda}$ are separated with $\Sigma_{\lambda}$ being an
attractor, and $\Gamma_{\lambda}\rightarrow 0$ as
$\lambda\rightarrow\lambda_0$.}\la{f8.13}
 \end{SCfigure}

\bp
Let  $v=y\cos\theta +z\sin\theta$, 
$u=v+\Phi (y,z)$,  and  $\Phi$ is the center manifold function. Then the
reduced equations of (\ref{8.58}) with (\ref{8.106}) are given by
\begin{equation}
\left.
\begin{aligned} 
&\frac{dy}{dt}=\beta_1(\lambda
)y+\frac{1}{\pi}\int^{r_1+1}_{r_1}\int^{2\pi}_0[\gamma_2\Delta
v^2+\gamma_3\Delta v^3+\gamma_2\Delta u^2]e^1_1drd\theta ,\\
&\frac{dz}{dt}=\beta_1(\lambda
)z+\frac{1}{\pi}\int^{r_1+1}_{r_1}\int^{2\pi}_0[\gamma_2\Delta
v^2+\gamma_3\Delta v^3+\gamma_2\Delta u^2]e^2_1drd\theta .
\end{aligned}
\right.\label{8.109}
\end{equation}
Direct computation shows that  (\ref{8.109})  can be  rewritten as
\begin{equation}
\left.
\begin{aligned} 
&\frac{dy}{dt}
=\beta_1y-\frac{1}{\pi
r^2_0}\left[\frac{3\pi}{4}\gamma_3y(y^2+z^2)
+
\int^{2\pi}_0 2\gamma_2 \Phi [y \cos^2\theta
 + z \sin\theta\cos\theta] d\theta
\right],\\
&\frac{dz}{dt}=\beta_1z-\frac{1}{\pi
r^2_0}\left[\frac{3\pi}{4}\gamma_3z(y^2+z^2)+
\int^{2\pi}_02\gamma_2 \Phi [ z\sin^2\theta
 + y \cos\theta\sin\theta] d\theta \right],
\end{aligned}
\right.\label{8.110}
\end{equation}
and the center manifold function $\Phi =\Phi (y,z)$ is
$$
\Phi =\frac{2\gamma_2}{\beta_2(\lambda )r^2_0}(y^2-z^2)\cos 2\theta
+\frac{4\gamma_2}{\beta_2(\lambda )r^2_0}yz\sin 2\theta,
$$ 
where
$$
\beta_2(\lambda )=\frac{4}{r^2_0}(\lambda -\frac{4}{r^2_0}).
$$
Putting $\Phi$ into (\ref{8.110}), we obtain the approximate
equations of (\ref{8.109}) as follows
\begin{equation}
\left.
\begin{aligned}
&\frac{dy}{dt}=\beta_1y-\frac{1}{r^2_0}\left(\frac{3}{4}\gamma_3+\frac{\gamma^2_2}{2(\lambda
-4r^{-2}_0)}\right)y(y^2+z^2),\\
&\frac{dz}{dt}=\beta_1z-\frac{1}{r^2_0}\left(\frac{3}{4}\gamma_3+\frac{\gamma^2_2}{2(\lambda
-4r^{-2}_0)}\right)z(y^2+z^2).
\end{aligned}
\right.\label{8.111}
\end{equation}
At $\lambda =\lambda_0=r^{-2}_0$, we have
$$\frac{3}{4}\gamma_3+\frac{\gamma^2_2}{2(\lambda
-4r^{-2}_0)}=\frac{3}{4}\gamma_3-\frac{1}{6}r^2_0\gamma^2_2.$$

In the same fashion as used in Theorem \ref{t8.6}, we derive from (\ref{8.111}) the assertions of this theorem. Here the statement that the 
bifurcated cycle $\Sigma_{\lambda}=S^1$ consists of singular
points  can proved by that the equation (\ref{8.58}) with
(\ref{8.106}) is invariant under the translation
$$u(\theta ,r)\rightarrow u(\theta +\theta_0,r) \ \ \ \
\forall\theta_0\in \R^1,$$ which ensures that the singular points
of (\ref{8.58}) with (\ref{8.106}) arise as a cycle $S^1$. Thus,
the theorem is proved. 
\ep

\bigskip

Now, we consider the problem that the equation (\ref{8.58}) is
defined in the whole space $\Omega =\R^n$  $(n\geq 2)$, with the
periodic boundary condition
\begin{equation}
u(x+2K\pi )=u(x)\qquad \forall  K=(k_1,\cdots,k_n)\in \Z^n.\label{8.112}
\end{equation}
In this case, the eigenvalues and eigenfunctions of $L_{\lambda}$
are given by
\begin{align*}
& 
\beta_K(\lambda )=|K|^2(\lambda -|K|^2),&& K=(k_1,\cdots,k_n),\quad  |K|^2=k^2_1+\cdots+k^2_n, \\
& e^1_K=\cos (k_1x_1+\cdots+k_nx_n),&& e^2_K=\sin (k_1x_1+\cdots+k_nx_n).
\end{align*}
 It is clear that the first eigenvalue $\beta_1(\lambda
)=\lambda -1$ of $L_{\lambda}$ has multiplicity $2n$, and the
first eigenfunctions are 
$$e^1_j=\cos x_j,\ \ \ \ e^2_j=\sin x_j\qquad  \forall 1\leq j\leq n.$$

\bt\la{t8.8}
\begin{enumerate}

\item If
$$\gamma_3>\frac{14}{27}\gamma^2_2,$$ 
then the phase transition of
(\ref{8.58}) with (\ref{8.112}) at $\lambda_0=1$ is Type-I. Moreover, 

\begin{enumerate}
\item the problem bifurcates
from $(u,\lambda )=(0,1)$  to an attractor $\Sigma_{\lambda}$
homeomorphic to a $(2n-1)$-dimensional sphere $S^{2n-1}$, and 
\item 
for each $k\ (0\leq
k\leq n-1)$, the attractor $\Sigma_{\lambda}$ contains
$C^k_n$-dimensional tori $\T^{n-k}$
 consisting of singular points.
\end{enumerate}

\item 
If 
$$\gamma_3<\frac{14}{27}\gamma^2_2,$$ 
 then the transition  is Type-II. 
\end{enumerate}
\et

\bp
We only have to prove Assertion (2), as the remaining part of the theorem is 
essentially the same as  the proof for Theorem \ref{t8.6}. 

Since the space of all even functions is an invariant space of
$L_{\lambda}+G$ defined by (\ref{8.59}), the problem (\ref{8.58})
with (\ref{8.112}) has solutions given in (\ref{8.102}) with
$m=n$ in the space of even functions.

By the translation invariance of (\ref{8.58}) and (\ref{8.112}),
for each $k$  $(0\leq k\leq n-1)$ and a fixed index
$(j_1,\cdots,j_k)$, the steady state solution associated with
(\ref{8.102}) generates an $(n-k)$-dimensional torus  $\T^{n-k}$ which
consists of steady state solutions of (\ref{8.58}) and
(\ref{8.112}). For example if $(j_1,\cdots,j_k)=(1,\cdots,k)$, the
$(n-k)$-dimensional singularity torus $\T^{n-k}$ is
$$\T^k=\left\{u(x+\theta )=\sum^n_{j=k+1}y_j\cos (x_j+\theta_j)+o(|y|),\
\ \ \ \forall (\theta_{n+1},\cdots,\theta_n)\in \R^{n-k}\right\},
$$ 
where
$u(x)$ is the steady state solution of (\ref{8.58}) with
(\ref{8.112}) associated with (\ref{8.102}) with $y_1=\cdots
=y_k=0, y_{k+1}=\cdots=y_n$.

Obviously, for a fixed $(j_1,\cdots,j_k)$, the $2^{n-k}$ steady
state solutions of (\ref{8.58}) and (\ref{8.112}) associated with
(\ref{8.102}) are in the same singular torus $\T^{n-k}$.
Furthermore, for two different index $k$-tuples $(j_1,\cdots,j_k)$
and $(i_1,\cdots,i_k)$, the two associated singularity tori are
different. Hence, for each $0\leq k\leq n-1$, there are exactly
$C^k_n$  $(n-k)$-dimensional singularity tori in $\Sigma_{\lambda}$.
Thus the proof is complete.
\ep

\section{Cahn-Hilliard Equations Coupled with Entropy}
\la{s8.2.6}

When a phase separation takes place in a binary system, the entropy
varies, and if the phase transition is Type-II, it will yield
latent heat. Hence, it is necessary to discuss the equations
(\ref{8.51}), which are called the Cahn-Hilliard equations coupled
with entropy.

To make the equations (\ref{8.51}) non-dimensional, let
\begin{align*}
&x=lx^{\prime},&&  t=\frac{1}{k_2}l^4t^{\prime},&& 
u=u_0u^{\prime},&& S=S_0S^{\prime},\\
&\mu =\frac{k_1}{k_2}l^2, &&  \alpha_1=\frac{1}{k_2}l^4a_1,&& \alpha_2=\frac{l^2}{k_2S_0}u^2_0a_2,&& \lambda =-\frac{l^2}{k_2}b_1,\\
&
\gamma_1=\frac{l^2}{k_2}S_0b_0,&&
\gamma_2=\frac{l^2}{k_2}u_0b_2,&&
\gamma_3=\frac{l^2}{k_2}u^2_0b_3.
\end{align*}
Omitting the primes, equations (\ref{8.51}) are in the following
form
\begin{equation}
\left.\begin{aligned} &\frac{\partial S}{\partial t}=\mu\Delta
S-\alpha_1S-\alpha_2u^2,\\
&\frac{\partial u}{\partial t}=-\Delta^2u-\lambda\Delta u+\Delta
(\gamma_1Su+\gamma_2u^2+\gamma_3u^3),\\
&\int_{\Omega}u(x,t)dx=0,\\
&\frac{\partial }{\partial n} ( u,     \Delta u, S) =0 \qquad  \text{ on } \partial \Omega,\\
&u(x,0)=\varphi (x).
\end{aligned}
\right.\label{8.113}
\end{equation}

By assumptions (\ref{8.50}) and (\ref{8.52}), the coefficients
satisfy
$$\mu >0,\ \ \ \ \alpha_1>0,\ \ \ \ \alpha_2>0,\ \ \ \
\gamma_1>0,\ \ \ \ \gamma_3>0.$$

\bt\la{t8.9}
Let $\Omega
=\Pi^n_{k=1}(0,L_k)\subset \R^n$ satisfy that $L=L_1=\cdots
=L_m>L_j(1\leq m\leq n)$ for  any $j> m$. 

\begin{enumerate}

\item[(1)] For the case where  $m=1$, let 
$$\sigma
=\frac{3}{2}\gamma_3-\frac{\alpha_2\gamma_1}{\alpha_1}-\frac{\alpha_2\gamma_1L^2}{2(\alpha_1L^2+4\pi^2\mu
)}-\frac{L^2\gamma^2_2}{3\pi^2}.$$

\begin{enumerate}

\item  If $\sigma <0$, then  the phase transition of
(\ref{8.113}) at $\lambda =\pi^2/L^2$ is Type-II and Assertion (1)
in Theorem \ref{t8.5} holds  true.

\item If $\sigma >0$ the phase transition is
Type-I and Assertion (2) in Theorem \ref{t8.5} holds  true.

\end{enumerate}

\item[(2)] For the case where  $m\geq 2$, let 
$$\widetilde{\sigma}=\frac{9}{2}\gamma_3-\alpha_2\gamma_1\left(\frac{2}{\alpha_1}+\frac{L^2}{2(\alpha_1L^2+
4\pi^2\mu
)}+\frac{2L^2}{\alpha_1L^2+2\pi^2\mu}\right)-\frac{13L^2\gamma^2_2}{3\pi^2}.$$

\begin{enumerate}

\item If
$\widetilde{\sigma}>0$,  the phase transition of (\ref{8.113}) at
$\lambda =\pi^2/L^2$ is Type-I and Assertions (1) and (2) in
Theorem \ref{t8.6} hold true. 

\item If $\widetilde{\sigma}<0$, then  the phase transition
is Type-II and Assertion (3) in  Theorem \ref{t8.6} holds  true.
\end{enumerate}

\end{enumerate}
\et

\bp
 It suffices to compute the reduced equations of
(\ref{8.113}) on the center manifold. Similar to (\ref{8.89}), the
second order approximation of the reduced equation can be
expressed as
\begin{align}
\frac{dy_i}{dt}=&\beta_1y_i-\frac{\pi^2}{2L^2}[\sigma_1y^3_i+\sigma_2y_i\sum_{j\neq
i}y^2_j]\label{8.114}\\
&-\frac{2\pi^2}{L_1\cdots
L_nL^2}\gamma_1\sum^m_{j=1}y_j\int_{\Omega}\Phi_1(y)\cos\frac{\pi
x_i}{L}\cos\frac{\pi x_j}{L}dx,\nonumber
\end{align}
where $\beta_1,\sigma_1$ and $\sigma_2$ are as in (\ref{8.89}),
and the center manifold function $\Phi_1(y)$ derived
from the first equation in (\ref{8.113}) can be expressed as
\begin{equation}
\left.
\begin{aligned} 
&\Phi_1(y)=\sum^{\infty}_{|K|\geq
0}\Phi_K(y)\varphi_K\\
&\Phi_K(y)=\frac{-\alpha_2}{\lambda_K\|\varphi_K\|^2}\int_{\Omega}
\left(\sum^m_{i=1}y_i\cos
\frac{\pi x_i}{L}\right)^2\varphi_Kdx
\end{aligned}
\right.\label{8.115}
\end{equation}
where $\lambda_K$ and $\varphi_K$ are the eigenvalues and
eigenfunctions of the following equation
\begin{align*}
&-\mu\Delta\varphi_K+\alpha_1\varphi_K=\lambda_K\varphi_K,\\
&\frac{\partial\varphi_K}{\partial n}|_{\partial\Omega}=0,
\end{align*} 
which are given by 
\begin{align*}
& \lambda_K=\alpha_1+\mu K^2\pi^2,\\
& \varphi_0=1,\ \ \ \ \varphi_K=\cos\frac{k_1\pi
x_1}{L^2_1}\cdots\cos\frac{k_n\pi x_n}{L^2_n}, \\
& K=(k_1/L_1,\cdots,k_n/L_n),\quad   |K|^2=\sum^n_{i=1}k^2_i/L^2_i&& \text{ for } k_i\in \Z,\ \ \ \ 1\leq i\leq n.
\end{align*}

Let 
$$K_i=(\delta_{i1}/L_1,\cdots,\delta_{in}/L_n).$$
Then  we find
\begin{align*}
\Phi_0 =&  \frac{-\alpha_2}{\alpha_1L_1\cdots L_n}\int_{\Omega}\left(\sum^n_{i=1}y_i\cos\frac{\pi
x_i}{L}\right)^2dx=\frac{-\alpha_2}{2\alpha_1}\sum^m_{j=1}y^2_j, 
\\
\Phi_k=&\frac{-\alpha_2}{\lambda_K\|\varphi_K\|^2}\int_{\Omega}\left(\sum^m_{i=1}y_i\cos\frac{\pi
x_i}{L}\right)^2dx\\
=&\left\{
  \begin{aligned} 
     &  0&& \text{ if }  K\neq K_l+K_r,\\
     &\frac{-\alpha_2}{\lambda_K\|\varphi_K\|^2}\sum^m_{i,j=1}y_iy_j\int_{\Omega}\cos\frac{\pi
x_i}{L}\cos\frac{\pi x_j}{L}\varphi_Kdx&& \text{ if }  K=K_l+K_r,
\end{aligned}
\right.
\end{align*}
for some $1\leq r, l\leq m$. Thus we derive that
\begin{align*}
\Phi_K=
&\left\{
\begin{aligned}
  &\frac{-\alpha_2}{\lambda_K\|\varphi_K\|^2}y^2_j\int_{\Omega}\cos^2\frac{\pi
x_j}{L}e_k&& \text{ if }  K=K_j+K_j, \ \ 1\leq j\leq m\\
&\frac{-2\alpha_2}{\lambda_K\|\varphi_K\|^2}y_iy_j\int_{\Omega}\cos\frac{\pi
x_i}{L}\cos\frac{\pi x_j}{L}e_K && \text{ if } K=K_i+K_j, \ \ i\neq j
\end{aligned}
\right.\\
=&\left\{\begin{aligned} &\frac{-\alpha_2}{2(\alpha_1+4\pi^2\mu
/L^2)}y^2_j  \qquad \qquad && \text{ if }K=2K_j, \ \ 1\leq j\leq m,\\
&\frac{-2\alpha_2}{\alpha_1+2\pi^2\mu /L^2}y_iy_j  && \text{ if }
K=K_i+K_j, \ \ i\neq j,\  \ 1\leq i,j\leq m.
\end{aligned}
\right.
\end{align*}
Putting $\Phi_0$ and $\Phi_K$ in (\ref{8.115}) we obtain
\begin{align*}
\Phi_1=&\frac{-\alpha_2}{2\alpha_1}\sum^n_{j=1}y^2_j-\frac{\alpha_2}{2(\alpha_1+4\pi^2\mu
/L^2)}\sum^n_{j=1}y^2_j\cos\frac{2\pi x_j}{L}\\
&-\frac{2\alpha_2}{\alpha_1+2\pi^2\mu
/L^2}\sum_{i<j}y_iy_j\cos\frac{\pi x_i}{L}\cos\frac{\pi x_j}{L}.
\end{align*}

Then, inserting $\Phi_1$ into (\ref{8.114}),  we derive the following reduced
equations:
\begin{align}
\frac{dy_i}{dt}=
&\beta_1(\lambda)y_i-\frac{\pi^2}{2L^2}\left[\left(\sigma_1-\frac{\alpha_2\gamma_1}{\alpha_1}-\frac{\alpha_2\gamma_1}
{2(\alpha_1+4\pi^2\mu /L^2}\right)y^3_i\right.\label{8.116}\\
&+ \left. \left(\sigma_2-\frac{\alpha_2\gamma_1}{\alpha_1}-\frac{2\alpha_2\gamma_1}{\alpha_1+2\pi^2\mu
/L^2}\right)y_i\sum_{j\neq i}y^2_j \right],\ \ \ \ 1\leq i\leq
m. \nonumber
\end{align}
Then the remaining part of the proof can be achieved in the same fashion as the proofs for Theorems \ref{t8.5} and \ref{t8.6}. The proof is complete.
\ep

\br\la{r8.7}
{\rm
From the phenomenological viewpoint, the
coefficient $\mu >0$ in (\ref{8.113}) is small. If let $\mu =0$,
then in the equilibrium state we have that
$S=-\frac{\alpha_2}{\alpha_1}u^2$. In this case, (\ref{8.113}) are
referred to the original Cahn-Hilliard equation (\ref{8.58}) with
$\gamma^{\prime}_3=\gamma_3-\alpha_2\gamma_1/\alpha_1$ as  
the coefficient of the cubic term, and the
criterion  $\sigma =0$ in Assertions (1) and (2) of Theorem \ref{t8.9}
are respectively equivalent to
$$\gamma^{\prime}_3=\frac{2}{9}\frac{L^2\gamma^2_2}{\pi^2}\ \ \ \
\text{and}\ \ \ \
\gamma^{\prime}_3=\frac{26}{27}\frac{L^2\gamma^2_2}{\pi^2},$$
which coincide with these in Theorems \ref{t8.5} and \ref{t8.6}. Hence, if we
consider the coefficient $\mu >0$ small, then the criterion
$\sigma =0$ are respectively equivalent to
$$\gamma^{\prime}_3=\frac{2}{9}\frac{L^2\gamma^2_2}{\pi^2}-\frac{4}{3}\frac{\pi^2\mu\alpha_2\gamma_1}{\alpha^2_1L^2},
\ \ \ \ \text{and}\ \ \ \ \
\gamma^{\prime}_3=\frac{26}{27}\frac{L^2\gamma^2_2}{\pi^2}-\frac{4}{3}\frac{\pi^2\mu\alpha_2\gamma_1}{\alpha^2_1L^2}.$$
Hence the item $-(4\pi^2\mu\alpha_2\gamma_1)/3\alpha^2_1L^2$ is
the effect yielded by $\mu\Delta S$.
\qed
}\er

\section{Physical remarks}
\la{s8.2.7}

We now address  the physical significance for the phase
transition theorems obtained in the previous sections.

\subsection{Equation of critical parameters}For a binary system, the equation describing the control
parameters $T,p,\Omega$ at the critical states is simple. 

We first
consider the critical temperature $T_c$. There are two different
critical temperatures $T_1$ and $T_0$ in the Cahn-Hilliard
equation. $T_1$ is the one given by (\ref{8.56}),  at which the
coefficient $b_1(T,p)$ or $\lambda =-l^2b_1(T,p)/k$ will change its
sign, and $T_0$ satisfies that $\lambda_0<\lambda_1$, and for
fixed $p$,
\begin{equation}
\lambda (T)\left\{
\begin{aligned} 
&  <\rho_1&& \text{ if } T>T_0,\\
& =\rho_1  && \text{ if } T=T_0,\\
& >\rho_1 && \text{ if } T<T_0,
\end{aligned}
\right.
\label{8.117}
\end{equation}
where $\rho_1$ is the first eigenvalue of (\ref{8.61}), which
depends on the geometrical properties of the material such as the size of the container of the sample $\Omega$. When $\Omega =(0,L)^m\times D\subset \R^n$ is a
rectangular domain with $L>$ diameter of $D, \rho_1=\pi^2/L^2$.
Hence, in general at the critical temperature $T_1$ a binary
system does not undergo any phase transitions, but the phase transition  does occur at $T=T_0$.

At $T_1$ and $T_0$ we know that
\begin{equation}
\lambda (T_1)=0,\ \ \ \ \lambda (T_0)=\rho_1.\label{8.118}
\end{equation}
For a rectangular domain, $\rho_1=\pi^2/L^2$, therefore from
(\ref{8.118}) we see that $T_1$ is a limit of the critical
temperature $T_0$ of phase transition as the size of $\Omega$
tends to infinite.

In fact,  for a general domain, it is easy to see that the first eigenvalue $\rho_1$ of the Laplace operator
is inversely  proportional to the square of  the  maximum diameter of
$\Omega$:
\begin{equation}
\rho_1\sim\frac{1}{L^2},\label{8.120}
\end{equation}
where $L$ represents the diameter scaling of $\Omega$. 

Thus the equation  of critical parameters in the
Cahn-Hilliard equation, by (\ref{8.118}) and (\ref{8.120}), is
given by
\begin{equation}
\lambda (T,p)=\frac{C}{L^2},\label{8.121}
\end{equation}
where $C>0$ is a constant depending on the geometry of $\Omega$.
According to the Hildebrand theory (see Reichl \cite{reichl}), the function
$\lambda (T,p)$ can be expressed in a explicit formula. If
regardless of the term $|\nabla u|^2$, the molar Gibbs free energy
takes the following form
\begin{equation}
g=\mu_A(1-u)+\mu_Bu+RT(1-u)\ln (1-u)+RTu\ln
u+au(1-u),\label{8.122}
\end{equation}
where $\mu_A,\mu_B$ are the chemical potential of $A$ and $B$
respectively, $R$ the molar gas constant, $a>0$ the measure of
repel action between $A$ and $B$. Therefore, the coefficient $b_1$
in (\ref{8.55}) with constant $p$ is
$$b_1=\frac{\partial^2g}{\partial
u^2}|_{u=u_0}=\frac{1}{u_0(1-u_0)}RT-a\ \ \ \ (a=a(p)),$$ where
$u_0=\bar{u}_B$ is the constant concentration of $B$. Hence
$$\lambda
(T,p)=-\frac{l^2}{k}b_1=\frac{2al^2}{k}-\frac{l^2R}{ku_0(1-u_0)}T.$$
Thus, equation (\ref{8.121}) is expressed as
\begin{equation}
\frac{l^2R}{ku_0(1-u_0)}T=\frac{2al^2}{K}-\frac{C}{L^2}.\label{8.123}
\end{equation}
Equation (\ref{8.123}) gives the critical parameter curve of a
binary system with constant pressure for temperature $T$ and
diameter scaling $L$ of container $\Omega$.  Because $T\geq 0$,
from (\ref{8.122}) we can deduce the following physical
conclusion.

\bigskip

\noindent
{\bf Physical Conclusion 7.1.}  
{\it 
Under constant pressure, for any
binary system with given geometrical shape of the container $\Omega$,
there is a value $L_0>0$ such that as the diameter scaling
$L<L_0$, no phase separation takes place at all temperature $T\geq
0$, and as $L>L_0$ phase separation will occur at some critical
temperature $T_0>0$ satisfying (\ref{8.123}).
}

\medskip

We shall see later that it is a universal property  that the dynamical
properties of phase transitions depend on the geometrical shape
and size of the container or sample $\Omega$.

\subsection{Physical explanations of phase transition theorems}

We first briefly recall the classical
thermodynamic theory for a binary system. Physically, phase
separation processes taking place in an unstable state are called
spinodal decompositions; see Cahn and Hilliard \cite{CH57} and Onuki \cite{onuki}. 
When consider the concentration $u$ as homogeneous in $\Omega$,
then by (\ref{8.122}) the dynamic equation of a binary system is
an ordinary differential equation:
\begin{equation}
\frac{du}{dt} = -\frac{dg}{du}\label{8.124}=
2au-RT\ln\frac{u}{1-u}+\mu_A-\mu_B-a.
\end{equation}

Let $u_0$  $(0<u_0<1)$ be the steady state solution of (\ref{8.124}).
Then, by the Taylor expansion at $u=u_0$, omitting the $n$th order
terms with $n\geq 4$, (\ref{8.124}) can be rewritten as
\begin{equation}
\frac{dv}{dt}=\lambda v+b_2v^2+b_3v^3,\label{8.125}
\end{equation}
where 
\begin{align*}
& v=u-u_0,   &&   \lambda =2a-\frac{1}{u_0(1-u_0)}RT,\\
& b_2=\frac{1-2u_0}{2u^2_0(1-u_0)^2}RT, &&b_3=-\frac{1}{3}\frac{1-u_0-2u^2_0+3u^3_0}{u^3_0(1-u_0)^4}RT.
\end{align*}
It is easy to see that
\begin{align*}
&b_2\left\{
  \begin{aligned} 
     & =0&& \text{ if } u_0=\frac{1}{2},\\
&   \neq 0&& \text{ if } u_0\neq\frac{1}{2},
\end{aligned}
\right.\\
&b_3<0 \qquad   \forall 0<u_0<1.
\end{align*}

It is clear that the critical parameter curve $\lambda=0$  in the $T-u_0$ plane is given by 
$$T_0=2a u_0 (1-u_0)/R,$$
which is schematically illustrated in the classical phase diagram; see the dotted line in Figure~\ref{f8.14}. We obtain from (\ref{8.125}) the following  transition steady states: 
$$
v^\pm = \frac{-1}{2b_3}(b_2 \pm \sqrt{b_2^2 -4b_3 \lambda})\qquad \text{ for } b_2^2 -4b_3 \lambda>0.
$$
By Theorem~\ref{t8.4}, we see that there is $T^*=T^*(u_0)$   satisfying 
that $b_2^2-4b_3 \lambda=0$; namely, 
\begin{equation}
 T^* = \frac{ 2 a u_0 (1-u_0)}{R(1-\beta(u_0))}, \qquad 
 \beta(u_0)= \frac{3 (1-2u_0)^2(1-u_0)}{16(1-u_0-2u_0^2 + 3u_0^3)},
 \label{8.126-1}
\end{equation}
such that if $T_0 < T < T^*$, 
\begin{align*}
&
v^+ \text{ is } \left\{
\begin{aligned}
& \text{ locally stable (metastable) } && \text{ for } 0 < u_0 < \frac12, \\
& \text{ unstable } && \text{ for } \frac12 < u_0 < 1, 
\end{aligned}
\right.  \\
&
v^- \text{ is } \left\{
\begin{aligned}
& \text{ locally stable (metastable) } && \text{ for }   \frac12< u_0 <1, \\
& \text{ unstable } && \text{ for } 0 < u_0 <  \frac12.
\end{aligned}
\right.  
\end{align*}
Here 
$$T^*(u_0) =\frac{T_0(u_0)}{1-\beta_0(u_0)} \ge T_0(u_0)$$
is illustrated by the solid line in Figure~\ref{f8.14}. 
This shows that the region $T_0(u_0) < T < T^*(u_0)$  is metastable, which is marked 
by the shadowed region in Figure \ref{f8.14}. See, among others,  Reichl \cite{reichl}, 
 Novick-Cohen and Segal \cite{NS84} ,  and Langer \cite{langer71} 
for the phase transition diagram from the classical thermodynamic theory.

\begin{SCfigure}[25][t]
  \centering
  \includegraphics[width=0.5\textwidth]{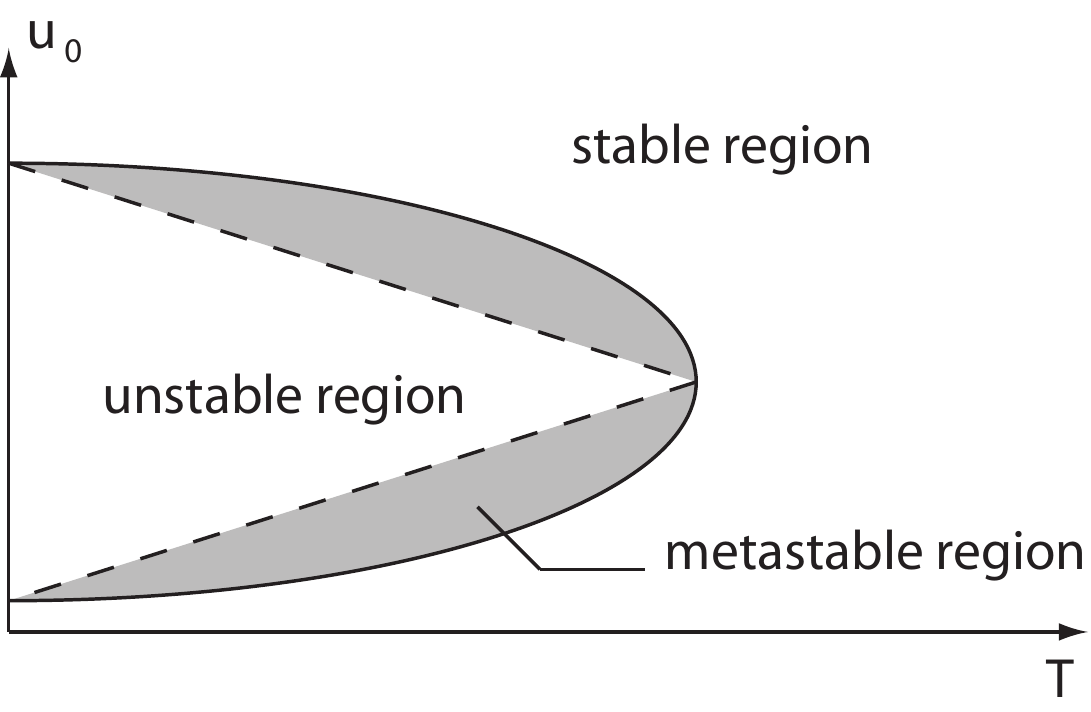}
  \caption{Typical phase diagram from classical
thermodynamic theory.}
  \la{f8.14}
 \end{SCfigure}

In the following we shall discuss the spinodal decomposition in a unified fashion by
applying the phase transition theorems presented in the previous sections.

As mentioned in the Introduction, phase separation processes
of binary systems occur in two ways, one of which proceeds
continuously depending on $T$, and the other one does not.
Obviously, the classical theory does not explain these phenomena.
In fact, the first one can be described by the Type-I phase
transition, and the second one can be explained by the Type-II and
Type-III phase transitions.

We first consider the case where the container $\Omega
=\Pi^n_{i=1}(0,L_i)$ with $L=L_1=\cdots=L_m>L_j(j>m)$ is a
rectangular  domain. Thus, by Theorems \ref{t8.5} and \ref{t8.6} 
(or Theorem \ref{t8.9})
there are only two phase transition types: Type-I  and Type-II, 
with the type of transition  depending on $L$. We see that if
$$L^2<\left\{\begin{aligned}
&\frac{9}{2}\frac{\pi^2\gamma_3}{\gamma^2_2} && \text{for}\
m=1,\\
&\frac{27}{26}\frac{\pi^2\gamma_3}{\gamma^2_2}  && \text{for}\
m\geq 2, \end{aligned} \right.
$$ 
then the transition  is Type-I, i.e., the phase
pattern formation gradually varies as the  temperature
decreases. In this case, no meta-stable states and no latent heat
appear. The phase diagram is given by Figure \ref{f8.15}, 
where the solid lines $u_i^T$ ($i=1,2$) represent   the transition solutions.
\begin{SCfigure}[25][t]
  \centering
  \includegraphics[width=0.4\textwidth]{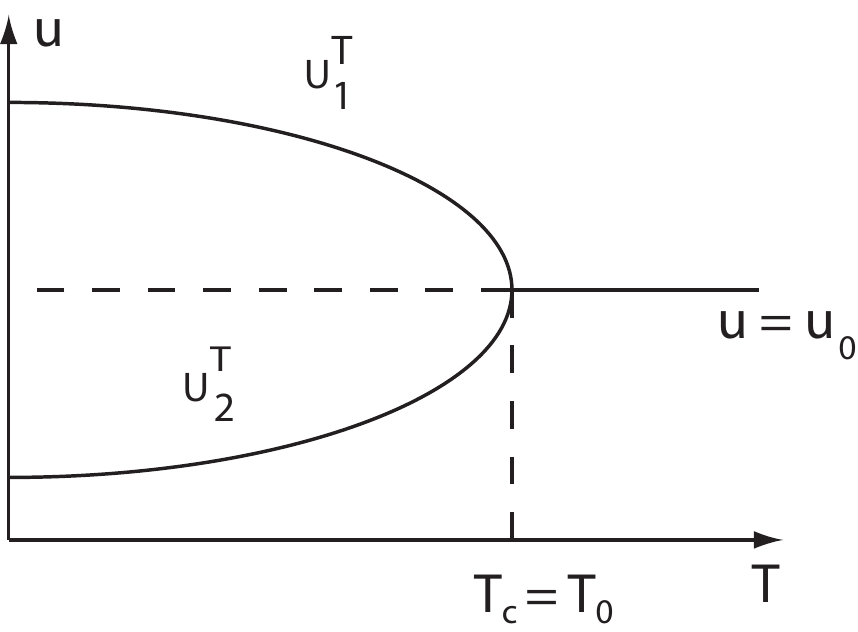}
  \caption{The state $u_0=\bar{u}_B$ is
stable  if $T_c=T_0 < T$, and  the state $u_0$ is unstable  if $T < T_0$, where $T_0$
is as in (\ref{8.117}).}
  \la{f8.15}
 \end{SCfigure}

If $L$ satisfies that
$$L^2>\left\{\begin{aligned}
&\frac{9}{2}\frac{\pi^2\gamma_3}{\gamma^2_2} &&  \text{for}\
m=1,\\
&\frac{27}{26}\frac{\pi^2\gamma_3}{\gamma^2_2} && \text{for}\
m\geq 2,
\end{aligned}
\right.
$$ 
then the phase transition is Type-II. Namely,  there is a
leaping change in phase pattern formation at the critical temperature
$T_c$. The phase diagram for Type-II  transition is given by Figure \ref{f8.16}.
\begin{SCfigure}[25][t]
  \centering
  \includegraphics[width=0.4\textwidth]{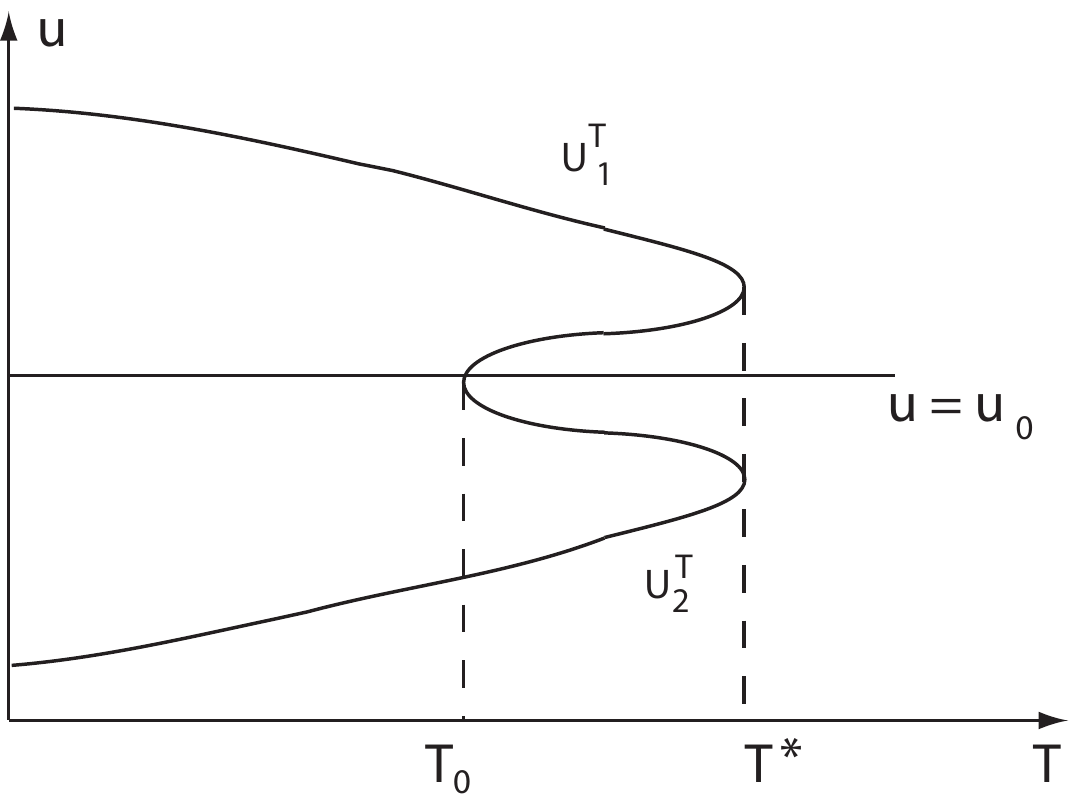}
  \caption{A Type-II phase transition.}
\la{f8.16}
 \end{SCfigure}

In Figure \ref{f8.16}, 
$T_0$ is the critical temperature as in (\ref{8.117}), 
$T^*$  is  defined by (\ref{8.126-1}) and is the saddle-node bifurcation point of (\ref{8.125}).
The constant concentration $u=u_0$ is stable in $T_c<T$,  is
meta-stable in $T_0<T<T_c$, and is unstable in $T<T_0$.
The two bifurcated states $U^T_1$
and $U^T_2$  from $T^*$ are meta-stable in $T_0<T<T^*$, and  are stable in $T<T_0$. 
Here for $i=1,2$, $U_i^T=u^T_i + u_0$, and $u_i$ are the separated solutions of 
(\ref{8.58}) with (\ref{8.62}) from $T^*$.

There is a remarkable difference between Type-I and Type-II transitions.  The
Type-I phase transition occurs at $T=T_0$ and Type-II does in
$T_0<T<T^*$. Furthermore, latent heat is 
accompanied  the Type-II phase transition. Actually, when a binary
system undergoes a transition  from $u_0$ to $U^T_i$  $(i=1,2)$, 
there is a gap  $|U^T_i-u_0|^2=|u_i^T|^2>\varepsilon >0$  for any $T_0<T<T^*$. 
By the first
equation in (\ref{8.113}) it yields a jump of entropy between
$u_0$ and $U^T_i$:
$$
\delta S_i=\int_{\Omega}Sdx
=-\frac{\alpha_2}{\alpha_1}\int_{\Omega}|u^T_i|^2dx<0,
$$
where $S=S_i-\bar{S}_0$ represents the entropy density deviation.
Hence the latent heat is given by
$$\delta H=T\delta
S_i=-\frac{\alpha_2T}{\alpha_1}\int_{\Omega}|u^T_i|^2dx<0,$$
which implies that the process from $u_0$ to $U^T_i$ is
exothermic, and the process from $U^T_i$ to $u_0$ is endothermic.

Now, we consider the case where the container $\Omega$ is
non rectangular. Thus, by Theorem \ref{t8.4} the transition is Type-III,   and its
phase diagram is given by Figure \ref{f8.17}.
\begin{SCfigure}[25][t]
  \centering
  \includegraphics[width=0.4\textwidth]{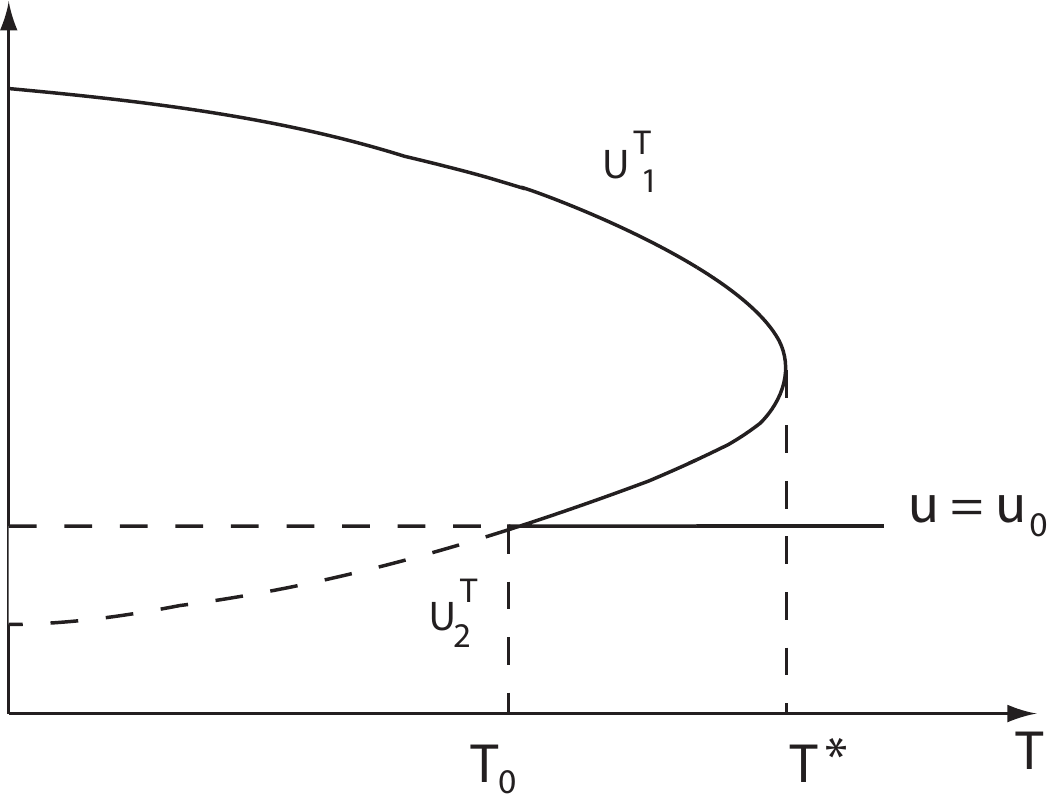}
  \caption{A Type-III phase transition.}\la{f8.17}
 \end{SCfigure}

In Figure \ref{f8.17},  $T_0$  and $T^*$ are the same as those in Figure \ref{f8.16}. The state $u_0$ is stable in
$T^*<T$, is metastable in $T_0<T<T^*$, and unstable in $T<T_0$.
The equilibrium state $U^T_1$ separated from $T^*$ is metastable in
$T_0<T<T^*$, and is stable in $T<T_0$. However, the equilibrium
state $U^T_2$ separated from $T^*$ is unstable in $T_0<T<T^*$, and
is metastable in $T<T_0$.

Similar to the Type-II, the Type-III phase transition has also
latent heat, which occurs in $T_0<T<T^*$. But the difference
between Type-II and Type-III is that Type-II has $2m $  $(m\geq 1)$
stable equilibrium states separated from $T=T^*$, but Type-III has
just one. The $2m$ stable states of a Type-II transition  are of some symmetry
caused by $\Omega$, and we shall investigate it later. A
particular aspect of Type-III is that there is a  state
$U^T_2$ bifurcated from $(u,T)=(u_0,T_0)$, which is rarely
observed in experiments.

\subsection{Symmetry and periodic structure}Physical experiments have shown that in pattern formation via
phase separation,  periodic or semi-periodic structure appears. From
Theorems \ref{t8.7} and \ref{t8.8} we  see that for the loop domains and bulk
domains which can be considered as $\R^n$ or $\R^m\times D$  $(D\subset
\R^{n-m})$ the steady state solutions of the Cahn-Hilliard equation
are periodic, and for rectangular domains they are semi-periodic,  and the periodicity is associated with the mirror image symmetry.

Let $\Omega =(0,L)^m\times D$  $(m\geq 1)$. By Remark  \ref{r8.5},
Theorem \ref{t8.5} is valid for $\Omega$. Actually, in this case the
following space
$$
\widetilde{H}=\left\{u\in L^2(\Omega )|\
u=\sum^{\infty}_{|K|=1}y_k\cos\frac{k_1\pi}{L}x_1\cdots\cos\frac{k_m\pi}{L}x_m
\right\}\subset
H
$$ 
is invariant for the Cahn-Hilliard equation (\ref{8.58}) and
(\ref{8.62}). All separated equilibrium states in Theorem \ref{t8.5} are
in $\widetilde{H}$. From the physical viewpoint, all equilibrium states
$u(x)$ and their mirror image states
$u(L-x^{\prime},x^{\prime\prime})$ are the same to describe  the pattern
formation. Mathematically, under the mirror image transformation
$$
x\rightarrow (L-x^{\prime},x^{\prime\prime}),\ \ \ \
x^{\prime}=(x_1,\cdots,x_m),\ \ \ \
x^{\prime\prime}=(x_{m+1},\cdots,x_n),
$$ 
the Cahn-Hilliard equation (\ref{8.58}) with (\ref{8.62}) is invariant. Hence,
 the steady state solutions will appear in pairs. In
particular, for  Type-I phase transition, there is a remarkable mirror
image symmetric. We address this problem as follows.

Let $m=1$ in $\Omega =(0,L)^m\times D$. By (\ref{8.73}) there are
two bifurcated stable equilibrium states, and their projections on
the first eigenspace are
\begin{eqnarray*}
&&u_1=y\cos\frac{\pi x_1}{L},\ \ \ \ u_2=-y\cos\frac{\pi
x_1}{L},\\
&&y=\sqrt{2(\lambda
-\pi^2/L^2)}/\left(\frac{3}{2}\gamma_3-\frac{L^2}{3\pi^2}\gamma^2_2\right).
\end{eqnarray*}
It is clear that $u_2(x_1)=u_1(L-x_1)$.

Let $m=2$. By Theorem \ref{t8.6} the bifurcated attractor
$\Sigma_{\lambda}$ contains 8 equilibrium states, whose projections
are given by
\begin{eqnarray*}
&&u^{\pm}_1=\pm y_0e_1,\ \ \ \ u^{\pm}_2=\pm y_1(e_1+e_2),\\
&&u^{\pm}_3=\pm y_0e_2,\ \ \ \ u^{\pm}_4=\pm y_1(-e_1+e_2),
\end{eqnarray*}
where $e_1=\left(\cos\frac{\pi
x_1}{L},0\right)$  and $e_2=\left(0,\cos\frac{\pi x_2}{L}\right)$ form an
orthogonal basis in $\R^2$, and 
\begin{eqnarray*}
&&y_0=\sqrt{2(\lambda -\pi^2/L^2)}/\sqrt{\sigma_1},\\
&&y_1=\sqrt{2(\lambda -\pi^2/L^2)}/\sqrt{\sigma_1+\sigma_2}\ \ \ \ \text{ with }
\sigma_1,\sigma_2\ \text{as\ in\ (\ref{8.90})}.
\end{eqnarray*}
These eight equilibrium states constitute an octagon in $\R^2$, as
shown in Figure \ref{f8.18}, and they are divided into two classes:
${\mathcal{A}}_1=\left\{u^{\pm}_1,u^{\pm}_3\right\}$ and
${\mathcal{A}}_3=\left\{u^{\pm}_2,u^{\pm}_4\right\}$ by the
$\frac{\pi}{2}$-rotation group $G(\frac{\pi}{2})$. 
Namely,  with the action of $G(\frac{\pi}{2}), {\mathcal{A}}_i$  $(i=1,2)$ are
invariant:
$$Bu\in{\mathcal{A}}_i,\ \ \ \ \forall u\in{\mathcal{A}}_i\
\text{and}\ B\in G(\frac{\pi}{2}),$$ 
where $G(\frac{\pi}{2})$ consists of
the orthogonal matrices
$$B^{\pm}_1=\pm\left(\begin{matrix}
1&0\\
0&1
\end{matrix}
\right),\ \ \ \ 
B^{\pm}_2=\pm\left(\begin{matrix} 1&0\\
0&-1
\end{matrix}
\right),\ \ \ \ 
B^{\pm}_3=\pm\left(\begin{matrix} 0&-1\\
1&0
\end{matrix}
\right),\ \ \ \ 
B^{\pm}_4=\pm\left(\begin{matrix} 0&1\\
1&0
\end{matrix}
\right).$$ 
The stability of the equilibrium states $u_k$ is associated
with  both classes ${\mathcal{A}}_1$ and ${\mathcal{A}}_2$,
i.e., either the elements in ${\mathcal{A}}_1$ are stable, or those
in ${\mathcal{A}}_2$ are stable; see Figure \ref{f8.9-1}. By (\ref{8.104})
we can derive  the criterion as follows
\begin{align*}
&  u^{\pm}_{2k+1}\in{\mathcal{A}}_1 
\text{ are stable }  &&  \Leftrightarrow \quad \frac{22}{9}\frac{L^2 \gamma^2_2}{\pi^2}>\gamma_3>\frac{26}{27}\frac{L^2\gamma^2_2}{\pi^2} && \text{ for } k=0,1, \\
& u^{\pm}_{2k}\in{\mathcal{A}}_2  \text{ are stable } && \Leftrightarrow \quad \gamma_3>\frac{22}{9}\frac{L^2 \gamma^2_2}{\pi^2} && \text{ for } k=1, 2,
\end{align*}
\begin{SCfigure}[25][t]
  \centering
  \includegraphics[width=0.4\textwidth]{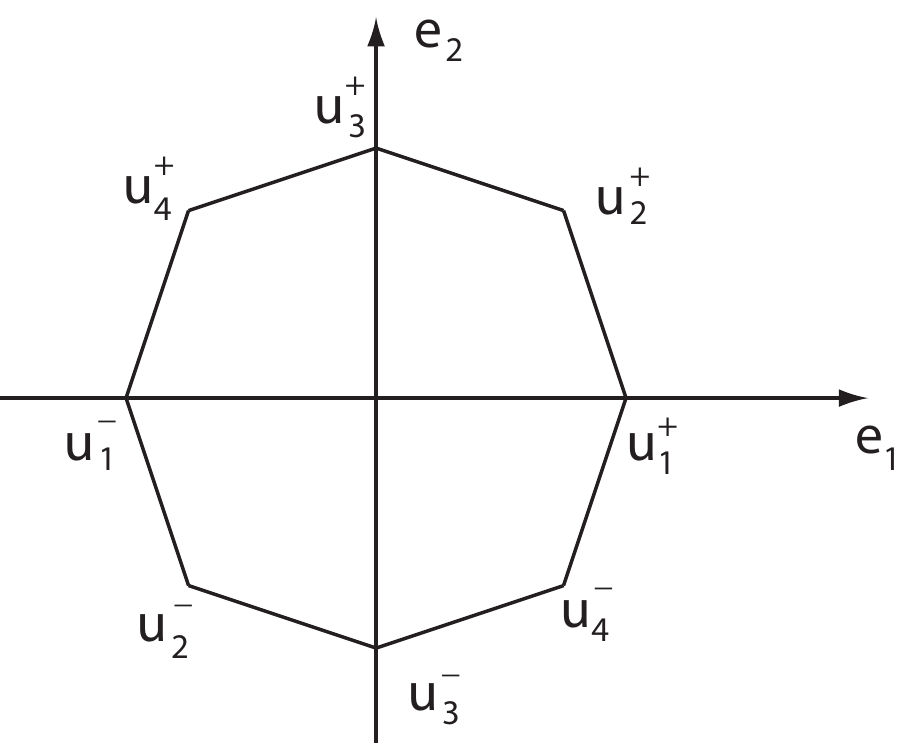}
  \caption{The eight equilibrium states in the case where $m=2$ 
  given by Theorem~\ref{t8.6}.}\la{f8.18}
 \end{SCfigure}

In Figure \ref{f8.18}, we see that elements  in ${\mathcal{A}}_1$ and
${\mathcal{A}}_2$ have a $\frac{\pi}{4}$ difference in their phase
angle. However, in their pattern structure,
$u^{\pm}_{2K+1}\in{\mathcal{A}}_1$ and
$u^{\pm}_{2k}\in{\mathcal{A}}_2$ also have a $\pi /4$ deference at
the angle between the lines of $u^{\pm}_{2K+1}=0$ and
$u^{\pm}_{2K}=0$. In fact, the lines that
\begin{eqnarray*}
&&u^+_1=-u^-_1=y_0\cos\frac{\pi x_1}{L}=0,\ \ \ \ \text{and}\\
&&u^+_3=-u^-_3=y_0\cos\frac{\pi x_2}{L}=0
\end{eqnarray*}
are given by $x_1=L/2$ and $x_2=L/2$ respectively, as shown in
Figure \ref{f8.19}(a), and the lines
\begin{eqnarray*}
&&u^+_2=-u^-_2=y_1\left(\cos\frac{\pi x_1}{L}+\cos\frac{\pi
x_2}{L}\right)=0,\ \ \ \ \text{and}\\
&&u^+_4=-u^-_4=y_1\left(-\cos\frac{\pi x_1}{L}+\cos\frac{\pi
x_2}{L}\right)=0
\end{eqnarray*}
are given by $x_2=L-x_1$ and $x_2=x_1$ respectively as shown in
Figure \ref{f8.19}(b).
\begin{figure}[hbt]
  \centering
  \includegraphics[width=0.28\textwidth]{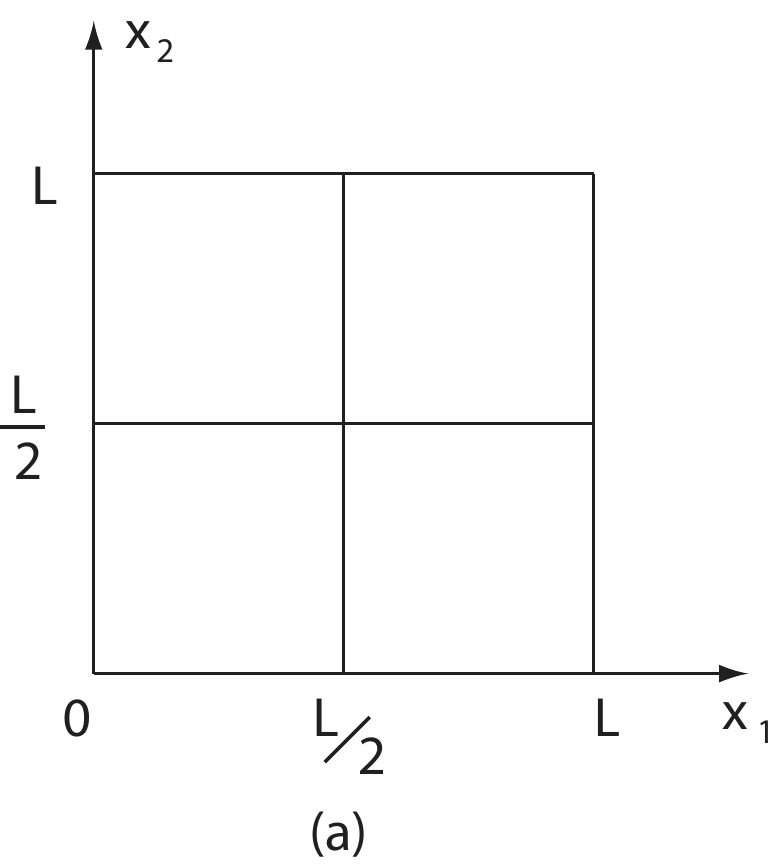}\qquad
  \includegraphics[width=0.28\textwidth]{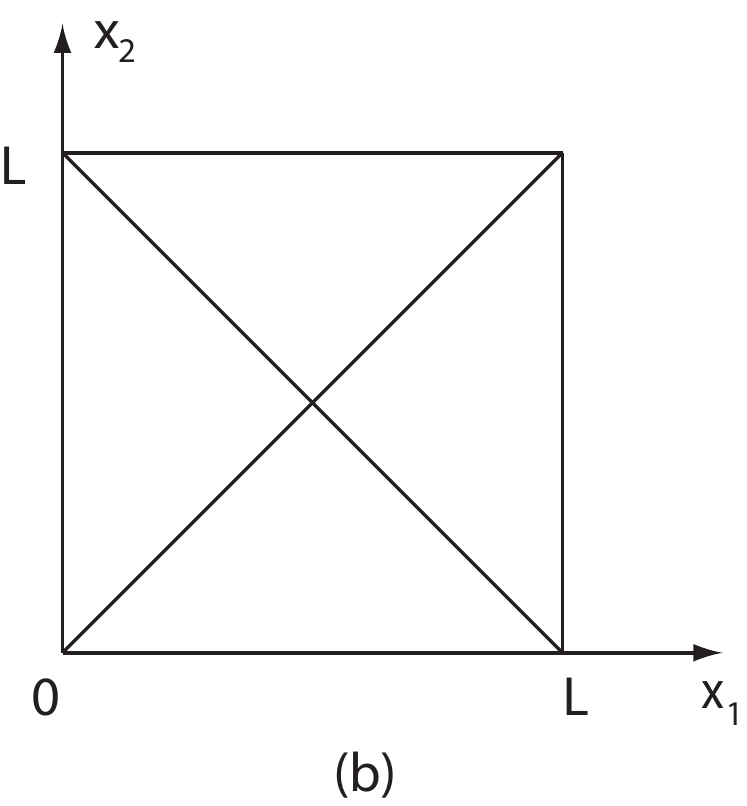}
  \caption{}\la{f8.19}
 \end{figure}

Let $m=3$. Then the bifurcated attractor $\Sigma_{\lambda}$
contains 26 equilibrium states which can by divide into three
classes by the 3-dimensional
$\left(\frac{\pi}{2},\frac{\pi}{2},\frac{\pi}{2}\right)$-rotation
group $G\left(\frac{\pi}{2},\frac{\pi}{2},\frac{\pi}{2}\right)$ as
follows
\begin{align*}
&{\mathcal{A}}_1=\left\{u^{\pm}_1=\pm y_0e_1,u^{\pm}_2=\pm
y_0e_2,u^{\pm}_3=\pm y_0e_3\right\}, \\
&{\mathcal{A}}_2=\left\{u^{\pm}_4=\pm y_1(e_1+e_2),u^{\pm}_5=\pm
y_1(e_1+e_3),u^{\pm}_6=\pm y_1(e_2+e_3),\right.\\
&\ \ \ \ u^{\pm}_7=\pm y_1(-e_1+e_2),u^{\pm}_8=\pm
y_1(-e_1+e_3),u^{\pm}_9=\pm y_1(-e_2+e_3)\}, \\
&{\mathcal{A}}_3=\{u^{\pm}_{10}=\pm
y_1(e_1+e_2+e_3),u^{\pm}_{11}=\pm y_1(-e_1+e_2+3_3),\\
&\ \ \ \ u^{\pm}_{12}=\pm y_1(e_1-e_2+e_3),u^{\pm}_{13}=\pm
y_1(e_1+e_2-e_3)\}.
\end{align*}
Only these elements in ${\mathcal{A}}_1$ or in ${\mathcal{A}}_3$
are stable, and  they are determined by the
following criterion
\begin{align*}
&\text{elements in}\ {\mathcal{A}}_1\ \text{is\
stable}\Leftrightarrow\frac{22}{9}\frac{L^2\gamma^2_2}{\pi^2}>\gamma_3>\frac{26}{27}\frac{L^2\gamma^2_2}{\pi^2},\\
&\text{elements\ in}\ {\mathcal{A}}_3\ \text{is\
stable}\Leftrightarrow \gamma_3>\frac{22}{9}\frac{L^2\gamma^2_2}{\pi^2}.
\end{align*}

\subsection{Critical exponents}
From (\ref{8.73}) and (\ref{8.105}) we see that for Type-I phase
transition of a binary system the critical exponent $\beta
=\frac{1}{2}$. In this case, it is a second order phase transition with the Ehrenfest classification scheme,
and there is a gap in heat capacity at critical temperature $T_0$.
To see this, by (\ref{8.73}) and (\ref{8.105}) we have
$$u^T=\left\{\begin{aligned}
& 0 &&\text{if}\ T_0<T,\\
&\alpha (\lambda (T)-\pi^2/L^2)^{{1}/{2}}e_1+o(|\lambda
-\pi^2/L^2|^{{1}/{2}})&&\text{if}\ T<T_0,
\end{aligned}
\right.
$$ 
and the free energy for (\ref{8.58}) at $u^T$ is
\begin{align*}
F(u^T)=&\int_{\Omega}\left[\frac{1}{2}|\nabla
u^T|^2-\frac{\lambda}{2}|u^T|^2+o(|u^T|^2)\right]dx\\
=&\int_{\Omega}\frac{1}{2}[-\Delta u^T-\lambda
u^T]u^T+\frac{1}{3}\gamma_2(u^T)^3+\frac{1}{4}\gamma_3(u^T)^4dx\\
=&\left\{\begin{aligned} 
&0 &&  \text{if}\ T_0<T,\\
&-\frac{\alpha^2}{2}(\lambda
-\pi^2/L^2)^2\cdot\int_{\Omega}[e^2_1+\frac{\alpha^2}{4}\gamma_3e^4_1]dx+o(|\lambda
-\pi^2/L^2|^2) && \text{if} \ T > T_0.
\end{aligned}
\right.
\end{align*}
Thus, the heat capacity $C$ at $T=T_0$ satisfies
$$C^+-C^-=-T_0\frac{\partial^2F(u^T)}{\partial
T^2}\Big|_{T_0^+}+T_0\frac{\partial^2F(u^T)}{\partial
T^2}\Big|_{T_0^-}=\alpha_1T_0\left(\frac{d\lambda}{dT}\right)^2|_{T=T_0}.$$
It is known that $d\lambda /dT\neq 0$; hence the heat capacity at
$T=T_0$ has a finite jump.

From (\ref{8.71}) we know that for the Type-III case, the critical
exponent $\beta =1$. Thus, it is not hard to  deduce
that the continuous phase transition in Type-III is of the 3rd order.

\appendix

\section{Dynamic Transition Theory for Nonlinear Systems}
In this appendix we recall some basic elements of the dynamic transition theory developed by the authors \cite{b-book, chinese-book}, which are used to carry out the dynamic transition analysis for the binary systems in this article. 

\subsection{New classification scheme}
Let $X$  and $ X_1$ be two Banach spaces,   and $X_1\subset X$ a compact and
dense inclusion. In this chapter, we always consider the following
nonlinear evolution equations
\begin{equation}
\left. 
\begin{aligned} 
&\frac{du}{dt}=L_{\lambda}u+G(u,\lambda),\\
&u(0)=\varphi ,
\end{aligned}
\right.\label{5.1}
\end{equation}
where $u:[0,\infty )\rightarrow X$ is unknown function,  and 
$\lambda\in \R^1$  is the system parameter.

Assume that $L_{\lambda}:X_1\rightarrow X$ is a parameterized
linear completely continuous field depending continuously on
$\lambda\in \R^1$, which satisfies
\begin{equation}
\left. 
\begin{aligned} 
&L_{\lambda}=-A+B_{\lambda}   && \text{a sectorial operator},\\
&A:X_1\rightarrow X   && \text{a linear homeomorphism},\\
&B_{\lambda}:X_1\rightarrow X&&  \text{a linear compact  operator}.
\end{aligned}
\right.\label{5.2}
\end{equation}
In this case, we can define the fractional order spaces
$X_{\sigma}$ for $\sigma\in \R^1$. Then we also assume that
$G(\cdot ,\lambda ):X_{\alpha}\rightarrow X$ is $C^r(r\geq 1)$
bounded mapping for some $0\leq\alpha <1$, depending continuously
on $\lambda\in \R^1$, and
\begin{equation}
G(u,\lambda )=o(\|u\|_{X_{\alpha}}),\ \ \ \ \forall\lambda\in
\R^1.\label{5.3}
\end{equation}

Hereafter we always assume the conditions (\ref{5.2}) and
(\ref{5.3}), which represent that the system (\ref{5.1}) has
a dissipative structure.

\bd\la{d5.1}
We say that the system (\ref{5.1}) has a
transition of equilibrium from $(u,\lambda )=(0,\lambda_0)$ on
$\lambda >\lambda_0$ (or $\lambda <\lambda_0)$ if  the following two conditions are 
satisfied:

\begin{itemize}

\item[(1)] when $\lambda <\lambda_0$ (or $\lambda >\lambda_0),
u=0$ is locally asymptotically stable for (\ref{5.1}); and

\item[(2)] when $\lambda >\lambda_0$ (or $\lambda <\lambda_0)$,
there exists a neighborhood $U\subset X$ of $u=0$ independent of
$\lambda$, such that for any $\varphi\in U \setminus \Gamma_{\lambda}$ the
solution $u_{\lambda}(t,\varphi )$ of (\ref{5.1}) satisfies that
$$\left. 
\begin{aligned}
&\limsup_{t\rightarrow\infty}\|u_{\lambda}(t,\varphi
)\|_X\geq\delta (\lambda )>0,\\
&\lim_{\lambda\rightarrow\lambda_0}\delta(\lambda )\geq 0,
\end{aligned}
\right.$$ 
where $\Gamma_{\lambda}$ is the stable manifold of
$u=0$, with  codim $\Gamma_{\lambda}\geq 1$ in $X$
for $\lambda >\lambda_0$ (or $\lambda <\lambda_0)$.
\end{itemize}
\ed

Obviously, the attractor bifurcation of (\ref{5.1}) is a type of
transition. However,  bifurcation and
transition are two different, but related concepts. 
Definition \ref{d5.1} defines the transition of (\ref{5.1}) from a stable
equilibrium point to other states (not necessary equilibrium state).
In general, we can define transitions from one attractor to
another as follows.

Let the eigenvalues (counting multiplicity) of $L_{\lambda}$ be given by
$$\{\beta_j(\lambda )\in \C\ \   |\ \ j=1,2,\cdots\}$$
Assume that
\begin{align}
&  \text{Re}\ \beta_i(\lambda )
\left\{ 
 \begin{aligned} 
 &  <0 &&    \text{ if } \lambda  <\lambda_0,\\
& =0 &&      \text{ if } \lambda =\lambda_0,\\
& >0&&     \text{ if } \lambda >\lambda_0,
\end{aligned}
\right.   &&  \forall 1\leq i\leq m,  \label{5.4}\\
&\text{Re}\ \beta_j(\lambda_0)<0 &&  \forall j\geq
m+1.\label{5.5}
\end{align}

The following theorem is a basic principle of transitions from
equilibrium states, which provides sufficient conditions and a basic
classification for transitions of nonlinear dissipative systems.
This theorem is a direct consequence of the center manifold
theorems and the stable manifold theorems; we omit the proof.

\bt\la{t5.1}
 Let the conditions (\ref{5.4}) and
(\ref{5.5}) hold true. Then, the system (\ref{5.1}) must have a
transition from $(u,\lambda )=(0,\lambda_0)$, and there is a
neighborhood $U\subset X$ of $u=0$ such that the transition is one
of the following three types:

\begin{itemize}
\item[(1)] {\sc Continuous Transition}: 
there exists an open and dense set
$\widetilde{U}_{\lambda}\subset U$ such that for any
$\varphi\in\widetilde{U}_{\lambda}$,  the solution
$u_{\lambda}(t,\varphi )$ of (\ref{5.1}) satisfies
$$\lim\limits_{\lambda\rightarrow\lambda_0}\limsup_{t\rightarrow\infty}\|u_{\lambda}(t,\varphi
)\|_X=0.$$ In particular, the attractor bifurcation of (\ref{5.1})
at $(0,\lambda_0)$ is a continuous transition.

\item[(2)] {\sc Jump Transition}: 
for any $\lambda_0<\lambda <\lambda_0+\varepsilon$ with some $\varepsilon >0$, there is an open
and dense set $U_{\lambda}\subset U$ such that 
for any $\varphi\in U_{\lambda}$, 
$$\limsup_{t\rightarrow\infty}\|u_{\lambda}(t,\varphi
)\|_X\geq\delta >0,$$ 
where $\delta >0$ is independent of $\lambda$. 
This type of transition  is also called the discontinuous 
transition. 

\item[(3)] {\sc Mixed Transition}: 
for any $\lambda_0<\lambda <\lambda_0+\varepsilon$  with some $\varepsilon >0$, 
$U$ can be decomposed into two open sets
$U^{\lambda}_1$ and $U^{\lambda}_2$  ($U^{\lambda}_i$ not necessarily
connected):
$$\bar{U}=\bar{U}^{\lambda}_1+\bar{U}^{\lambda}_2,\ \ \
\ U^{\lambda}_1\cap U^{\lambda}_2=\emptyset ,$$ 
such that
\begin{align*}
&\lim\limits_{\lambda\rightarrow\lambda_0}\limsup_{t\rightarrow\infty}\|u(t,\varphi
)\|_X=0   &&   \forall\varphi\in U^{\lambda}_1,\\
& \limsup_{t\rightarrow\infty}\|u(t,\varphi
)\|_X\geq\delta >0 && \forall\varphi\in U^{\lambda}_2.
\end{align*}
\end{itemize}
\et

With this theorem in our disposal, we are in position to give a new dynamic classification scheme for dynamic phase transitions.

\begin{definition}[Dynamic Classification of Phase Transition]
The phase transitions for  (\ref{5.1}) at $\lambda =\lambda_0$ is classified using  their  dynamic properties: continuous, jump, and mixed as given in Theorem~\ref{t5.1}, which are called Type-I, Type-II and Type-III respectively.
\end{definition}

An important aspect of the  transition theory is to determine which 
of the three types of transitions given by Theorem \ref{t5.1} occurs in
a specific  problem. By  reduction to the center manifold of
(\ref{5.1}), we know that the type of transitions for (\ref{5.1}) at
$(0,\lambda_0)$ is completely dictated  by its reduction equation
near $\lambda =\lambda_0$, which can be
expressed as:
\begin{equation}
\frac{dx}{dt}=J_{\lambda}x+PG(x + \Phi (x,\lambda ),\lambda ) \ \ \
\ \text{ for } x\in \R^m,\label{5.6}
\end{equation}
where $J_{\lambda}$ is the $m\times m$ order Jordan matrix
corresponding to the eigenvalues given by (\ref{5.4}),  $\Phi
(x,\lambda )$ is the center manifold function of (\ref{5.1}) near
$\lambda_0$, $P:X\rightarrow E_{\lambda}$ is the canonical
projection, and 
$$
E_{\lambda}=\cup_{1\leq i\leq m}\cup_{k\in N}\{u\in X_1|\ \
(L_{\lambda}-\beta_i(\lambda ))^ku=0\}
$$ 
is the eigenspace of $L_{\lambda}$.

By the spectral theorem, (\ref{5.6}) can be expressed into the
following explicit form
\begin{equation}
\frac{dx}{dt}=J_{\lambda}x+g(x,\lambda ),\label{5.7}
\end{equation}
where
\begin{equation}
\left. 
\begin{aligned} 
&g(x,\lambda )=(g_1(x,\lambda ),\cdots
,g_m(x,\lambda )),\\
&g_j(x,\lambda )= <G(\sum^m_{i=1}x_ie_i + \Phi (x,\lambda ),\lambda
), e^*_j> \quad   \forall 1\leq j\leq m.
\end{aligned}
\right.\label{5.8}
\end{equation}
Here $e_j$ and $e^*_j$  $(1\leq j\leq m)$ are the eigenvectors of
$L_{\lambda}$ and $L^*_{\lambda}$ respectively corresponding to the
eigenvalues $\beta_j(\lambda )$ as in (\ref{5.4}).

In particular, if $G(u,\lambda )$ has the Taylor expansion
\begin{equation}
G(u,\lambda )=G_k(u,\lambda )+o(\|u\|^k_{X_{\alpha}}),\label{5.9}
\end{equation}
for some $k\geq 2$, where $G_k(u,\lambda )$ is a $k$-multilinear operator, then (\ref{5.7})  can be rewritten as
\begin{equation}
\frac{dx}{dt}=J_{\lambda}x+g_k(x,\lambda
)+o(|x|^k),\label{5.10}
\end{equation}
where
\begin{align*}
&g_k(x,\lambda )=(g_{k1}(x,\lambda ),\cdots ,g_{km}(x,\lambda)),\\
&g_{kj}(x,\lambda )=<G_k(\sum^n_{i=1}x_ie_i,\lambda ),e^*_j>
&&    \forall 1\leq j\leq m.
\end{align*}

When $x=0$ is an isolated singular point of $g_k(x,\lambda )$, in
general the transition of (\ref{5.1}) is determined by the
first-order approximate bifurcation equation of (\ref{5.10}) as follows:
\begin{equation}
\frac{dx}{dt}=J_{\lambda}x+g_k(x,\lambda ).\label{5.11}
\end{equation}

The following theorem is useful to distinguish the transition
types of (\ref{5.1}) at $(u,\lambda )=(0,\lambda_0)$.

\bt\la{t5.2}
Let the conditions (\ref{5.4}) and
(\ref{5.5}) hold true, and $U\subset \R^m$ be a neighborhood of $x=0$.
Then we have the following assertions:

\begin{itemize}

\item[(1)] If the transition of (\ref{5.1}) at $(0,\lambda_0)$ is
continuous, then there is an open and dense set
$\widetilde{U}\subset U$ such that for any $x_0\in\widetilde{U}$
the solution $x(t,x_0)$ of (\ref{5.7}) at $\lambda =\lambda_0$
with $x(0,x_0)=x_0$ satisfies that
$$\lim\limits_{t\rightarrow\infty}x(t,x_0)=0.$$

\item[(2)] If there exists an open and dense set
$\widetilde{U}\subset U$ such that for any $x_0\in\widetilde{U}$
the solution $x(t,x_0)$ of (\ref{5.7}) at $\lambda =\lambda_0$
satisfies
$$\limsup_{t\rightarrow\infty}|x(t,x_0)|\neq 0,$$
then the transition is a jump transition. 

\item[(3)] If the transition is
mixed, then there exists an open set $\widetilde{U}\subset U$
such that for any $x_0\in\widetilde{U}$ the solution $x(t,x_0)$ of
(\ref{5.7}) at $\lambda =\lambda_0$ with $x(0,x_0)=x_0$ satisfies
$$\lim\limits_{t\rightarrow\infty}x(t,x_0)=0.$$

\item[(4)] If the vector field in (\ref{5.7}) at $\lambda
=\lambda_0$ satisfies
$$<J_{\lambda_0}x+g(x,\lambda_0),x><0\quad \forall x\in U,\ x\neq 0,$$
then the transition of (\ref{5.1}) at $(0,\lambda_0)$ is an
$S^{m-1}$-attractor bifurcation, and  the transition  is continuous. 

\item[(5)] If the vector field $g(x,\lambda_0)$ given by (\ref{5.8})
satisfies
$$<g(x,\lambda_0),x>>0 \ \ \ \ \forall x\in U,\ \ \ \ x\neq 0,$$
then the transition is jump.
\end{itemize}
\et

In general, the conditions in Assertions (1)-(3) of Theorem \ref{t5.2}
are not sufficient. They, however, do give sufficient conditions when (\ref{5.1}) has
a variational structure. To see this, let (\ref{5.1}) be a gradient-type equation. Under the conditions (\ref{5.4}) and (\ref{5.5}), in a
neighborhood $U\subset X$ of $u=0$, the center manifold $M^c$ in
$U$ at $\lambda =\lambda_0$ consists three subsets
$$M^c=W^u+W^s+D,$$
where $W^s$ is the stable set, $W^u$ is the unstable set, and $D$ is  the
hyperbolic set of (\ref{5.7}). Then we have the following
theorem.

\bt\la{t5.3} 
Let (\ref{5.1}) be a gradient-type equation,
and the conditions (\ref{5.4}) and (\ref{5.5}) hold true. If $u=0$
is an isolated singular point of (\ref{5.1}) at $\lambda
=\lambda_0$, then we have the following assertions:

\begin{itemize}

\item[(1)] The transition of (\ref{5.1}) at $(u,\lambda)=(0,\lambda_0)$ is continuous if and only if $u=0$ is locally asymptotically stable at $\lambda =\lambda_0$, i.e., the center
manifold is stable: $M^c=W^s$. Moreover, (\ref{5.1}) 
bifurcates from $(0,\lambda_0)$ to minimal attractors consisting  of
singular points of (\ref{5.1}). 

\item[(2)] If the stable set $W^s$ of (\ref{5.1}) has no interior points in $M^c$, i.e.,
$M^c=\bar{W}^u+\bar{D}$, then the transition is jump.
\end{itemize}
\et

\subsection{Transitions from simple eigenvalues}
We consider the transition of (\ref{5.1}) from a simple critical
eigenvalue. Let the eigenvalues $\beta_j(\lambda )$ of
$L_{\lambda}$ satisfy (\ref{5.4})  and (\ref{5.5}) with $m=1$.
Then the first eigenvalue $\beta_1(\lambda )$ must be a real eigenvalue. 
Let $e_1(\lambda )$ and $e^*_1(\lambda )$   be  the eigenvectors of
$L_{\lambda}$ and $L^*_{\lambda}$ respectively corresponding to
$\beta_1(\lambda )$ with
$$L_{\lambda_0}e_1=0,\ \ \ \ L^*_{\lambda_0}e^*_1=0,\ \ \ \
<e_1,e^*_1>=1.$$ 
Let $\Phi (x,\lambda )$    be the center manifold
function of (\ref{5.1}) near $\lambda =\lambda_0$. We assume that
\begin{equation}
<G(xe_1+\Phi (x,\lambda_0),\lambda_0),e^*_1>=\alpha
x^k+o(|x|^k),\label{5.36}
\end{equation}
where $k\geq 2$ an integer and $\alpha\neq 0$ a real number.
\begin{figure}
  \centering
  \includegraphics[width=0.35\textwidth]{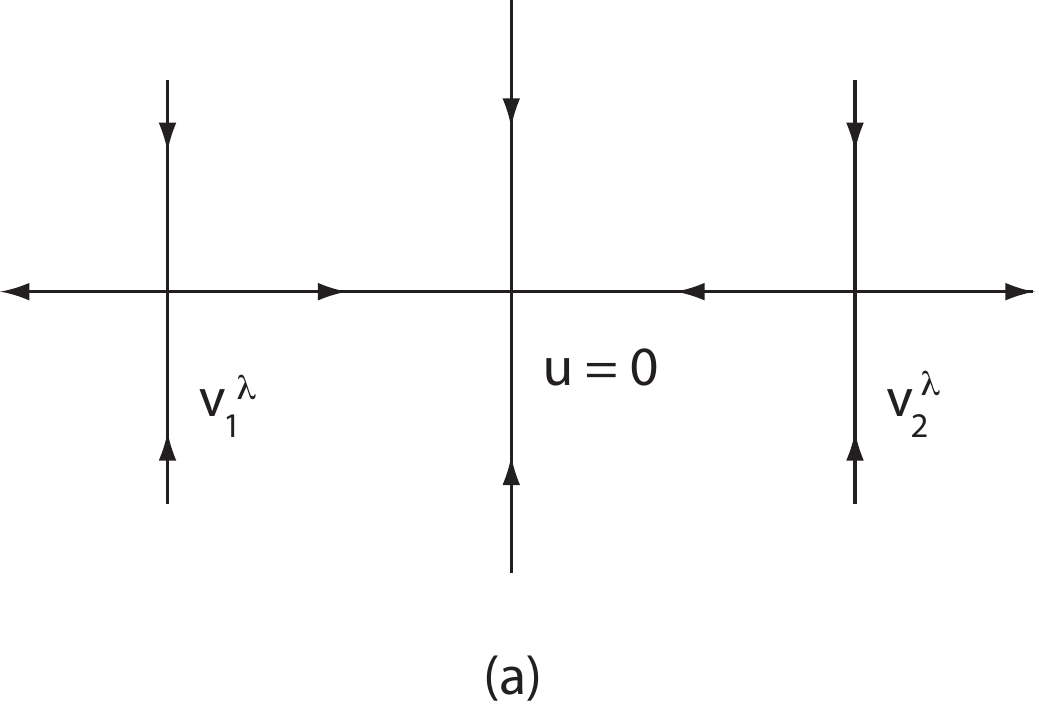} \quad 
  \includegraphics[width=0.2\textwidth]{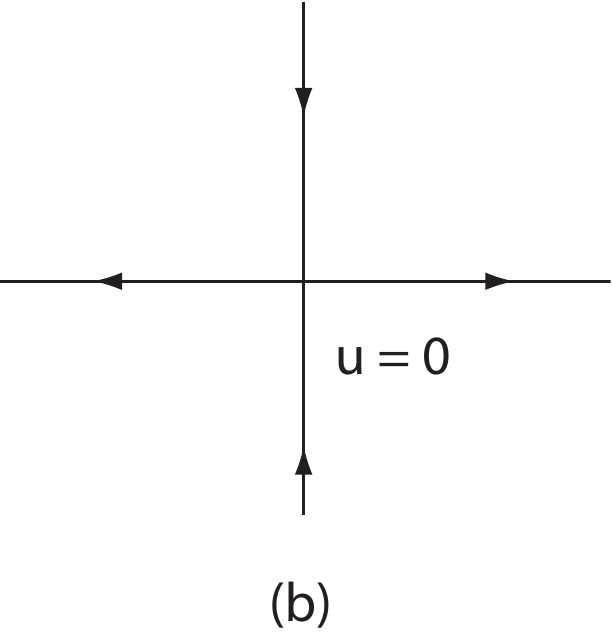}
  \caption{Topological structure of the jump transition of
(\ref{5.1}) when $k$=odd and $\alpha >0$: (a) $\lambda
<\lambda_0$; (b) $\lambda\geq\lambda_0$. Here the horizontal line
represents the center manifold.}\la{f5.5}
 \end{figure}
 \begin{figure}
  \centering
  \includegraphics[width=0.23\textwidth]{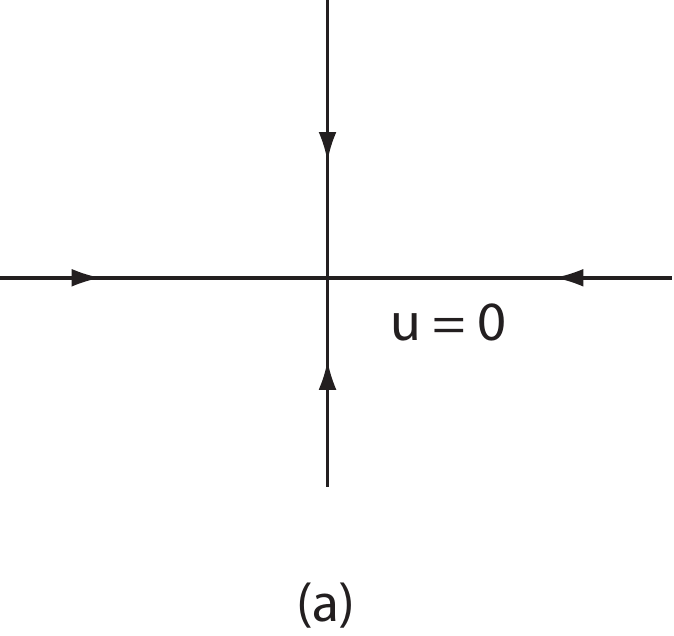}
  \includegraphics[width=0.35\textwidth]{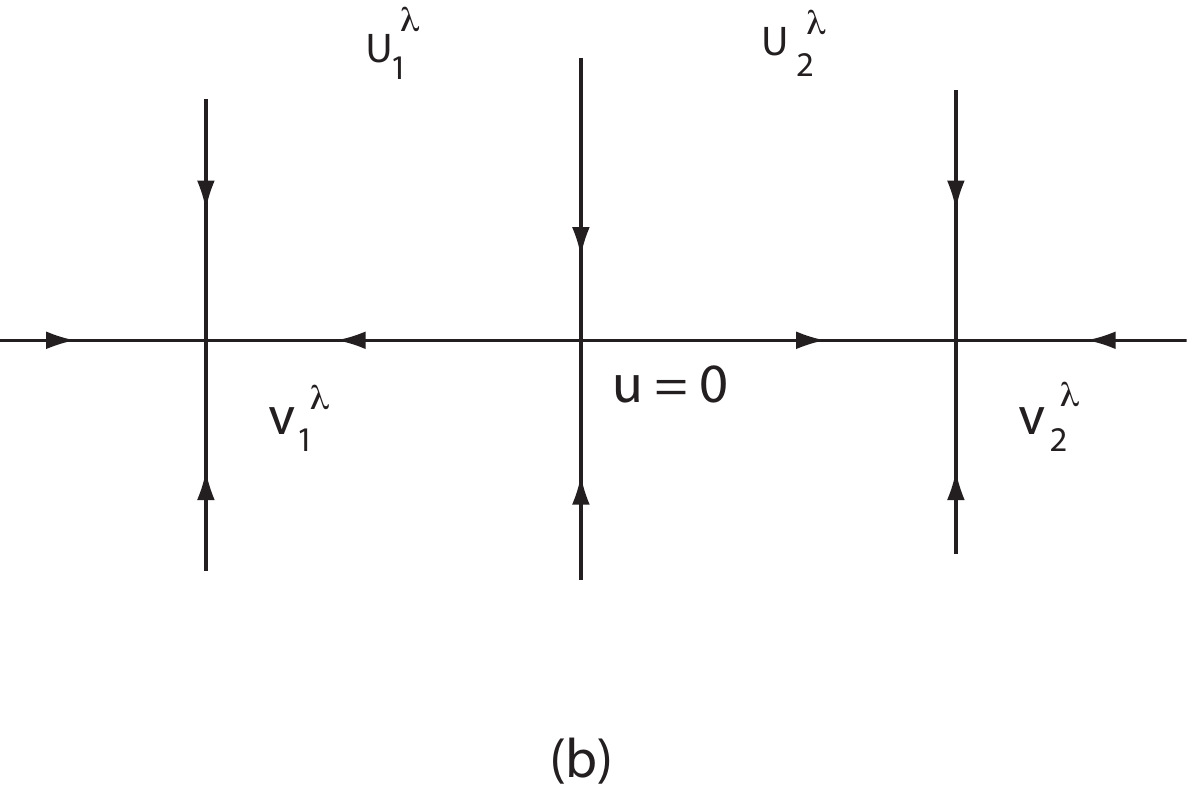}
  \caption{Topological structure of the continuous transition
of (\ref{5.1}) when $k$=odd and $\alpha <0$: (a)
$\lambda\leq\lambda_0$; (b) $\lambda >\lambda_0$.}\la{f5.6}
 \end{figure}
 \begin{figure}
  \centering
  \includegraphics[width=0.32\textwidth]{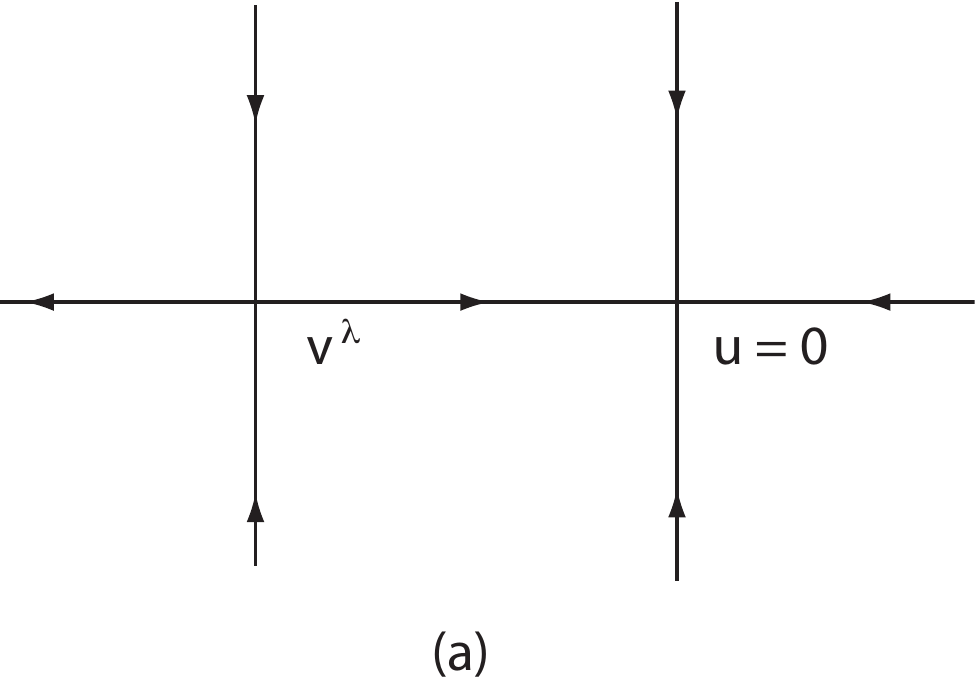}
   \includegraphics[width=0.22\textwidth]{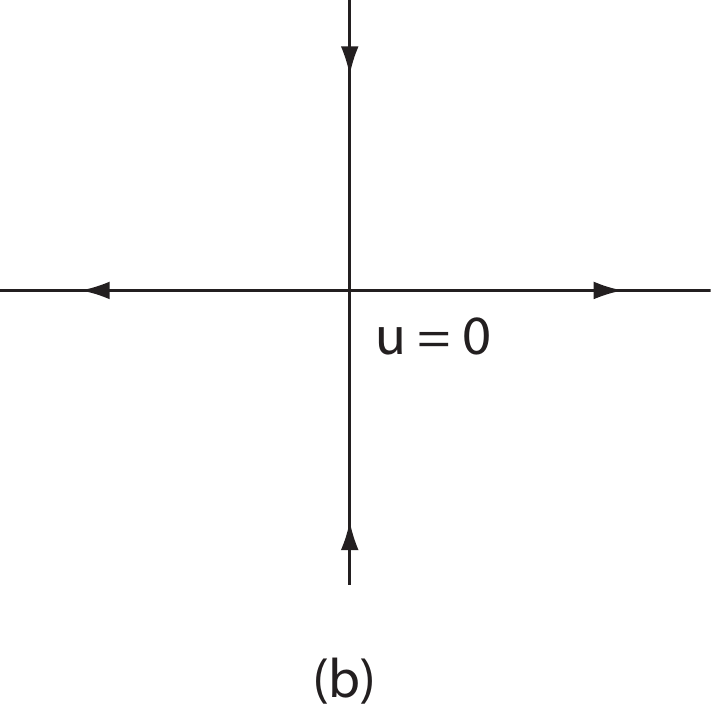} 
   \includegraphics[width=0.32\textwidth]{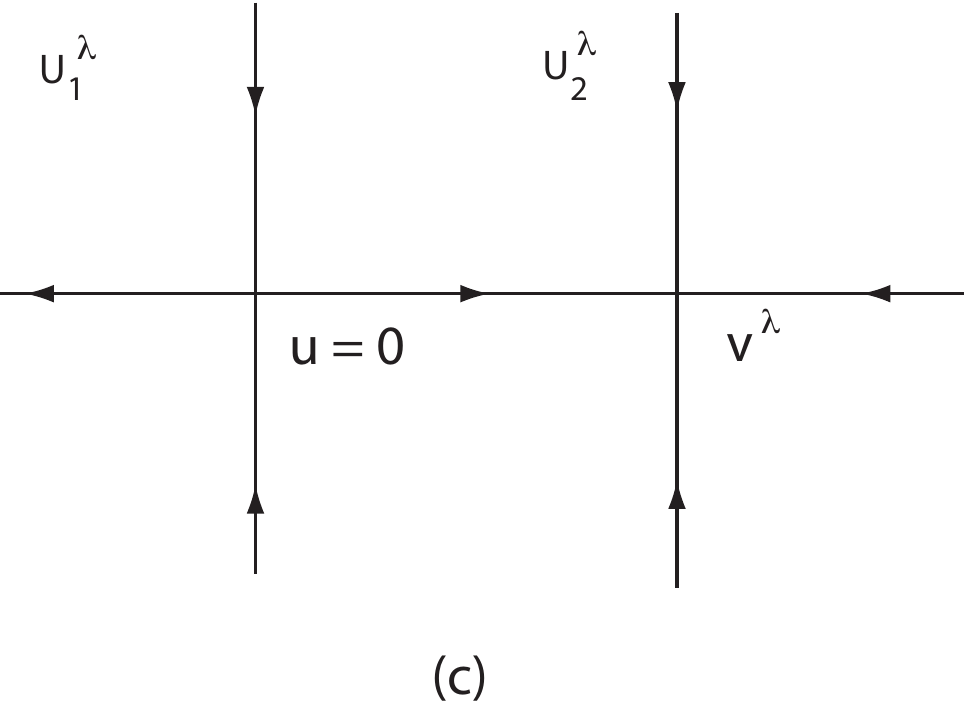}
  \caption{Topological structure of the mixing transition of
(\ref{5.1}) when $k$=even and $\alpha\neq 0$: (a) $\lambda
<\lambda_0$; (b) $\lambda =\lambda_0$; (c) $\lambda >\lambda_0$. Here
$U^{\lambda}_1$ is the unstable domain, and $U^{\lambda}_2$ the
stable domain.}\la{f5.7}
 \end{figure}

\bt\la{t5.8}
Assume  (\ref{5.4})  and (\ref{5.5}) with $m=1$, and (\ref{5.36}).  If $k$=odd and $\alpha\neq 0$ in (\ref{5.36}) then
the following assertions hold true:

\begin{itemize}

\item[(1)] If $\alpha >0$,  then (\ref{5.1}) has a jump
transition from $(0,\lambda_0)$, and bifurcates on $\lambda
<\lambda_0$ to exactly two saddle points $v^{\lambda}_1$ and
$v^{\lambda}_2$ with the Morse index one, as shown in Figure \ref{f5.5}.

\item[(2)] If $\alpha <0$,  then (\ref{5.1}) has a continuous
transition from $(0,\lambda_0)$, which is an attractor bifurcation
 as shown in Figure \ref{f5.6}. 

\item[(3)] The bifurcated singular points $v^{\lambda}_1$ and $v^{\lambda}_2$ 
in the above cases can
be expressed in the following form
$$v^{\lambda}_{1,2}=\pm |\beta_1(\lambda )/\alpha
|^{{1}/{k-1}}e_1(\lambda )+o(|\beta_1|^{{1}/{k-1}}).$$

\end{itemize}
\et

\bt\la{t5.9}
 Assume  (\ref{5.4})  and (\ref{5.5}) with $m=1$,  and
(\ref{5.36}). If $k$=even and $\alpha\neq 0$, then we have the
following assertions:

\begin{enumerate}

\item (\ref{5.1}) has a mixed transition from
$(0,\lambda_0)$. More precisely, there exists a neighborhood
$U\subset X$ of $u=0$ such that $U$ is separated into two disjoint
open sets $U^{\lambda}_1$ and $U^{\lambda}_2$ by the stable
manifold $\Gamma_{\lambda}$ of $u=0$ satisfying the following properties:

\begin{enumerate}

\item $U=U^{\lambda}_1+U^{\lambda}_2+\Gamma_{\lambda}$,

\item the transition in $U^{\lambda}_1$ is jump, and 

\item the transition in $U^{\lambda}_2$ is
continuous. The local transition structure is as shown in Figure \ref{f5.7}.

\end{enumerate}

\item (\ref{5.1}) bifurcates in $U^{\lambda}_2$ to a unique
singular point $v^{\lambda}$ on $\lambda >\lambda_0$, which is an
attractor such that for any $\varphi\in U^{\lambda}_2$, 
$$\lim\limits_{t\rightarrow\infty}\|u(t,\varphi
)-v^{\lambda}\|_X=0,$$
where $u(t,\varphi )$ is the solution of (\ref{5.1}). 

\item (\ref{5.1})\ bifurcates on $\lambda <\lambda_0$ to a unique saddle
point $v^{\lambda}$ with the Morse index one. 

\item The bifurcated singular point $v^{\lambda}$ can be expressed as
$$v^{\lambda}=-(\beta_1(\lambda )/\alpha
)^{{1}/{(k-1)}}e_1+o(|\beta_1|^{{1}/{(k-1)}}).$$
\end{enumerate}
\et

\subsection{Singular Separation}
\la{s6.2}
In this section, we study  an important
problem associated with the discontinuous transition of
(\ref{5.1}), which we call  the singular separation.

\bd\la{d6.1}
\begin{enumerate}

\item 
An invariant set $\Sigma$ of (\ref{5.1}) is called a singular
element if $\Sigma$ is either a singular point or a periodic
orbit. 

\item Let $\Sigma_1\subset X$ be a singular
element of (\ref{5.1}) and $U\subset X$ a neighborhood of
$\Sigma_1$. We say that (\ref{5.1}) has a singular separation of
$\Sigma$ at $\lambda =\lambda_1$ if 

\begin{enumerate}

\item (\ref{5.1}) has no singular
elements in $U$ as $\lambda <\lambda_1$ (or $\lambda >\lambda_1$),
and generates a singular element $\Sigma_1\subset  U$ at $\lambda
=\lambda_1$,  and 

\item there are branches of singular elements
$\Sigma_{\lambda}$, which are  separated from $\Sigma_1$ for $\lambda
>\lambda_1$ (or $\lambda <\lambda_1$), i.e.,
$$\lim\limits_{\lambda\rightarrow\lambda_1}\max_{x\in\Sigma_{\lambda}}\text{dist}(x,\Sigma_1)=0.$$
\end{enumerate}

\end{enumerate}
\ed

A special case of singular separation is the saddle-node
bifurcation defined as follows.

\bd\la{d6.2}
Let $u_1\in X$ be a singular point of
(\ref{5.1}) at $\lambda =\lambda_1$ with $u_1\neq 0$. We say that
(\ref{5.1}) has a saddle-node bifurcation at $(u_1,\lambda_1)$ if

\begin{enumerate}

\item the index of $L_{\lambda}+G$ at $(u_1,\lambda_1)$ is zero, i.e.,
ind$(-(L_{\lambda_1}+G),u_1)=0$, 

\item  there are two branches
$\Gamma_1(\lambda )$ and $\Gamma_2(\lambda )$ of singular points
of (\ref{5.1}), which  are separated from $u_1$ for $\lambda >\lambda_1$
(or $\lambda <\lambda_1)$,  i.e.,  for any
$u_{\lambda}\in\Gamma_i(\lambda )$ $(i=1,2)$ we have
$$u_{\lambda}\rightarrow u_1\ \text{in}\ X\ \ \ \ \text{as}\
\lambda\rightarrow\lambda_1, $$ 
and

\item  the indices of
$u^i_{\lambda}\in\Gamma_i(\lambda )$ are as follows
$$\text{ind}(-(L_{\lambda}+G),u_{\lambda})=
\left\{
\begin{aligned}
&  1  &&  \text{ if } u_{\lambda}\in\Gamma_2(\lambda ),\\
&  -1  &&  \text{ if } u_{\lambda}\in\Gamma_1(\lambda ).
\end{aligned}
\right.$$
\end{enumerate}
\ed

Intuitively, the saddle-node bifurcation is
schematically shown as in Figure~\ref{f6.1}, where the singular points in
$\Gamma_1(\lambda )$ are saddle points and in $\Gamma_2(\lambda )$
are nodes, and the singular separation of periodic orbits is as in
shown Figure~\ref{f6.2}.
\begin{figure}[hbt]
  \centering
  \includegraphics[width=0.35\textwidth]{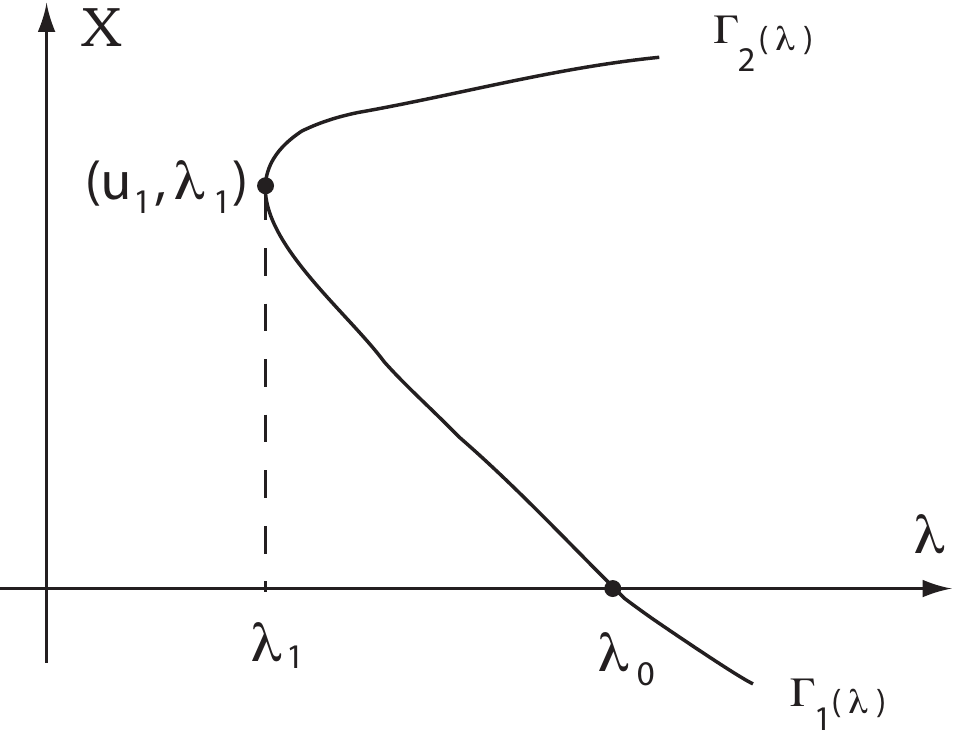}
  \caption{Saddle-node bifurcation.}\la{f6.1}
 \end{figure}
\begin{figure}[hbt]
  \centering
  \includegraphics[width=0.35\textwidth]{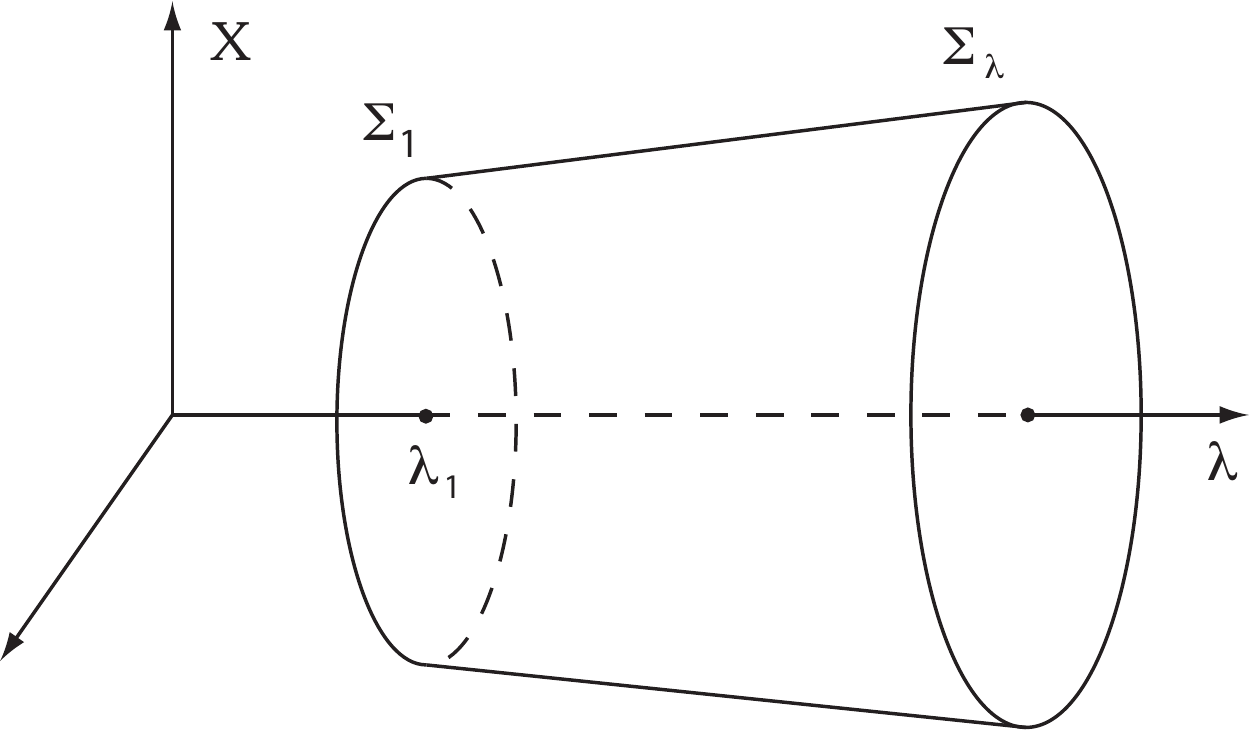}
  \caption{Singular separation of periodic orbits.}\la{f6.2}
 \end{figure}

For the singular separation we can give a general principle as
follows, which provides a basis for singular separation theory.

\bt\la{t6.4}
Let the conditions (\ref{5.4}) and
(\ref{5.5}) hold true. Then we have the following assertions.

\begin{enumerate}

\item[(1)] If (\ref{5.1}) bifurcates from $(u,\lambda
)=(0,\lambda_0)$   to a branch $\Sigma_{\lambda}$ of singular elements on
$\lambda <\lambda_0$ which is bounded in $X\times (-\infty
,\lambda_0)$ then (\ref{5.1}) has a singular separation of
singular elements at some $(\Sigma_0,\lambda_1)\subset X\times
(-\infty ,\lambda_0)$. 

\item[(2)] If the bifurcated branch
$\Sigma_{\lambda}$ consists of singular points which has index $-1$,
i.e., 
$$\text{ind}(-(L_{\lambda}+G),u_{\lambda})=-1 \ \ \ \ \forall u_{\lambda}\in
E_{\lambda},\ \ \ \ \lambda <\lambda_0,$$ then the singular
separation is a saddle-node bifurcation from some
$(u_1,\lambda_1)\in X\times (-\infty ,\lambda_0).$
\end{enumerate}
\et

We consider the equation (\ref{5.1}) defined on the Hilbert spaces
$X=H, X_1=H_1$. Let $L_{\lambda}=-A+\lambda B$. For $L_{\lambda}$
and $G(\cdot ,\lambda ):H_1\rightarrow H$, we assume that
$A:H_1\rightarrow H$ is symmetric, and
\begin{eqnarray}
&&<Au,u>_H\geq c\|u\|^2_{H_{{1}/{2}}},\label{6.46}\\
&&<Bu,u>_H\geq c\|u\|^2_H,\label{6.47}\\
&&<Gu,u>_H\leq -c_1\|u\|^p_H+c_2\|u\|^2_H,\label{6.48}
\end{eqnarray}
where $p>2, c, c_1, c_2>0$ are constants.

\bt\la{t6.5}
 Assume the conditions (\ref{5.3}),
(\ref{5.4}) and (\ref{6.46})-(\ref{6.48}), then (\ref{5.1}) has a
transition at $(u,\lambda )=(0,\lambda_0)$, and the following
assertions hold true:

\begin{enumerate}
\item[(1)] If $u=0$ is an even-order nondegenerate singular point
of $L_{\lambda}+G$ at $\lambda =\lambda_0$, then (\ref{5.1}) has a
singular separation of singular points at some $(u_1,\lambda_1)\in
H\times (-\infty ,\lambda_0)$. 

\item[(2)]  If $m=1$ and $G$
satisfies (\ref{5.36}) with $\alpha >0$ if $k$=odd and $\alpha\neq
0$ if $k$=even, then (\ref{5.1}) has a saddle-node bifurcation at
some singular point $(u_1,\lambda_1)$ with $\lambda_1<\lambda_0$.
\end{enumerate}
\et

\subsection{Transition and Singular Separation of Perturbed Systems}
We consider the following perturbed equation of (\ref{5.1}): 
\begin{equation}
\frac{du}{dt}=(L_{\lambda}+S^{\varepsilon}_{\lambda})u+G(u,\lambda
)+T^{\varepsilon}(u,\lambda ),\label{6.56}
\end{equation}
where $L_{\lambda}$ and $G_{\lambda}$ are as in (\ref{5.1}),
$S^{\varepsilon}_{\lambda}:X_{\sigma}\rightarrow X$ is a linear
perturbed operator,
$T^{\varepsilon}_{\lambda}:X_{\sigma}\rightarrow X$ a $C^1$
nonlinear perturbed operator,  and  $X_{\sigma}$ the fractional order
space, $0\leq\sigma <1$. Also assume that 
$G_{\lambda},T^{\varepsilon}_{\lambda}$ are
$C^3$ on $u$, and
\begin{equation}
\left.
\begin{aligned} 
& \|S^{\varepsilon}_{\lambda}\|<\varepsilon,\\
& \|T^{\varepsilon}_{\lambda}\|<\varepsilon ,\\
& T^{\varepsilon}(u,\lambda )=o(\|u\|_{X_{\alpha}}).
\end{aligned}
\right.\label{6.57}
\end{equation}

Let  (\ref{5.4})  and (\ref{5.5}) with $m=1$ hold true, $G(u,\lambda )=G_2(u,\lambda
)+o(\|u\|^2_{X_1})$, where $G_2(\cdot ,\lambda )$ is a bilinear
operator, and
\begin{equation}
b=<G_2(e,\lambda_0),e^*>\neq 0,\label{6.65}
\end{equation}
where $e\in X$ and $e^*\in X^*$ are the eigenvectors of
$L_{\lambda}$ and $L^*_{\lambda}$ corresponding to
$\beta_1(\lambda )$ at $\lambda =\lambda_0$ respectively.

We now consider the transition associated with the saddle-node
bifurcation of the perturbed system (\ref{6.56}). Let $h(x,\lambda
)$ be the center manifold function of (\ref{5.1}) near $\lambda
=\lambda_0$. Assume that
\begin{equation}
<G(xe+h(x,\lambda_0),\lambda_0),e^*>=b_1x^3+o(|x|^3),\label{6.66}
\end{equation}
where $b_1\neq 0$, and $e$ and $e^*$ are as in (\ref{6.65}).

Then we have the following theorems.

\bt\la{t6.10}
 Let the conditions (\ref{5.4})  and (\ref{5.5}) with $m=1$, and
(\ref{6.66}) hold true, and $b_1<0$. Then there is an
$\varepsilon >0$ such that if  $S^{\varepsilon}_{\lambda}$ and
$T^{\varepsilon}_{\lambda}$ satisfy (\ref{6.57}), then the transition of
(\ref{6.56}) is either continuous or mixed. If the transition is
continuous, then Assertions (2) and (3) of Theorem \ref{t5.8} are valid
for (\ref{6.56}). If the transition is mixed, then the
following assertions hold true:

\begin{enumerate}

\item[(1)] (\ref{6.56}) has a saddle-node bifurcation at some
point $(u_1,\lambda_1)\in X\times (-\infty
,\lambda^{\varepsilon}_0)$, and there are exactly two branches
$$\Gamma^{\lambda}_i=\{(u^{\lambda}_i,\lambda )|\ \lambda_1<\lambda
<\lambda^{\varepsilon}_0+\delta\} \qquad   i=1,2, $$ 
separated from
$(u_1,\lambda_1)$ as shown in Figure~\ref{f6.1}, which satisfy that 
\begin{align*}
&  \|u^{\lambda}_2\|_X\neq 0  &&\forall (u^{\lambda}_2,\lambda
)\in\Gamma^{\lambda}_2,\ \ \ \ \lambda_1<\lambda
<\lambda^{\varepsilon}_0+\delta ,\\
& \lim_{\lambda\rightarrow\lambda^{\varepsilon}_0}\|u^{\lambda}_1\|_X=0 &&
\forall  (u^{\lambda}_1,\lambda )\in\Gamma^{\lambda}_1.
\end{align*}

\item[(2)] There is a neighborhood $U\subset X$ of
$u=0$, such that for each $\lambda$ with $\lambda_1<\lambda
<\lambda^{\varepsilon}_0+\delta$ and
$\lambda\neq\lambda^{\varepsilon}_0, U$ contains only two
nontrivial singular points $u^{\lambda}_1$ and $u^{\lambda}_2$ of
(\ref{6.56}). 

\item[(3)] For each $\lambda_1<\lambda
<\lambda^{\varepsilon}_0+\delta , U$ can be decomposed into two
open sets $\bar{U}=\bar{U}^{\lambda}_1+\bar{U}^{\lambda}_2$ with
$U^{\lambda}_1\cap U^{\lambda}_2=\emptyset$, such that

\begin{enumerate}

\item if $  \lambda_1<\lambda <\lambda^{\varepsilon}_0$, 
$$0\in U^{\lambda}_1,\ \ \ \ u^{\lambda}_2\in U^{\lambda}_2,\ \ \
\ u^{\lambda}_1\in\partial U^{\lambda}_1\cap\partial
U^{\lambda}_2,$$ with $u=0$ and $u^{\lambda}_2$ being
attractors which attract $U^{\lambda}_1$ and $U^{\lambda}_2$
respectively, and

\item if  $ \lambda^{\varepsilon}_0<\lambda <\lambda^{\varepsilon}_0+\delta$,
$$u^{\lambda}_1\in U^{\lambda}_1,\ \ \ \ u^{\lambda}_2\in
U^{\lambda}_2,\ \ \ \ 0\in\partial U^{\lambda}_1\cap\partial
U^{\lambda}_2, $$ 
with $u^{\lambda}_1$ and
$u^{\lambda}_2$ being attractors which attract $U^{\lambda}_1$ and
$U^{\lambda}_2$ respectively. 

\end{enumerate}

\item[(4)] Near $(u,\lambda)=(0,\lambda^{\varepsilon}_0),u^{\lambda}_1$ 
and $u^{\lambda}_2$
can be expressed as 
\begin{equation} 
\left.
\begin{aligned}
& u^{\lambda}_1=\alpha_1(\lambda ,\varepsilon )e+o(|\alpha_1|),\\
& u^{\lambda}_2=\alpha_2(\lambda ,\varepsilon )e+o(|\alpha_2|),\\
& \lim_{  \lambda\rightarrow\lambda^{\varepsilon}_0  }  
\alpha_1(\lambda ,\varepsilon )= 0,\\
& \alpha_2(\lambda^{\varepsilon}_0,\varepsilon )\neq 0,
\end{aligned}
\right.\label{6.67}
\end{equation}
where $e$ is as in (\ref{6.66}).
\end{enumerate}
\et

\bt\la{t6.11} Assume the conditions (\ref{5.4})  and (\ref{5.5}) with $m=1$, and
(\ref{6.66}) with $b_1>0$. Then, there is an $\varepsilon >0$ such
that when $S^{\varepsilon}_{\lambda}$ and
$T^{\varepsilon}_{\lambda}$ satisfy (\ref{6.57}), the transition of
(\ref{6.56}) is either jump or mixed. If it is jump transition,  then
Assertions (1) and  (3) of Theorem \ref{t5.8} are valid for (\ref{6.56}). 
If it is mixed, then the following assertions hold true:

\begin{enumerate}

\item[(1)] (\ref{6.56}) has a saddle-node bifurcation at some
point $(u_1,\lambda_1)\in X\times (\lambda^{\varepsilon}_0,+\infty
)$, and there are exactly two branches
$$\Gamma^{\lambda}_i=\{(u^{\lambda}_i,\lambda )|\
\lambda^{\varepsilon}_0-\delta <\lambda <\lambda_1\} \qquad   (i=1,2), $$
separated from $(u_1,\lambda_1)$, which satisfy
\begin{align*}
&  \|u^{\lambda}_2\|_X=0 &&\forall (u^{\lambda}_2,\lambda)\in\Gamma^*_2,\ \ \ \ \lambda^{\varepsilon}_0-\varepsilon
<\lambda <\lambda_1,\\
& \lim\limits_{\lambda\rightarrow\lambda^{\varepsilon}_0}\|u^{\lambda}_1\|_X=0
&& \forall 
(u^{\lambda}_1,\lambda )\in\Gamma^{\lambda}_1.
\end{align*}

\item[(2)] There is a neighborhood $U\subset X$ of
$u=0$, such that for each $\lambda$ with
$\lambda^{\varepsilon}_0-\delta <\lambda <\lambda_1, U$ contains
only two nontrivial singular points $u^{\lambda}_1$ and
$u^{\lambda}_2$ of (\ref{6.56}).

\item[(3)] For every
$\lambda^{\varepsilon}_0-\delta <\lambda <\lambda_1, U$ can be
decomposed into three open sets
$\bar{U}=\bar{U}_0+\bar{U}_1+\bar{U}_2$ with $U_i\cap
U_j=\emptyset$ $ (i\neq j)$ such that

\begin{enumerate}

\item  if   $ \lambda^{\varepsilon}_0-\delta <\lambda <\lambda^{\varepsilon}_0$, 
then 
$$u=0\in U^{\lambda}_0,\ \ \ \ u^{\lambda}_i\in\partial
U^{\lambda}_i\cap\partial U^{\lambda}_0(i=1,2),$$ 
with $u=0$ being an attractor which
attracts $U^{\lambda}_0$ and $U^{\lambda}_i(i=1,2)$ two saddle
points with the Morse index one, and

\item  if   $ \lambda^{\varepsilon}_0<\lambda <\lambda_1$,  then 
$$u^{\lambda}_1\in U^{\lambda}_1,\ \ \ \ u^{\lambda}_2\in\partial
U^{\lambda}_2\cap\partial U^{\lambda}_1,\ \ \ \ 0\in\partial
U^{\lambda}_1\cap\partial U^{\lambda}_0,$$ 
with $u^{\lambda}_1$
being an attractor which attracts $U^{\lambda}_1$ and
$u^{\lambda}_2$ and $u=0$ being saddle points with the Morse index
one. 

\end{enumerate}

\item[(4)] Near $(0,\lambda^0_{\varepsilon}),u^{\lambda}_1$
and $u^{\lambda}_2$ can be expressed by (\ref{6.67}).
\end{enumerate}
\et

\section{Ginzburg-Landau Models}
\label{s7.2.2}
In this subsection, we introduce the time-dependent Ginzburg-Landau model for equilibrium phase transitions. 

We start with thermodynamic potentials and the Ginzburg-Landau free energy. As we know, four thermodynamic potentials-- internal energy, the enthalpy, the Helmholtz free energy and the Gibbs free energy--are useful in the chemical thermodynamics of reactions and non-cyclic processes. 

Consider a thermal system, its order parameter $u$ changes in
$\Omega\subset \R^n$ $(1\leq n\leq 3)$. In this situation,  the free energy of this system is of the form
\begin{equation}
{\mathcal{H}}(u,\lambda
)= {\mathcal{H}}_0 
 +\int_{\Omega}\Big[\frac{1}{2}\sum^m_{i=1}\mu_i|\nabla
u_i|^2  +g(u,\nabla u,\lambda )\Big]dx\label{7.28}
\end{equation}
where $N\geq 3$ is an integer,
$u=(u_1,\cdots,u_m),\mu_i=\mu_i(\lambda )>0$, and $g(u,\nabla
u,\lambda )$ is a $C^r(r\geq 2)$ function of $(u,\nabla u)$ with
the Taylor expansion \begin{equation} g(u,\nabla u,\lambda
)=\sum\alpha_{ijk}u_iD_ju_k+\sum^N_{|I|=1}\alpha_Iu^I+o(|u|^N)-fX,\label{7.29}
\end{equation}
where $I=(i_1,\cdots,i_m),i_k\geq 0$ are integer,
$|I|=\sum^m_{k=1}i_k$, the coefficients $\alpha_{ijk}$ and
$\alpha_I$ continuously depend on $\lambda$, which are determined
by the concrete physical problem, $u^I=u^{i_1}_1\cdots u^{im}_m$
and $fX$ the generalized work.

A thermal system is controlled by some parameter $\lambda$. When
$\lambda$ is far from the critical point $\lambda_0$ the system
lies on a stable equilibrium state $\Sigma_1$, and when $\lambda$
reaches or exceeds $\lambda_0$ the state $\Sigma_1$ becomes unstable, and 
meanwhile the system will undergo a transition  from $\Sigma_1$ to another stable
state $\Sigma_2$. The basic principle is that there often exists  fluctuations in the system leading  to a deviation from the equilibrium states, and  the phase transition process is a
dynamical behavior, which should be described by a time-dependent
equation.

To derive a general time-dependent model, first we recall that  the classical  
le Ch\^atelier  principle amounts to saying that 
for a stable  equilibrium state of a system $\Sigma$, when the system deviates from
$\Sigma$ by a small perturbation or fluctuation, there will be a
resuming force to restore this system to return to the stable state
$\Sigma$.
Second, we know that  a stable equilibrium state of a thermal system must
be the minimal value point of the thermodynamic potential. 

By the mathematical characterization of gradient systems and the le Ch\^atelier principle, for a system with
thermodynamic potential ${\mathcal{H}}(u,\lambda )$, the governing
equations are essentially determined by the functional
${\mathcal{H}}(u,\lambda )$.
When the order parameters $(u_1,\cdots,u_m)$ are nonconserved
variables, i.e., the integers
$$\int_{\Omega}u_i(x,t)dx=a_i(t)\neq\text{constant}.$$
then the time-dependent equations are given by
\begin{equation}
\left.
\begin{aligned} 
&\frac{\partial u_i}{\partial
t}=-\beta_i\frac{\delta}{\delta u_i}{\mathcal{H}}(u,\lambda
)+\Phi_i(u,\nabla u,\lambda ),\\
&\frac{\partial u}{\partial n}|_{\partial\Omega}=0\ \ \ \
(\text{or}\ u|_{\partial\Omega}=0),\\
&u(x,0)=\varphi (x),
\end{aligned}
\right.\label{7.30}
\end{equation}
for any $1 \le i \le m$, 
where $\delta /\delta u_i$ are the variational derivative,
$\beta_i>0$ and $\Phi_i$ satisfy
\begin{equation}
\int_{\Omega}\sum_i\Phi_i\frac{\delta}{\delta
u_i}{\mathcal{H}}(u,\lambda )dx=0.\label{7.31}
\end{equation}
The condition (\ref{7.31})  is  required by
the Le Ch\^atelier principle. In the concrete problem, the terms
$\Phi_i$ can be determined by physical laws and (\ref{7.31}).

When the order parameters are the number density and the system
has no material exchange with the external, then $u_j$  $(1\leq j\leq
m)$ are conserved, i.e.,
\begin{equation}
\int_{\Omega}u_j(x,t)dx=\text{constant}.\label{7.32}
\end{equation}
This conservation law requires a continuous equation
\begin{equation}
\frac{\partial u_j}{\partial t}=-\nabla\cdot J_j(u,\lambda
),\label{7.33}
\end{equation}
where $J_j(u,\lambda )$ is the flux of component $u_j$. In
addition, $J_j$ satisfy
\begin{equation}
J_j=-k_j\nabla (\mu_j-\sum_{i\neq j}\mu_i),\label{7.34}
\end{equation}
where $\mu_l$ is the chemical potential of component $u_l$, 
\begin{equation}
\mu_j-\sum_{i\neq j}\mu_i=\frac{\delta}{\delta
u_j}{\mathcal{H}}(u,\lambda )-\phi_j(u,\nabla u,\lambda
), \label{7.35}
\end{equation}
and  $\phi_j(u,\lambda )$ is a function depending on the other
components $u_i$ $(i\neq j)$. When $m=1$, i.e., the system consists of
two components $A$ and $B$, this term $\phi_j=0$. Thus, from
(\ref{7.33})-(\ref{7.35}) we obtain the dynamical equations as
follows
\begin{equation}
\left.
\begin{aligned} &\frac{\partial u_j}{\partial
t}=\beta_j\Delta\left[\frac{\delta}{\delta
u_j}{\mathcal{H}}(u,\lambda )-\phi_j(u,\nabla u,\lambda )\right],\\
&\frac{\partial u}{\partial n}|_{\partial\Omega}=0,\ \ \ \
\frac{\partial\Delta u}{\partial n}|_{\partial\Omega}=0,\\
&u(x,0)=\varphi (x),
\end{aligned}
\right.\label{7.36}
\end{equation}
for $1 \le j \le m$, 
where $\beta_j>0$ are constants, $\phi_j$ satisfy
\begin{equation}
\int_{\Omega}\sum_j\Delta\phi_j\cdot\frac{\delta}{\delta
u_j}{\mathcal{H}}(u,\lambda )dx=0.\label{7.37}
\end{equation}

If the order parameters $(u_1,\cdots,u_k)$ are coupled to the
conserved variables $(u_{k+1},\cdots,u_m)$, then the dynamical
equations are
\begin{equation}
\left.
\begin{aligned} 
&\frac{\partial u_i}{\partial t}
   =-\beta_i\frac{\delta}{\delta u_i}{\mathcal{H}}(u,\lambda)+\Phi_i(u,\nabla u,\lambda ),\\
& \frac{\partial u_j}{\partial t}
  =\beta_j\Delta\left[\frac{\delta}{\delta u_j}{\mathcal{H}}(u,\lambda )
    -\phi_j(u,\nabla u,\lambda )\right],\\
&\frac{\partial u_i}{\partial n}|_{\partial\Omega}=0\ \ \ \
(\text{or}\ u_i|_{\partial\Omega}=0),\\
&\frac{\partial u_j}{\partial n}|_{\partial\Omega}=0,\ \ \ \
\frac{\partial\Delta u_j}{\partial n}|_{\partial\Omega}=0,\\
&u(x,0)=\varphi (x).
\end{aligned}
\right.\label{7.38}
\end{equation}
for $1 \le i \le k$  and $k+1 \le j \le m$.

The model (\ref{7.38}) gives a general form of the governing
equations to thermodynamic phase transitions.  Hence, the dynamics of
equilibrium phase transition in statistic physics is based on the new Ginzburg-Landau formulation  (\ref{7.38}).

Physically, the initial value condition $u(0)=\varphi$ in
(\ref{7.38}) stands for the fluctuation of system or perturbation
from the external. Hence, $\varphi$ is generally small. However,
we can not exclude the possibility of a bigger noise $\varphi$.

\bibliographystyle{siam}
\def\cprime{$'$}

\end{document}